\documentclass[]{style}

\definecolor{lastauthor}{RGB}{143, 68, 115}

\title{Dive into Claude Code: The Design Space of Today's and Future AI Agent Systems}

\author[1,*]{{Jiacheng Liu}}
\author[1,*]{{Xiaohan Zhao}}
\author[1,2]{{Xinyi Shang}}
\author[1,\dagger]{{Zhiqiang Shen}}

\affiliation[1]{VILA Lab, Mohamed bin Zayed University of Artificial Intelligence}
\affiliation[2]{University College London}

\contribution[*]{These authors contributed equally to this work}
\contribution[\dagger]{Corresponding author}

\usepackage[table]{xcolor}

\usepackage{amsmath}
\usepackage{amssymb}
\usepackage{graphicx}
\usepackage{booktabs}
\usepackage{xcolor}
\usepackage{hyperref}
\usepackage{cleveref}
\usepackage{natbib}
\usepackage{enumitem}
\usepackage{multirow}
\usepackage{subcaption}
\usepackage{microtype}
\usepackage{tabularx}

\usepackage{makecell}
\newcommand{\eg}{\textit{e.g.}}

\newcommand{\code}[1]{\texttt{#1}}
\newcommand{\file}[1]{\texttt{#1}}
\newcommand{\func}[1]{\texttt{#1()}}

\newcommand{\tierA}{\textsc{Tier~A}}
\newcommand{\tierB}{\textsc{Tier~B}}
\newcommand{\tierC}{\textsc{Tier~C}}

\graphicspath{{resources/figures/}}

\hypersetup{
  colorlinks=true,
  linkcolor=blue!70!black,
  citecolor=green!50!black,
  urlcolor=blue!70!black,
}

\abstract{
Claude Code is an agentic coding tool that can run shell commands, edit files, and call external services on behalf of the user.
This study describes its comprehensive architecture by analyzing the publicly available TypeScript source code\footnote{v2.1.88, \href{https://github.com/chauncygu/collection-claude-code-source-code?tab=readme-ov-file\#1-original-source-code}{link}. {\bf Disclaimer}: All materials used in this work are obtained from publicly available online sources. We have not used any private, confidential, or unauthorized materials, and we do not intend to infringe any copyright or intellectual property rights. The original intellectual property rights to the source code belong to Anthropic.} and further comparing it with two independent open-source AI agent systems, \texttt{OpenClaw} and \texttt{Hermes Agent}, that answer many similar or even the same design questions from different deployment contexts.
Our analysis identifies five human values, philosophies, and needs that motivate the architecture: human decision authority, safety, security, and privacy, reliable execution, capability amplification, and contextual adaptability. We then trace them through thirteen design principles to specific implementation choices.
The core of the system is a simple while-loop that calls the model, runs tools, and repeats.
Most of the code, however, lives in the systems around this loop: a permission system with seven modes and an ML-based classifier, a five-layer compaction pipeline for context management, four extensibility mechanisms (MCP, plugins, skills, and hooks), a subagent delegation and orchestration mechanism, and append-oriented session storage.
Comparisons with OpenClaw (a multi-channel personal assistant gateway) and Hermes Agent (a single-process, multi-surface assistant) show that the same design questions produce different answers across three deployment contexts. Claude Code emphasizes per-action safety, OpenClaw emphasizes perimeter-level access, and Hermes renders per-action approvals across many surfaces. At the runtime layer, Claude Code uses a single CLI loop, OpenClaw embeds the runtime within a gateway control plane, and Hermes uses one process whose role is set by its entry point. At the context and extension layer, Claude Code extends the context window, OpenClaw registers gateway-wide capabilities, and Hermes provides pluggable memory and model backends. We finally identify six open design directions for future agent systems, grounded in recent empirical, architectural, and policy literature. Our GitHub is available at: \url{https://github.com/VILA-Lab/Dive-into-Claude-Code}.
}

\correspondence{Zhiqiang Shen (\email{Zhiqiang.Shen@mbzuai.ac.ae})}

\usepackage[table]{xcolor}  

\usepackage{epigraph}
\usepackage[normalem]{ulem}
\definecolor{quotemark}{gray}{0.7}
\makeatletter
\def\fquote{%
    \@ifnextchar[{\fquote@i}{\fquote@i[]}%]
           }%
\def\fquote@i[#1]{%
    \def\tempa{#1}%
    \@ifnextchar[{\fquote@ii}{\fquote@ii[]}%]
                 }%
\def\fquote@ii[#1]{%
    \def\tempb{#1}%
    \@ifnextchar[{\fquote@iii}{\fquote@iii[]}%]
                      }%
\def\fquote@iii[#1]{%
    \def\tempc{#1}%
    \vspace{1em}%
    \noindent%
    \begin{list}{}{%
         \setlength{\leftmargin}{0.05\textwidth}%
         \setlength{\rightmargin}{0.05\textwidth}%
                  }%
         \item[]%
         \begin{picture}(0,0)%
         \put(-8,-5){\makebox(0,0){\scalebox{2}{\textcolor{quotemark}{``}}}}%
         \end{picture}%
         \begingroup\itshape}%
 \def\endfquote{%
 \endgroup\par%
 \makebox[0pt][l]{%
 \hspace{0.27\textwidth}%
 \begin{picture}(0,0)(0,0)%
 \put(60,20){\makebox(0,0){%
 \scalebox{2}{\color{quotemark}''}}}%
 \end{picture}}%
 \ifx\tempa\empty%
 \else%
    \ifx\tempc\empty%
       \hfill\rule{110pt}{0.5pt}\\\mbox{}\hfill\tempa,\ \emph{\tempb}%
   \else%
       \hfill\rule{100pt}{0.5pt}\\\mbox{}\hfill\tempa,\ \emph{\tempb},\ \tempc%
   \fi\fi\par%
   \vspace{0.5em}%
 \end{list}%
 }%
 \makeatother
\newcommand*\circled[1]{\tikz[baseline=(char.base)]{\node[shape=circle,draw,inner sep=.3pt] (char) {#1};}}

\definecolor{orange}{RGB}{178,92,35}
\definecolor{green1}{RGB}{95,145,51}

\definecolor{red1}{RGB}{197,64,57}
\definecolor{blue1}{RGB}{59,130,220}
\definecolor{green2}{RGB}{82,181,150}
\definecolor{purple1}{RGB}{105,93,223}
\definecolor{orange1}{RGB}{164, 47, 54}
\definecolor{green3}{RGB}{94,145,51}

\definecolor{c_step1}{RGB}{60, 105, 199}
\definecolor{c_step2}{RGB}{159,37,28}
\definecolor{c_step3}{RGB}{107,40,157}

\newenvironment{talign*}
{\csname align*\endcsname}
{\endalign}

\usepackage[utf8]{inputenc}         %
\usepackage[T1]{fontenc}            %
\usepackage{url}                    %
\usepackage{booktabs}               %
\usepackage{amsfonts}               %
\usepackage{nicefrac}               %
\usepackage{microtype}              %
\usepackage{algorithm}
\usepackage{algorithmic}
\usepackage{graphicx}
\usepackage{subcaption}
\usepackage[flushleft]{threeparttable}
\usepackage{float}
\usepackage{multirow}
\usepackage{xspace}
\usepackage{enumitem}
\usepackage[font=small]{caption}
\usepackage{autobreak}
\usepackage{sidecap}
\usepackage{wrapfig}
\usepackage{bbding}
\usepackage[toc, page, header]{appendix}
\usepackage{tikz}
\usepackage{pifont}
\usepackage{mdframed}
\usepackage{colortbl}

\definecolor{coral}{RGB}{255,127,80}
\definecolor{darkgreen}{RGB}{0,100,0}
\definecolor{darkyellow}{RGB}{204,153,0}
\definecolor{salmon}{RGB}{250,128,114}
\definecolor{darkred}{RGB}{150,0,0}
\newcommand{\darkredtext}[1]{{\color{darkred}#1}}

\newcommand{\secref}[1]{\hyperref[#1]{\darkredtext{Sec.~\ref*{#1}}}}
\newcommand{\thmref}[1]{\hyperref[#1]{\darkredtext{Thm.~\ref*{#1}}}}
\newcommand{\defref}[1]{\hyperref[#1]{\darkredtext{Def.~\ref*{#1}}}}
\newcommand{\propref}[1]{\hyperref[#1]{\darkredtext{Prop.~\ref*{#1}}}}
\newcommand{\assumpref}[1]{\hyperref[#1]{\darkredtext{Assump.~\ref*{#1}}}}
\newcommand{\remarkref}[1]{\hyperref[#1]{\darkredtext{Rem.~\ref*{#1}}}}
\newcommand{\hypref}[1]{\hyperref[#1]{\darkredtext{Hyp.~\ref*{#1}}}}
\newcommand{\conjref}[1]{\hyperref[#1]{\darkredtext{Conj.~\ref*{#1}}}}
\newcommand{\lemref}[1]{\hyperref[#1]{\darkredtext{Lem.~\ref*{#1}}}}
\newcommand{\corref}[1]{\hyperref[#1]{\darkredtext{Cor.~\ref*{#1}}}}
\newcommand{\noteref}[1]{\hyperref[#1]{\darkredtext{Nota.~\ref*{#1}}}}
\newcommand{\claimref}[1]{\hyperref[#1]{\darkredtext{Clm.~\ref*{#1}}}}
\newcommand{\obsref}[1]{\hyperref[#1]{\darkredtext{Obs.~\ref*{#1}}}}
\newcommand{\algref}[1]{\hyperref[#1]{\darkredtext{Alg.~\ref*{#1}}}}
\newcommand{\figref}[1]{\hyperref[#1]{\darkredtext{Fig.~\ref*{#1}}}}
\newcommand{\tabref}[1]{\hyperref[#1]{\darkredtext{Tab.~\ref*{#1}}}}
\newcommand{\appref}[1]{\hyperref[#1]{\darkredtext{App.~\ref*{#1}}}}
\renewcommand{\eqref}[1]{\hyperref[#1]{\darkredtext{Eq.~\ref*{#1}}}}

\usepackage{listings}
\usepackage{xcolor}
\usepackage{booktabs}
\usepackage{amssymb}      % for \cup
\usepackage{tikz}         % for \circled (remove if you use the fallback below)

\renewcommand{\circled}[1]{%
  \tikz[baseline=(char.base)]{%
    \node[shape=circle,draw,inner sep=1pt,minimum size=1.1em] (char) {\scriptsize #1};}}
\lstdefinestyle{agentloop}{
  basicstyle=\ttfamily\scriptsize,
  commentstyle=\color{gray!70!black}\itshape,
  keywordstyle=\color{blue!55!black}\bfseries,
  showstringspaces=false,
  columns=fullflexible,
  keepspaces=true,
  escapeinside={(*@}{@*)},
  morekeywords={while,if,not,is,continue,and,or,True,False,None,def,return},
  frame=tb,
  framerule=0.5pt,
  rulecolor=\color{gray!60},
  framesep=5pt,
  backgroundcolor=\color{gray!4},
  xleftmargin=8pt,
  xrightmargin=4pt,
  aboveskip=2pt,
  belowskip=2pt,
  breaklines=false,
}

\definecolor{darkred}{RGB}{165,42,42}
\newcommand{\hi}[1]{\textcolor{darkred}{#1}}

\begin{document}

\maketitle

\section{Introduction}
\label{sec:intro}

AI-assisted software development has evolved from autocomplete-style tools such as GitHub Copilot \citep{chen2021evaluating}, through IDE-integrated assistants like Cursor~\citep{cursor2026official}, to fully agentic systems that autonomously plan multi-step modifications, execute shell commands, read and write files, and iterate on their own outputs. Claude Code~\citep{anthropic2026claudecode} is an agentic coding tool released by Anthropic~\citep{anthropic2026github}. Its official documentation describes an ``agentic loop'' that plans and executes actions toward accomplishing a goal and can call tools, evaluate results, and continue until the task is done~\footnote{\url{https://code.claude.com/docs/en/how-claude-code-works}.}. This shift from suggestion to autonomous action introduces architectural requirements that have no counterpart in completion-based tools. These requirements define a design space, a set of recurring questions spanning topics such as safety, context management, extensibility, and delegation that every coding agent must navigate. This study uses source-level analysis of Claude Code to show how one production system answers these questions.

Despite growing adoption, Anthropic publishes user-facing documentation for Claude Code but not detailed architectural descriptions. This study uses source code analysis to describe architectural design decisions. Anthropic's internal survey of 132 engineers and researchers~\citep{anthropic2025internal} reports that about 27\% of Claude Code-assisted tasks were work that would not have been attempted without the tool, suggesting that the architecture enables qualitatively new workflows rather than merely accelerating existing ones.

In this work, we first identify five human values/philosophies and thirteen design principles that motivate the architecture (\Cref{sec:values}), then organize the analysis in three parts:

\begin{enumerate}
  \item \textbf{Design-space analysis.} We identify recurring design questions (where reasoning lives, how the iteration loop is structured, what safety posture to adopt, how the extension surface is partitioned, how context is managed, how work is delegated across subagents, and how sessions persist) and analyze Claude Code's answers through a 7-component high-level structure and a 5-layer subsystem architecture, tracing each choice to specific source files (\Cref{sec:arch}). The analysis aims to build a deep understanding of the system mechanism, with the goal of informing the design of better and more powerful agent systems.
  \item \textbf{Architectural contrast with OpenClaw and Hermes Agent.}
  Beyond analyzing Claude Code itself, we also compare its design philosophy with that of two open-source agent systems, OpenClaw~\citep{openclaw2026} (a multi-channel personal assistant gateway) and Hermes Agent~\citep{hermesagent2026} (a single-process, multi-surface assistant), across six design dimensions to show how the same recurring questions produce different answers under different deployment contexts (\Cref{sec:compare}), in order to highlight both the common principles and the key differences between commercial and open-source software. This comparison helps reveal how deployment setting, product goals, safety requirements, and user assumptions shape architectural choices in different ways. By examining where these systems converge and where they diverge, our study aims to provide useful guidance and practical insights for the design of future, more capable agent systems.

  \item \textbf{Open directions for future agent systems.} Building on the design-space analysis and the OpenClaw and Hermes contrasts, \Cref{sec:future} identifies six open directions spanning the observability-evaluation gap, cross-session persistence, harness boundary evolution, horizon scaling, governance, and long-term developer capability, each drawing on empirical, architectural, and policy literature. The long-term capability question also reveals an open concern: while the Claude Code agent system amplifies the short-term capabilities of programmers and end users, it offers limited mechanisms that explicitly support long-term human improvement, deeper understanding, and sustained codebase coherence. 

\end{enumerate}

The core agent loop is a while-true cycle with state management. The surrounding subsystems for safety, extensibility, context management, delegation, and persistence make up the bulk of the implementation. Source-level analysis\footnote{Our analysis is grounded primarily in the source code, supplemented by official Anthropic documentation and selected community analysis. The appendix details the evidence base and methodology.} allows us to identify design choices, subsystem boundaries, and implementation trade-offs directly from the system itself rather than inferring them solely from product descriptions.

\paragraph{Running example.} 
To keep the architecture concrete, we trace the task ``Fix the failing test in \texttt{auth.test.ts}'' through \Cref{sec:arch,sec:turn,sec:auth,sec:ext,sec:context,sec:subagent,sec:persist}. This example illustrates how a seemingly simple user request activates multiple architectural layers, including tool invocation, permission checks, context selection, iterative repair, delegation, and session persistence.

\paragraph{Paper organization.}
\Cref{sec:values} identifies the human values and design principles that motivate the architecture. \Cref{sec:arch} introduces the high-level architecture and the design questions it answers. \Cref{sec:turn,sec:auth,sec:ext,sec:context,sec:subagent,sec:persist} each analyze a major subsystem's design choices. \Cref{sec:compare} contrasts the analysis with OpenClaw and Hermes Agent, \Cref{sec:related} positions the work against prior agent and software-engineering literature, \Cref{sec:discuss} provides discussion, and \Cref{sec:future} surveys open questions for future agent systems. \Cref{sec:conclude} concludes. The appendix describes the evidence base and methodology, package-structure notes, community reimplementations, and a companion map from new agent-system signals to the paper's design-space questions.

\section{Design Philosophies, Design Principles and Architectural Motivations}
\label{sec:values}

Production coding agents are built by humans, for humans, and the architectural decisions they embed reflect what their creators believe matters. This section identifies the human values that motivate Claude Code's design, traces them through recurring design principles, and frames the design-space questions that organize the analysis in \Cref{sec:arch,sec:turn,sec:auth,sec:ext,sec:context,sec:subagent,sec:persist}.

Anthropic's framework for safe agents states a central tension: ``Agents must be able to work autonomously; their independent operation is exactly what makes them valuable. But humans should retain control over how their goals are pursued''~\citep{anthropic2025agents}. Claude's Constitution resolves this not through rigid decision procedures but by cultivating ``good judgment and sound values that can be applied contextually''~\citep{anthropic2026constitution}. These commitments, together with empirical findings about how developers actually use the tool~\citep{anthropic2025internal,anthropic2026autonomy}, point to five human values that shape the architecture.

\subsection{Five Values and Philosophies}
\label{sec:values:five}

\paragraph{Human Decision Authority.}
The human retains ultimate decision authority over what the system does, organized through a principal hierarchy (Anthropic, then operators, then users) that formalizes who holds authority over what~\citep{anthropic2026constitution}. The system is designed so that humans can exercise informed control: they can observe actions in real time, approve or reject proposed operations, interrupt compatible in-progress operations, and audit after the fact. When Anthropic found that users approve 93\% of permission prompts~\citep{anthropic2026automode}, the response was not to add more warnings but to restructure the problem: defined boundaries (sandboxing, auto-mode classifiers) within which the agent can work freely, rather than per-action approvals that users stop reviewing once habituated~\citep{anthropic2025sandboxing}.

\paragraph{Safety, Security, and Privacy.}
The system protects humans, their code, their data, and their infrastructure from harm, even when the human is inattentive or makes mistakes. This is distinct from Human Decision Authority: where authority is about the human's \emph{power to choose}, safety is about the system's \emph{obligation to protect even when that power lapses}. Anthropic's safe-agents framework separately identifies securing agent interactions and protecting privacy across extended interactions as core commitments~\citep{anthropic2025agents}. The auto-mode threat model~\citep{anthropic2026automode} explicitly targets four risk categories: overeager behavior, honest mistakes, prompt injection, and model misalignment.

\paragraph{Reliable Execution.}
The agent does what the human actually meant, stays coherent over time, and supports verifying its work before declaring success. This value spans both single-turn correctness (did it interpret the request faithfully?) and long-horizon dependability (does it remain coherent across context window boundaries, session resumption, and multi-agent delegation?). Anthropic's product documentation~\citep{anthropic2026howworks} describes a three-phase loop that the agent repeats until the task is complete: gather context, take action, and verify results. The agent design guidance~\citep{anthropic2024effective} further emphasizes that ``ground truth from the environment'' at each step assesses progress. The harness-design guidance~\citep{anthropic2026harness} likewise notes that ``agents tend to respond by confidently praising the work,'' even when quality is mediocre, motivating separation of generation from evaluation.

\paragraph{Capability Amplification.}
The system materially increases what the human can accomplish per unit of effort and cost. Anthropic's internal survey~\citep{anthropic2025internal}, discussed in
\Cref{sec:intro}, suggests that the architecture enables qualitatively new
workflows, not merely faster existing ones: approximately 27\% of tasks
represented work that would not otherwise have been attempted. The system is described by its creators as ``a Unix utility rather than a traditional product,'' built from the smallest building blocks that are ``useful, understandable, and extensible''~\citep{cherny2025latentspace}. The architecture invests in deterministic infrastructure (context management, tool routing, recovery) rather than decision scaffolding (explicit planners or state graphs), on the premise that increasingly capable models benefit more from a rich operational environment than from frameworks that constrain their choices.

\paragraph{Contextual Adaptability.}
The system fits the user's specific context (their project, tools, conventions, and skill level) and the relationship improves over time. The extension architecture (CLAUDE.md, skills, MCP, hooks, plugins) provides configurability at multiple levels of context cost (\Cref{sec:ext,sec:context}). Longitudinal data~\citep{anthropic2026autonomy} shows that the human-agent relationship evolves: auto-approve rates increase from approximately 20\% at fewer than 50 sessions to over 40\% by 750 sessions. This pattern, described as autonomy that is ``co-constructed by the model, the user, and the product,'' means the system is designed for trust trajectories rather than fixed trust states. MCP's donation to the Linux Foundation's Agentic AI Foundation~\citep{linuxfoundation2025aaif} reflects the ecosystem dimension of this value.

\subsection{Design Principles}
\label{sec:values:principles}

These values are operationalized through thirteen design principles, each answering a recurring question that production coding agents must resolve. \Cref{tab:principles} summarizes the principles; subsequent sections (\Cref{sec:arch} to \Cref{sec:persist}) trace each through specific implementation choices.

\begin{table*}[t]
\centering
\small
\caption{Design principles, the values they serve, and the design-space question each answers. Principles map to multiple values; implementations appear in the sections indicated.}
\label{tab:principles}
\begin{tabularx}{\textwidth}{@{}>{\raggedright\arraybackslash}m{3.8cm}>{\raggedright\arraybackslash}m{3.2cm}>{\raggedright\arraybackslash}X>{\centering\arraybackslash}m{1.2cm}@{}}
\toprule
\textbf{Principle} & \textbf{Values Served} & \textbf{Design Question} & \textbf{Sections} \\
\midrule
Deny-first with human escalation & Authority, Safety & Should unrecognized actions be allowed, blocked, or escalated to the human? & \ref{sec:auth}, \ref{sec:subagent}, \ref{sec:persist} \\
\midrule
Graduated trust spectrum & Authority, Adaptability & Fixed permission level, or a spectrum users traverse over time? & \ref{sec:auth} \\
\midrule
Defense in depth with layered mechanisms & Safety, Authority, Reliability & Single safety boundary, or multiple overlapping ones using different techniques? & \ref{sec:arch}, \ref{sec:auth} \\
\midrule
Externalized programmable policy & Safety, Authority, Adaptability & Hardcoded policy, or externalized configs with lifecycle hooks? & \ref{sec:auth}, \ref{sec:ext} \\
\midrule
Context as scarce resource with progressive management & Reliability, Capability & What is the binding resource constraint, and how to manage it: single-pass truncation or graduated pipeline? & \ref{sec:turn}, \ref{sec:ext}, \ref{sec:context}, \ref{sec:subagent} \\
\midrule
Append-only durable state & Reliability, Authority & Mutable state, checkpoint snapshots, or append-only logs? & \ref{sec:turn}, \ref{sec:persist} \\
\midrule
Minimal scaffolding, maximal operational harness & Capability, Reliability & Invest in scaffolding-side reasoning, or operational infrastructure that lets the model reason freely? & \ref{sec:arch}, \ref{sec:turn} \\
\midrule
Values over rules & Capability, Authority & Rigid decision procedures, or contextual judgment backed by deterministic guardrails? & \ref{sec:arch}, \ref{sec:auth}, \ref{sec:context} \\
\midrule
Composable multi-mechanism extensibility & Capability, Adaptability & One unified extension API, or layered mechanisms at different context costs? & \ref{sec:ext} \\
\midrule
Reversibility-weighted risk assessment & Capability, Safety & Same oversight for all actions, or lighter for reversible and read-only ones? & \ref{sec:turn}, \ref{sec:auth}, \ref{sec:subagent} \\
\midrule
Transparent file-based configuration and memory & Adaptability, Authority & Opaque database, embedding-based retrieval, or user-visible version-controllable files? & \ref{sec:context} \\
\midrule
Isolated subagent boundaries & Reliability, Safety, Capability & Subagents share the parent's context and permissions, or operate in isolation? & \ref{sec:subagent} \\
\midrule
Graceful recovery and resilience & Reliability, Capability & Fail hard on errors, or silently recover and reserve human attention for unrecoverable situations? & \ref{sec:turn}, \ref{sec:auth} \\
\bottomrule
\end{tabularx}
\end{table*}

These principles can be read against three major alternative design families. First, \emph{rule-based orchestration}: frameworks such as LangGraph~\citep{langgraph2024} encode decision logic as explicit state graphs with typed edges, choosing scaffolding over minimal harness. Second, \emph{container-isolated execution}: SWE-Agent and OpenHands~\citep{yang2024sweagent,wang2024openhands} rely on Docker isolation rather than layered policy enforcement. Third, \emph{version-control-as-safety}: tools like Aider~\citep{gauthier2024aider} use Git rollback as the primary safety mechanism rather than deny-first evaluation. Claude Code's principle set is distinctive in combining minimal decision scaffolding with layered policy enforcement, values-based judgment with deny-first defaults, and progressive context management with composable extensibility.

\subsection{From Values to Architecture}
\label{sec:values:bridge}

Each value traces through its principles to specific architectural decisions:

\begin{itemize}[nosep]
  \item \textbf{Human Decision Authority} motivates deny-first evaluation, the graduated trust spectrum, append-only state (auditable history), externalized programmable policy, and values-over-rules (\Cref{sec:auth,sec:persist,sec:ext,sec:context}).
  \item \textbf{Safety, Security, and Privacy} motivates defense in depth, deny-first defaults, reversibility-weighted assessment, externalized policy, and isolated subagent boundaries (\Cref{sec:auth,sec:subagent}).
  \item \textbf{Reliable Execution} motivates context-as-scarce-resource, append-only durable state, graceful recovery, isolated subagent boundaries, and defense in depth (\Cref{sec:turn,sec:context,sec:subagent,sec:persist}).
  \item \textbf{Capability Amplification} motivates minimal scaffolding, composable extensibility, reversibility-weighted risk, context management, and graceful recovery (\Cref{sec:turn,sec:ext,sec:auth}).
  \item \textbf{Contextual Adaptability} motivates transparent file-based memory, composable extensibility, the graduated trust spectrum, and externalized programmable policy (\Cref{sec:context,sec:ext,sec:auth}).
\end{itemize}

These mappings also reveal what the architecture does \emph{not} do: it does not impose explicit planning graphs on the model's reasoning, does not provide a single unified extension mechanism, and does not restore all session-scoped trust-related state across resume. These absences are consistent with the principle set above.

\subsection{A Cross-Cutting Question: Long-term Developer Capability}
\label{sec:values:lens}

The five values above describe what the architecture is designed to serve. This paper also asks a sixth question: whether the architecture helps developers preserve long-term understanding and capability. This concern is real: Anthropic's own study of 132 engineers and researchers~\citep{anthropic2025internal} documents a ``paradox of supervision'' in which overreliance on AI risks atrophying the skills needed to supervise it, and independent research~\citep{shen2026skill} finds that developers in AI-assisted conditions score 17\% lower on comprehension tests. However, this concern is not prominently reflected as a design driver in the architecture or in Anthropic's stated design values. We therefore treat it not as a co-equal value but as a cross-cutting concern: a question applied across all five values in \Cref{sec:discuss}, asking whether short-term amplification costs long-term human understanding, codebase coherence, and the developer pipeline.

\begin{figure}[!t]
    \centering
    \includegraphics[width=0.87\linewidth]{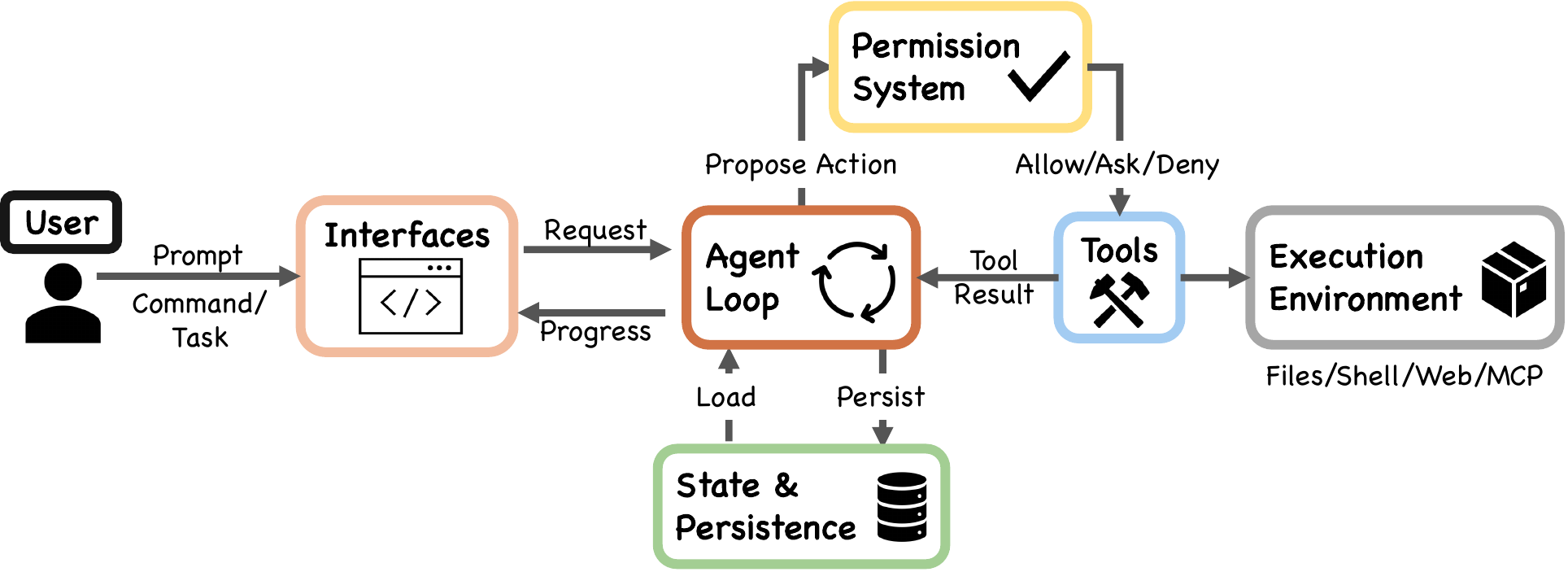}
    \caption{High-level system structure of Claude Code. The system decomposes into seven functional components: user, interfaces, the agent loop, a permission system, tools, state \& persistence, and an execution environment. All entry surfaces converge on the same agent loop.}
    \label{fig:high-level}
\end{figure}

\section{Architecture Overview}
\label{sec:arch}

Building a production coding agent requires answering several recurring design questions: where should reasoning live, how many execution engines are needed, what safety posture to adopt, and what resource to treat as the binding constraint. Claude Code's architecture can be read as one set of answers to these questions. At the implementation level, the system has seven components connected by a main data flow: a user submits a prompt through one of several interfaces, which feeds into a shared agent loop. The agent loop assembles context, calls the Claude model, receives responses that may include tool-use requests, routes those requests through a permission system, and dispatches approved actions to concrete tools that interact with the execution environment. Throughout this process, state and persistence mechanisms record the conversation transcript, manage session identity, and support resume, fork, and rewind operations.

\subsection{Design Questions and Running Example}
\label{sec:arch:principles}

The description is organized around four design questions that recur across production coding agents, each grounding one or more of the design principles identified in \Cref{tab:principles}. Each question is introduced here with Claude Code's answer, a note on plausible alternatives, and then demonstrated progressively through \Cref{sec:turn,sec:auth,sec:ext,sec:context,sec:subagent,sec:persist}.

\paragraph{Where does reasoning live?}
In Claude Code, the model reasons about what to do; the harness is responsible for executing actions. The model emits \code{tool\_use} blocks as part of its response, and the harness parses them, checks permissions, dispatches them to tool implementations, and collects results (\file{query.ts}). The model never directly accesses the filesystem, runs shell commands, or makes network requests. This separation has a security consequence: because reasoning and enforcement occupy separate code paths, a compromised or adversarially manipulated model cannot override the sandboxing, permission checks, or deny-first rules implemented in the harness. The model's only interface to the outside world is the structured \code{tool\_use} protocol, which the harness validates before execution. Community analysis of the extracted source estimates that only about 1.6\% of Claude Code's codebase constitutes AI decision logic, with the remaining 98.4\% being operational infrastructure, a ratio that illustrates how thin the core agent reasoning layer is. Alternative designs invest more in scaffolding-side reasoning: Devin maintains explicit planning and task-tracking structures, while LangGraph~\citep{langgraph2024} routes control flow through developer-defined state graphs.

\paragraph{How many execution engines?}
Claude Code uses a single \func{queryLoop} function that executes regardless of whether the user is interacting through an interactive terminal, a headless CLI invocation, the Agent SDK, or an IDE integration (\file{query.ts}). Only the rendering and user-interaction layer varies. Other systems use mode-specific engines: for example, an IDE integration may follow a different code path than a CLI tool, trading uniformity for surface-specific optimization.

\paragraph{What is the default safety posture?}
Claude Code's default safety posture is deny-first with human escalation: deny rules override ask rules override allow rules, and unrecognized actions are escalated to the user rather than allowed silently (\file{permissions.ts}). Multiple independent safety layers (permission rules, PreToolUse hooks, the auto-mode classifier when enabled, and optional shell sandboxing) are stacked, so any one can block an action (\Cref{sec:auth}). This combines the \emph{deny-first with human escalation} and \emph{defense in depth with layered mechanisms} principles from \Cref{tab:principles}. Alternative approaches shift the trust boundary elsewhere: SWE-Agent and OpenHands~\citep{yang2024sweagent,wang2024openhands} rely on container-based isolation to contain arbitrary execution, while Aider~\citep{gauthier2024aider} uses git-based rollback as its primary safety net.

\paragraph{What is the binding resource constraint?}
In Claude Code, the context window (200K for older models, 1M for the Claude 4.6 series) is the binding resource constraint. Five distinct context-reduction strategies execute before every model call (\file{query.ts}), and several other subsystem decisions (lazy loading of instructions, deferred tool schemas, summary-only subagent returns) exist to limit context consumption (\Cref{sec:context}). The five-layer pipeline exists because no single compaction strategy addresses all types of context pressure. Budget reduction targets individual tool outputs that overflow size limits. Snip handles temporal depth. Microcompact reacts to cache overhead. Context collapse manages very long histories. Auto-compact performs semantic compression as a last resort. Each layer operates at a different cost-benefit tradeoff, and earlier, cheaper layers run before costlier ones. Alternative architectures treat other resources as the primary bottleneck, for instance compute budget (limiting the number of model calls or tool invocations) or working memory (maintaining an explicit scratchpad rather than relying on the conversation history).

\paragraph{Running example.}
To ground these principles, we thread a single task through \Cref{sec:arch,sec:turn,sec:auth,sec:ext,sec:context,sec:subagent,sec:persist}: \emph{``Fix the failing test in auth.test.ts.''} In this section the user submits the prompt through one of Claude Code's interfaces. Subsequent sections trace the request through the query loop, permission gate, tool pool, context window, subagent delegation, and session persistence.

\subsection{High-Level System Structure}
\label{sec:arch:highlevel}

\begin{figure}[!t]
\centering
\includegraphics[width=0.65\textwidth]{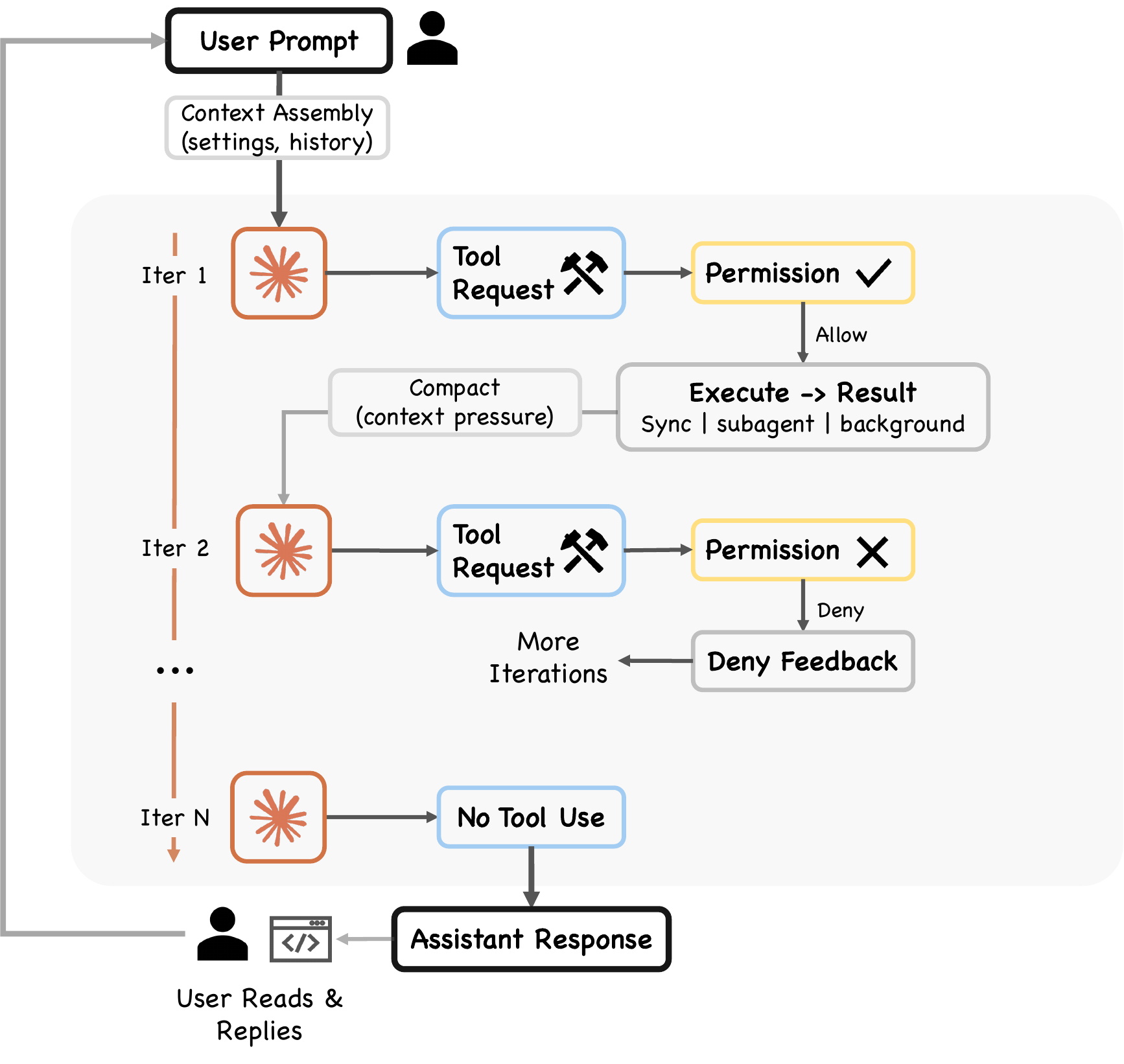}
\caption{Runtime turn flow showing the end-to-end execution of a single agentic turn: user prompt enters through context assembly, the model is called, tool requests pass through the permission gate, tool results feed back into the loop, and compaction manages context pressure.}
\label{fig:turn-flow}
\end{figure}

The seven-component model (\Cref{fig:high-level}) maps directly to source files:

\begin{enumerate}[nosep]
  \item \textbf{User}: Submits prompts, approves permissions, reviews output.
  \item \textbf{Interfaces}: Interactive CLI, headless CLI (\code{claude -p}), Agent SDK, and IDE/Desktop/Browser. All surfaces feed the same loop.
  \item \textbf{Agent loop}: The iterative cycle of model call, tool dispatch, and result collection, implemented as the \func{queryLoop} async generator in \file{query.ts}.
  \item \textbf{Permission system}: Deny-first rule evaluation (\file{permissions.ts}), the auto-mode ML classifier, and hook-based interception (\file{types/hooks.ts}).
  \item \textbf{Tools}: Up to 54 built-in tools (19 unconditional, 35 conditional on feature flags and user type) assembled via \func{assembleToolPool} (\file{tools.ts}), merged with MCP-provided tools. Plugins contribute indirectly through MCP servers and the skill/command registry.
  \item \textbf{State \& persistence}: Mostly append-only JSONL session transcripts (\file{sessionStorage.ts}), global prompt history (\file{history.ts}), and subagent sidechain files.
  \item \textbf{Execution environment}: Shell execution with optional sandboxing (\file{shouldUseSandbox.ts}), filesystem operations, web fetching, MCP server connections, and remote execution.
\end{enumerate}

The data flow follows a left-to-right spine: the user submits a request through an interface, which enters the agent loop. The loop proposes actions to the permission system; approved actions reach tools, which interact with the execution environment and return \code{tool\_result} messages back to the loop. State and persistence sit alongside the loop, recording transcripts and loading prior session data.

The application entry point \func{main} in \file{main.tsx} initializes security settings (including \code{NoDefaultCurrentDirectoryInExePath} to prevent Windows PATH hijacking), registers signal handlers for graceful shutdown, and dispatches to the appropriate execution mode.

\subsection{Layered Subsystem Decomposition}
\label{sec:arch:layers}

\begin{figure}[t]
\centering
\includegraphics[width=\textwidth]{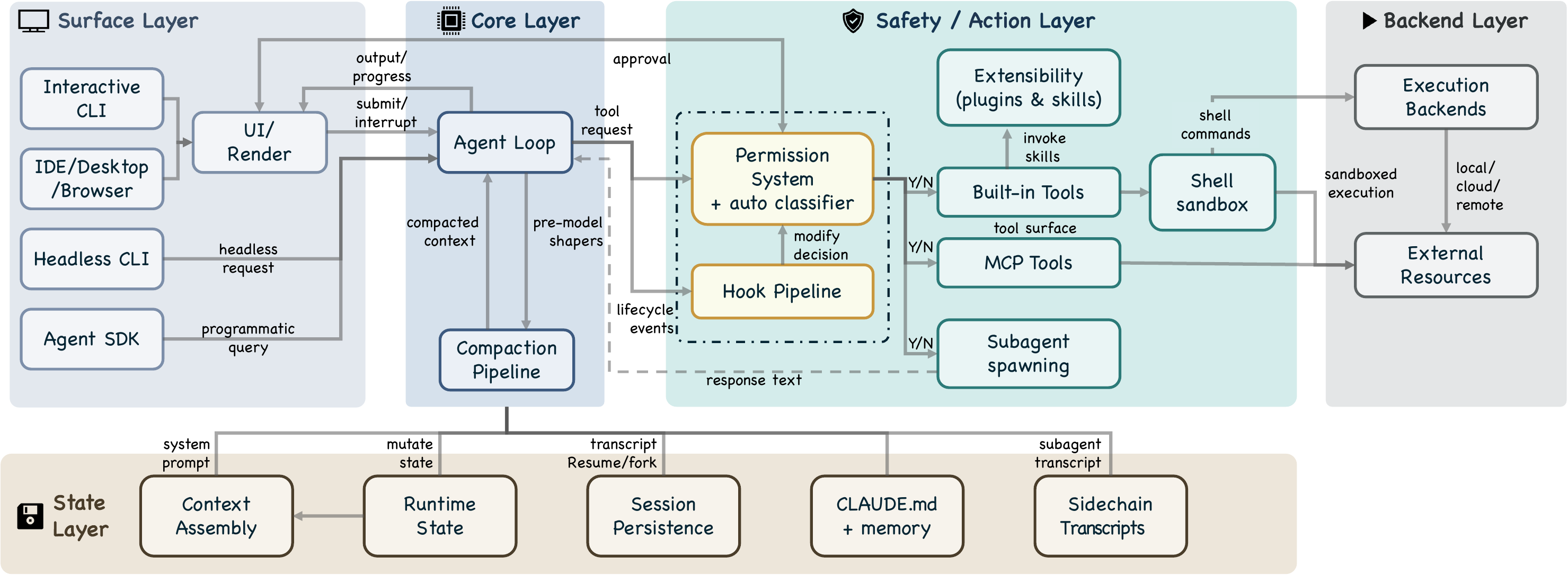}
\caption{Expanded layered architecture showing five subsystem layers: surface (Interactive CLI, Headless CLI, Agent SDK, IDE/Desktop/Browser, UI/renderer), core (agent loop, compaction pipeline), safety/action (permission system incl.\ auto-mode classifier, hook pipeline, extensibility, built-in tools, MCP tools, shell sandbox, subagent spawning), state (context assembly, runtime state, session persistence, CLAUDE.md + memory, sidechain transcripts), and backend (execution backends, external resources).}
\label{fig:layered}
\end{figure}

The five-layer decomposition (\Cref{fig:layered}) expands the seven-component model into a finer-grained view, mapping each layer to specific source directories.

\paragraph{Surface layer (entry points and rendering).}
The \file{src/entrypoints/} directory contains startup paths, including the SDK entry with \file{coreTypes.ts}, \file{controlSchemas.ts}, and \file{coreSchemas.ts}. The \file{src/screens/} directory composes full-screen layouts, and \file{src/components/} provides terminal UI building blocks via the ink framework. The interactive CLI launches a terminal UI with real-time streaming, permission dialogs, and progress indicators. The headless CLI (\code{claude -p}) creates a \code{QueryEngine} instance, a thin conversation wrapper around the shared query path (\Cref{sec:arch:queryengine}), for single-shot processing. The Agent SDK emits typed events via async generators.

\paragraph{Core layer (agent loop, compaction pipeline).}
The \func{queryLoop} async generator (\file{query.ts}) implements the iterative agent loop, consuming assembled context from the state layer and dispatching tool requests to the safety/action layer. Before every model call, a \emph{compaction pipeline} of five sequential shapers (\file{query.ts:365-453}) manages context pressure: budget reduction, snip, microcompact, context collapse, and auto-compact (\Cref{sec:turn:shapers,sec:context:compaction}).

\paragraph{Safety/action layer (permission system, hooks, extensibility, tools, sandbox, subagents).}
The \emph{permission system} (\file{permissions.ts}) implements deny-first rule evaluation with up to seven permission modes (if also counting internal-only \code{bubble} and feature-gated \code{auto}) (\file{types/permissions.ts}) and an integrated \emph{auto-mode ML classifier} (\file{yoloClassifier.ts}) that provides a two-stage fast-filter and chain-of-thought evaluation of tool safety (\Cref{sec:auth}). A \emph{hook pipeline} spanning 27 event types (\file{coreTypes.ts}; output schemas in \file{types/hooks.ts}) can block, rewrite, or annotate tool requests; of these, 5 are safety-related while the remaining 22 serve lifecycle and orchestration purposes (\Cref{sec:ext}). An \emph{extensibility} subsystem allows plugins and skills to register tools and hooks into the runtime. Tool pool assembly via \func{assembleToolPool} (\file{tools.ts}) merges built-in and MCP-provided tools. Approved shell commands pass through a \emph{shell sandbox} (\file{shouldUseSandbox.ts}) that restricts filesystem and network access independently of the permission system. \emph{Subagent spawning} via \code{AgentTool} (\file{AgentTool.tsx}, \file{runAgent.ts}) is dispatched through the same \func{buildTool} factory as all other tools, re-entering the \func{queryLoop} with an isolated context window and returning only a summary to the parent (\Cref{sec:subagent}).

\paragraph{State layer (context assembly, runtime state, persistence, memory, sidechains).}
\emph{Context assembly} is a memoized state loader, not a routing hub: \func{getSystemContext} (\file{context.ts}) computes session-level system context including git status, and \func{getUserContext} (\file{context.ts}) loads the CLAUDE.md hierarchy and current date. Both are cached for reuse: system context is appended to the system prompt, while user context is added as a user-context message. The \file{src/state/} directory manages runtime application state. Session transcripts are stored as mostly append-only JSONL files at project-specific paths (\file{sessionStorage.ts}). The \emph{CLAUDE.md + memory} subsystem provides a four-level instruction hierarchy (\file{claudemd.ts}) from managed settings to directory-specific files, plus auto-memory entries that Claude writes during conversations (\Cref{sec:context:claudemd}). \emph{Sidechain transcripts} (\file{sessionStorage.ts:247}) store each subagent's conversation in a separate file, preventing subagent content from inflating the parent context (\Cref{sec:subagent:sidechain}). Global prompt history is maintained in \file{history.jsonl} (\file{history.ts}). Resume and fork operations reconstruct session state from transcripts (\file{conversationRecovery.ts}).

\paragraph{Backend layer (execution backends, external resources).}
Shell command execution with optional sandboxing (\file{BashTool.tsx}, \file{PowerShellTool.tsx}), remote execution support (\file{src/remote/}), MCP server connections across multiple transport variants including stdio, SSE, HTTP, WebSocket, SDK, and IDE-specific adapters (\file{services/mcp/client.ts}), and 42 tool subdirectories in \file{src/tools/} implement concrete tool logic.

\subsection{QueryEngine: A Clarification}
\label{sec:arch:queryengine}

The class documentation at \file{QueryEngine.ts} states: ``QueryEngine owns the query lifecycle and session state for a conversation. It extracts the core logic from \code{ask()} into a standalone class that can be used by both the headless/SDK path and (in a future phase) the REPL.'' The class is a \emph{conversation wrapper} for non-interactive surfaces, not the engine itself. Its constructor accepts a \code{QueryEngineConfig} with initial messages, an abort controller, a file-state cache, and other per-conversation state. Its \func{submitMessage} method is an async generator that orchestrates a single turn. The shared query path lives in \func{query} (\file{query.ts}), which wraps an internal \func{queryLoop}; \code{QueryEngine} delegates to \func{query}.

This distinction matters architecturally: the interactive CLI also calls \func{query}, bypassing \code{QueryEngine} entirely. The shared code path is the loop function, not the engine class.

\subsection{Permission and Safety Layers}
\label{sec:arch:safety-preview}

The safety-by-default principle is implemented through seven independent layers. A request must pass through all applicable layers, and any single layer can block it:

\begin{enumerate}[nosep]
  \item \textbf{Tool pre-filtering} (\file{tools.ts}): Blanket-denied tools are removed from the model's view before any call, preventing the model from attempting to invoke them.
  \item \textbf{Deny-first rule evaluation} (\file{permissions.ts}): Deny rules always take precedence over allow rules, even when the allow rule is more specific.
  \item \textbf{Permission mode constraints} (\file{types/permissions.ts}): The active mode determines baseline handling for requests matching no explicit rule.
  \item \textbf{Auto-mode classifier}: In auto mode, an ML-based classifier can evaluate requests still requiring approval after rule evaluation and fast-path checks; selected allow rules that would bypass it (e.g., \code{Bash(*)}) are set aside while auto mode is active.
  \item \textbf{Shell sandboxing} (\file{shouldUseSandbox.ts}): Approved shell commands may still execute inside a sandbox restricting filesystem and network access.
  \item \textbf{Not restoring permissions on resume} (\file{conversationRecovery.ts}): Session-scoped permissions are not stored in the transcript, so a session resumed in a new process starts without them.
  \item \textbf{Hook-based interception} (\file{types/hooks.ts}): PreToolUse hooks can modify permission decisions; PermissionRequest hooks can resolve decisions asynchronously alongside the user dialog (or before it, in coordinator mode).
\end{enumerate}

These layers are described in detail in \Cref{sec:auth}.

\subsection{Context as Bottleneck: Beyond Compaction}
\label{sec:arch:context-preview}

Beyond the five-layer compaction pipeline (detailed in \Cref{sec:context}), several other subsystem decisions reflect the context-as-bottleneck constraint:

\begin{itemize}[nosep]
  \item \textbf{CLAUDE.md lazy loading}: The base CLAUDE.md hierarchy is loaded at session start, but additional nested-directory instruction files and conditional rules are loaded only when the agent reads files in those directories, preventing unused instructions from consuming context.
  \item \textbf{Deferred tool schemas}: When ToolSearch is enabled, some tools include only their names in the initial context; full schemas are loaded on demand.
  \item \textbf{Subagent summary-only return}: Subagents return only summary text to the parent, not their full conversation history (\Cref{sec:subagent}).
  \item \textbf{Per-tool-result budget}: Individual tool results are capped at a configurable size, preventing a single verbose output from consuming disproportionate context.
\end{itemize}

\section{Turn Execution: The Agentic Query Loop}
\label{sec:turn}

When the user submits ``Fix the failing test in auth.test.ts,'' the input enters a reactive loop, one of several possible orchestration patterns for coding agents. This section examines Claude Code's choice of a simple while-loop architecture and traces one turn of that loop end-to-end, illustrating three design principles from \Cref{tab:principles}: \emph{minimal scaffolding with maximal operational harness}, \emph{context as scarce resource with progressive management}, and \emph{graceful recovery and resilience}.

\subsection{The Query Pipeline}
\label{sec:turn:pipeline}

Each turn follows a fixed sequence (\Cref{fig:turn-flow}, \file{query.ts}):

\begin{enumerate}[nosep]
  \item \textbf{Settings resolution.} The \func{queryLoop} function destructures immutable parameters including the system prompt, user context, permission callback, and model configuration.
  \item \textbf{Mutable state initialization.} A single \code{State} object stores all mutable state across iterations, including messages, tool context, compaction tracking, and recovery counters. The loop's seven \code{continue} points (the ``continue sites'') each overwrite this object in one whole-object assignment rather than mutating fields individually.
  \item \textbf{Context assembly.} The function \func{getMessagesAfterCompactBoundary} retrieves messages from the last compact boundary forward, ensuring that compacted content is represented by its summary rather than the original messages.
  \item \textbf{Pre-model context shapers.} Five shapers execute sequentially (\Cref{sec:turn:shapers}).
  \item \textbf{Model call.} A \code{for await} loop over \code{deps.callModel()} streams the model's response, passing assembled messages (with user context prepended), the full system prompt, thinking configuration, the available tool set, an abort signal, the current model specification, and additional options including fast-mode settings, effort value, and fallback model.
  \item \textbf{Tool-use dispatch.} If the response contains \code{tool\_use} blocks, they flow to the tool orchestration layer (\Cref{sec:turn:dispatch}).
  \item \textbf{Permission gate.} Each tool request passes through the permission system (\Cref{sec:auth}).
  \item \textbf{Tool execution and result collection.} Tool results are added to the conversation as \code{tool\_result} messages, and the loop continues.
  \item \textbf{Stop condition.} If the response contains no \code{tool\_use} blocks (text only), the turn is complete.
\end{enumerate}

The \func{queryLoop} function is defined as an \code{AsyncGenerator}, yielding \code{StreamEvent}, \code{RequestStartEvent}, \code{Message}, \code{TombstoneMessage}, and \code{ToolUseSummaryMessage} events as it progresses. This generator-based design enables streaming output to the UI layer while maintaining a single synchronous control flow within the loop.

Claude Code's reactive loop follows the ReAct pattern~\citep{yao2022react}: the model generates reasoning and tool invocations, the harness executes actions, and results feed the next iteration. Alternative orchestration patterns include explicit graph-based routing~\citep{langgraph2024}, where control flow is defined as a state machine with typed edges, and tree-search methods~\citep{zhou2024lats} that explore multiple action trajectories before committing. Anthropic's own documentation~\citep{anthropic2024effective} identifies five composable workflow patterns (prompt chaining, routing, parallelization, orchestrator-workers, and evaluator-optimizer) of which Claude Code primarily uses the orchestrator-workers pattern for subagent delegation (\Cref{sec:subagent}) while keeping the core loop reactive. The reactive design trades search completeness for simplicity and latency: each turn commits to one action sequence without backtracking.

\subsection{Tool Dispatch and Streaming Execution}
\label{sec:turn:dispatch}

When the model response contains \code{tool\_use} blocks, the system chooses between two execution paths. The primary path uses \code{StreamingToolExecutor}, which begins executing tools as they stream in from the model response, reducing latency for multi-tool responses. The fallback path uses \func{runTools} in \file{toolOrchestration.ts}, which iterates over partitions produced by \func{partitionToolCalls}. Both paths classify tools as concurrent-safe or exclusive. Read-only operations can execute in parallel, while state-modifying operations like shell commands are serialized.

The \code{StreamingToolExecutor} (\file{StreamingToolExecutor.ts}) manages concurrent execution with two coordination mechanisms:

\begin{itemize}[nosep]
  \item \textbf{Sibling abort controller.} Fires when any Bash tool errors, immediately terminating other in-flight subprocesses rather than letting them run to completion.
  \item \textbf{Progress-available signal.} Wakes up the \func{getRemainingResults} consumer when new output is ready.
\end{itemize}

Results are buffered so that none is emitted ahead of an earlier non-concurrency-safe tool that is still running; results from concurrency-safe tools running in parallel are emitted as they complete. This concurrent-read, serial-write execution model occupies a middle ground between fully serial dispatch and more aggressive speculative approaches such as PASTE~\citep{sui2026paste}, which speculatively pre-executes predicted future tool calls while the model is still generating, hiding tool latency through speculation.

The tool result collection phase iterates over updates from either the streaming executor or the batched \func{runTools} async generator. Both are async iterables consumed by the same \code{for await} loop; the contrast is in whether tool execution starts before the model finishes streaming (streaming) or after partitioning into concurrency-safe groups (batched). Each update may carry a tool result, an attachment, or a progress event. A special check detects \code{hook\_stopped\_continuation} attachments: if a PostToolUse hook signals that the turn should not continue, a \code{shouldPreventContinuation} flag is set. Results are normalized for the Anthropic API via \func{normalizeMessagesForAPI}, filtering to keep only user-type messages.

\subsection{Pre-Model Context Shapers}
\label{sec:turn:shapers}

Five context shapers execute sequentially in \file{query.ts} before every model call, each operating on the \code{messagesForQuery} array. The five shapers run in sequence, with earlier steps applying lighter reductions before later steps apply broader compaction.

\paragraph{Budget reduction.} (\func{applyToolResultBudget}).
Enforces per-message size limits on tool results, replacing oversized outputs with content references. Exempt tools (those where \code{maxResultSizeChars} is not finite) retain their full output. Content replacements are persisted for agent and session query sources to enable reconstruction on resume. Budget reduction runs before microcompact because microcompact operates purely by \code{tool\_use\_id} and never inspects content; the two compose cleanly.

\paragraph{Snip.} (\func{snipCompactIfNeeded}, gated by \code{HISTORY\_SNIP}).
A lightweight trim that removes older history segments, returning \code{\{messages, tokensFreed, boundaryMessage\}}. The \code{snipTokensFreed} value is plumbed to auto-compact because the main token counter derives context size from the \code{usage} field on the most recent assistant message, and that message survives snip with its pre-snip \code{input\_tokens} still attached; snip's savings are therefore invisible to the counter unless passed through explicitly.

\paragraph{Microcompact.}
Fine-grained compression that always runs a time-based path and optionally a cache-aware path (gated by \code{CACHED\_MICROCOMPACT}). When the cached path is enabled, boundary messages are deferred until after the API response so they can use actual \code{cache\_deleted\_input\_tokens} rather than estimates. Returns \code{\{messages, compactionInfo\}} where \code{compactionInfo} may include \code{pendingCacheEdits}.

\paragraph{Context collapse.} Gated by \code{CONTEXT\_COLLAPSE}.
A read-time projection over the conversation history. The source comments explain: ``Nothing is yielded; the collapsed view is a read-time projection over the REPL's full history. Summary messages live in the collapse store, not the REPL array. This is what makes collapses persist across turns.'' Unlike the other shapers, context collapse does not mutate the REPL's stored history; it replaces the \code{messagesForQuery} array with a projected view via \func{applyCollapsesIfNeeded}, so the model sees the collapsed version while the full history remains available for reconstruction.

\paragraph{Auto-compact.}
The fifth shaper, triggering a full model-generated summary via \func{compactConversation} in \file{compact.ts}. This function executes \code{PreCompact} hooks, creates a summary request using \func{getCompactPrompt}, and calls the model to produce a compressed summary. The result feeds into \func{buildPostCompactMessages} (\file{compact.ts}). Auto-compact fires only when the context still exceeds the pressure threshold after all four previous shapers have run.

\subsection{Recovery Mechanisms}
\label{sec:turn:recovery}

The query loop implements several recovery mechanisms for edge cases:

\begin{itemize}[nosep]
  \item \textbf{Max output tokens escalation}: When the response hits the output token cap, the system can retry with an escalated limit, subject to a GrowthBook flag and the absence of an existing override or environment-variable cap. Up to three recovery attempts are allowed per turn (\code{MAX\_OUTPUT\_TOKENS\_RECOVERY\_LIMIT = 3}).
  \item \textbf{Reactive compaction} (gated by \code{REACTIVE\_COMPACT}): After a prompt-too-long or oversized-media error, the loop can attempt reactive compaction and retry the request if recovery succeeds. The \code{hasAttemptedReactiveCompact} flag ensures this fires at most once per turn.
  \item \textbf{Prompt-too-long handling}: If the API returns a \code{prompt\_too\_long} error, the loop first attempts context-collapse overflow recovery and reactive compaction. Only after these fail does it terminate with \code{reason: 'prompt\_too\_long'}.
  \item \textbf{Streaming fallback}: The \code{onStreamingFallback} callback handles streaming API issues, allowing the loop to retry with a different strategy.
  \item \textbf{Fallback model}: The \code{fallbackModel} parameter enables switching to an alternative model if the primary model fails.
\end{itemize}

\subsection{Stop Conditions}
\label{sec:turn:stop}

Multiple conditions can terminate the loop:

\begin{enumerate}[nosep]
  \item \textbf{No tool use}: The model produces only text content (the primary stop condition).
  \item \textbf{Max turns}: The configurable \code{maxTurns} limit is reached.
  \item \textbf{Context overflow}: The API returns \code{prompt\_too\_long}.
  \item \textbf{Hook intervention}: A PostToolUse hook sets \code{hook\_stopped\_continuation}.
  \item \textbf{Explicit abort}: The \code{abortController} signal fires.
\end{enumerate}

The turn pipeline determines \emph{how} tool requests are orchestrated and recovered. The next section examines the gate that determines \emph{whether} each request is permitted to execute at all.

\begin{figure}[!t]
  \centering
  \begin{minipage}[c]{0.54\linewidth}
      \centering
      \includegraphics[width=\linewidth]{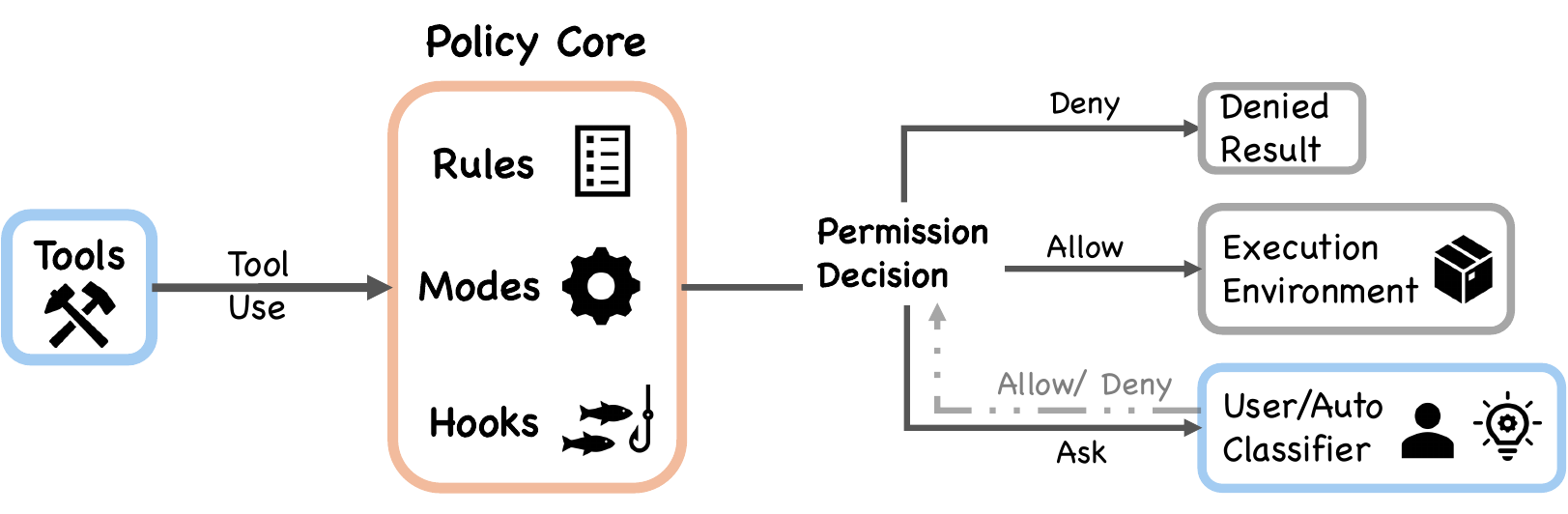}
  \end{minipage}%
  \hfill
  \begin{minipage}[c]{0.44\linewidth}
      \centering
      \scriptsize
      \setlength{\tabcolsep}{4pt}
      \renewcommand{\arraystretch}{1.1}
      \begin{tabular}{@{}>{\centering\arraybackslash}p{2.0cm}>{\raggedright\arraybackslash}p{4.2cm}@{}}
          \toprule
          \textbf{Principle} & \textbf{Description} \\
          \midrule
          Progressive Trust & The agent starts with minimal autonomy; users expand it by approving tool invocations that become permanent rules. \\
          \addlinespace
          Deny-First, Ask-by-Default & Deny rules always win, even under looser modes. In default and manual approval modes, unmatched risk-bearing actions ask the user instead of running silently; optional auto mode routes many such reviews through classifier-mediated checks. \\
          \addlinespace
          Composable Policy & Three mechanisms shape policy: declarative rules, global trust modes, and programmable hooks, each independently configurable. \\
          \bottomrule
      \end{tabular}
  \end{minipage}
  \caption{Permission gate overview and design principles.}
  \label{fig:permission}
\end{figure}

\section{Tool Authorization and Control Boundaries}
\label{sec:auth}

Production coding agents adopt different safety architectures: layered policy enforcement, OS-level sandboxing, or version-control-based rollback. Claude Code combines the first two, implementing four design principles from \Cref{tab:principles}: \emph{deny-first with human escalation}, \emph{graduated trust spectrum}, \emph{defense in depth with layered mechanisms}, and \emph{reversibility-weighted risk assessment}.

When Claude decides to execute a tool (for example, running \code{npm test} via BashTool to reproduce the auth test failure), the request enters the permission pipeline shown in \Cref{fig:permission}. Except in bypass mode, tool invocations pass through the permission system. In the default interactive posture, actions beyond read-only access ask for approval rather than run silently, while more permissive modes replace some prompts with scoped auto-approval, denial, or classifier-mediated review. This default is motivated by a documented behavioral pattern: Anthropic's auto-mode analysis~\citep{anthropic2026automode} found that users approve approximately 93\% of permission prompts, indicating that approval fatigue renders interactive confirmation behaviorally unreliable as a sole safety mechanism. Because users habitually approve without careful review, the system must maintain safety independently of human vigilance. This motivates the architectural commitment to deny-first evaluation, blanket-deny pre-filtering, and sandboxing as independent layers that operate regardless of user attentiveness.

\subsection{Permission Modes and Rule Evaluation}
\label{sec:auth:modes}

Seven permission modes exist across the type definitions (5 external modes at \file{types/permissions.ts}; \code{auto} added conditionally; \code{bubble} in the type union):

\begin{enumerate}[nosep]
  \item \code{plan}: The model must create a plan; execution proceeds only after user approval.
  \item \code{default}: Standard interactive use. Most operations require user approval.
  \item \code{acceptEdits}: Edits within the working directory and certain filesystem shell commands (mkdir, rmdir, touch, rm, mv, cp, sed) are auto-approved; other shell commands require approval.
  \item \code{auto}: An ML-based classifier evaluates requests that do not pass fast-path checks (gated by \code{TRANSCRIPT\_CLASSIFIER}).
  \item \code{dontAsk}: Prompts are suppressed, and actions that would otherwise prompt are auto-denied; explicit allow and deny rules still apply.
  \item \code{bypassPermissions}: Skips most permission prompts, but safety-critical checks and bypass-immune rules still apply.
  \item \code{bubble}: Internal-only mode for subagent permission escalation to the parent terminal.
\end{enumerate}

The five externally visible modes (\code{acceptEdits}, \code{bypassPermissions}, \code{default}, \code{dontAsk}, \code{plan}) are defined in the \code{EXTERNAL\_PERMISSION\_MODES} array. The \code{auto} mode is conditionally included only when the \code{TRANSCRIPT\_CLASSIFIER} feature flag is active. The \code{bubble} mode exists in the type union but not in either mode array; it is used internally for subagent permission escalation (\Cref{sec:subagent}).

Permission rules are evaluated in deny-first order (\file{permissions.ts}). The \func{toolMatchesRule} function checks deny rules first: a deny rule always takes precedence over an allow rule, even when the allow rule is more specific. A broad deny (``deny all shell commands'') cannot be overridden by a narrow allow (``allow \code{npm test}''). The rule system supports tool-level matching (by tool name) and content-level matching (matching specific tool input patterns, such as \code{Bash(prefix:npm)}).

The seven modes span a graduated autonomy spectrum, from \code{plan} (user approves all plans before execution) through \code{default} and \code{acceptEdits} to \code{bypassPermissions} (minimal prompting). This gradient reflects a recurring design tension: as autonomy increases, the system must shift from interactive approval to automated safety checks. Other agent systems resolve this tension differently. SWE-Agent and OpenHands~\citep{yang2024sweagent,wang2024openhands} use Docker container isolation, sandboxing the agent's entire execution environment rather than evaluating individual tool invocations. Aider~\citep{gauthier2024aider} relies on Git as a safety net, making all changes reversible through version control. Claude Code's approach layers multiple policy-enforcement mechanisms on top of an optional shell sandbox (\Cref{sec:auth:sandbox}), trading simplicity for fine-grained control over individual actions.

\subsection{The Authorization Pipeline}
\label{sec:auth:pipeline}

The full authorization pipeline proceeds through several stages:

\paragraph{Pre-filtering.}
Before any tool request reaches runtime evaluation, \func{filterToolsByDenyRules} (\file{tools.ts}) strips blanket-denied tools from the model's view entirely at tool pool assembly time. The documentation states: ``Uses the same matcher as the runtime permission check, so MCP server-prefix rules like \code{mcp\_\_server} strip all tools from that server before the model sees them.'' This prevents the model from attempting to invoke forbidden tools, so the model does not waste calls on them.

\paragraph{PreToolUse hook.}
Registered hooks fire as part of the permission pipeline. A PreToolUse hook can return a \code{permissionDecision} to deny or ask, or an \code{updatedInput} that modifies the tool's input parameters (\file{types/hooks.ts}). A hook \code{allow} does not bypass subsequent rule-based denies or safety checks. In the interactive path, the user dialog is queued first and hooks run asynchronously; coordinator and similar background-agent paths await automated checks before showing the dialog.

\paragraph{Rule evaluation.}
The deny-first rule engine evaluates the request. MCP tools are matched by their fully qualified \code{mcp\_\_server\_\_tool} name, and server-level rules match all tools from that server.

\paragraph{Permission handler.}
The handler in \file{useCanUseTool.tsx} branches into one of four paths based on runtime context:

\begin{enumerate}[nosep]
  \item \textbf{Coordinator}: For multi-agent coordination mode. Attempts automated resolution (classifier, hooks, rules) before falling back to user interaction.
  \item \textbf{Swarm worker}: Handles worker agents in a multi-agent swarm with their own resolution logic.
  \item \textbf{Speculative classifier}: When \code{BASH\_CLASSIFIER} is enabled and the tool is BashTool, a speculative classifier races a pre-started classification result against a timeout. If the classifier returns with high confidence, the tool is approved instantly without user interaction.
  \item \textbf{Interactive}: The fallback path. Presents the standard user approval dialog through the terminal UI.
\end{enumerate}

In coordinator and some background paths, automated resolution is attempted before user interaction. In the standard interactive path, the dialog can appear first while hooks or classifier checks continue in parallel. When the classifier or a deny rule blocks an action, the system treats the denial as a routing signal rather than a hard stop: the model receives the denial reason, revises its approach, and attempts a safer alternative in the next loop iteration. The \code{PermissionDenied} hook event (\Cref{sec:ext}) enables external code to observe and respond to these denials programmatically. This recovery-oriented design means that permission enforcement shapes the agent's behavior rather than simply halting it.

\subsection{Auto-Mode Classifier and Hook Lifecycle}
\label{sec:auth:classifier}

The auto-mode classifier (\file{yoloClassifier.ts}) participates in permission decisions when enabled. When \code{TRANSCRIPT\_CLASSIFIER} is enabled, the classifier loads three prompt resources:

\begin{itemize}[nosep]
  \item A base system prompt.
  \item An external permissions template.
  \item For Anthropic-internal users, a separate internal template.
\end{itemize}

The classifier evaluates the proposed tool invocation against the conversation transcript and the permission template, producing an allow, deny, or request for manual approval. The function \func{isUsingExternalPermissions} checks \code{USER\_TYPE} and a \code{forceExternalPermissions} config flag to select the appropriate template.

Of the 27 hook events defined in the source (\file{coreTypes.ts}), five participate directly in the permission flow, each with a specific Zod-validated output schema (\file{types/hooks.ts}):

\begin{itemize}[nosep]
  \item \textbf{PreToolUse}: Can return \code{permissionDecision} (deny or ask, but allow does not bypass subsequent checks), \code{permissionDecisionReason}, and \code{updatedInput} (modify parameters).
  \item \textbf{PostToolUse}: Can inject \code{additionalContext} and, for MCP tools, return \code{updatedMCPToolOutput} to modify results before they enter the context.
  \item \textbf{PostToolUseFailure}: Can inject \code{additionalContext} for error-specific guidance.
  \item \textbf{PermissionDenied}: Can provide \code{retry} guidance after auto-mode denials.
  \item \textbf{PermissionRequest}: Can return a \code{decision} of \code{allow} or \code{deny}. In coordinator and similar paths, this can resolve before the user dialog. In the standard interactive path, it can also run alongside the dialog.
\end{itemize}

For non-MCP tools, the \code{tool\_result} is emitted before the PostToolUse hook fires. For MCP tools, the result is delayed until after post hooks have run, enabling \code{updatedMCPToolOutput} to take effect.

\subsection{Shell Sandboxing}
\label{sec:auth:sandbox}

Shell sandboxing provides an additional layer of protection for Bash and PowerShell commands (\file{shouldUseSandbox.ts}). The \func{shouldUseSandbox} function checks whether sandboxing is globally enabled, whether the invocation has opted out, and whether the command matches any exclusion patterns.

When active, the sandbox provides filesystem and network isolation independent of the application-level permission model.\footnote{Independent security research reports a parser-differential gap in this network confinement during the analyzed window: the host allowlist used a JavaScript suffix match while name resolution truncated at an embedded null byte, so a crafted hostname could pass the suffix check yet resolve to a different, blocked host. The reported affected range (v2.0.24--v2.1.89) includes the v2.1.88 snapshot analyzed here, and no CVE was assigned~\citep{cybersecnews2026sandbox} (\tierC{}).} A command can be permission-approved but still sandboxed, or permission-denied and never reach the sandbox check. The two systems operate on different axes: authorization versus isolation.

The layered safety architecture rests on an independence assumption: if one layer fails, others catch the violation. However, several layers share common performance constraints. Security researchers~\citep{adversa2026bypass} have documented that commands with more than 50 subcommands fall back to a single generic approval prompt instead of running per-subcommand deny-rule checks, because per-subcommand parsing caused UI freezes. This example demonstrates that defense-in-depth can degrade when its layers share failure modes, a structural tension between safety and performance analyzed further in \Cref{sec:discuss:tradeoffs}.

The permission pipeline governs whether a tool request executes. The next section examines what determines which tools exist in the first place: the extensibility architecture that assembles the model's action surface.

\section{Extensibility: MCP, Plugins, Skills, and Hooks}
\label{sec:ext}

A recurring design question for coding agents is how to structure the extension surface: a single unified mechanism, a small number of specialized mechanisms, or a layered stack with different context costs. The analysis here illustrates two design principles from \Cref{tab:principles}: \emph{composable multi-mechanism extensibility} and \emph{externalized programmable policy}. Returning to the running example, once Claude is trying to repair \file{auth.test.ts} and the earlier \code{npm test} request has been mediated by the permission system (\Cref{sec:auth}), the next question is what extension-enabled action surface is available for the repair. When a turn begins in Claude Code, the model sees not just built-in tools like BashTool and FileReadTool, but also database query tools from an MCP server, a custom lint skill from \file{.claude/skills/}, and tools contributed by an installed plugin. These arrive through four mechanisms that extend the agent at different points of the loop: MCP servers provide external tool integration, plugins package and distribute bundles of components, skills inject domain-specific instructions, and hooks intercept the tool execution lifecycle. Anthropic's documentation~\citep{anthropic2026howworks} presents a broader view that includes CLAUDE.md (\Cref{sec:context}) and subagents (\Cref{sec:subagent}) alongside the four mechanisms analyzed here. We treat CLAUDE.md and subagents in their own sections because they operate in different subsystems (context construction and delegation, respectively), but the context-cost ordering is architecturally significant: it reveals how each extension point trades off expressiveness against the bounded context window.

\subsection{Four Extension Mechanisms}
\label{sec:ext:mechanisms}

\begin{figure*}[!t]
  \centering
  \begin{minipage}[t]{0.58\textwidth}
    \vspace{0pt}
\begin{lstlisting}[style=agentloop]
# one turn of Claude Code's agent loop
while not stopped:
    # (*@\hi{\circled{a}}@*) (*@\hi{assemble}@*): build what the model sees
    context = assemble(
        system_prompt,     # instructions header
        tool_schemas,      # callable tool signatures
        history,           # prior turn messages
        hook_additions,    # pushed in by hooks
    )
    # (*@\hi{\circled{b}}@*) (*@\hi{model}@*): pick the next action
    action = model(context, tools)    # flat tool pool
    if action.is_text_only():
        stopped = run_stop_hooks(action)    # may veto
        continue
    # (*@\hi{\circled{c}}@*) (*@\hi{execute}@*): authorize and run the tool call
    action, hook_decision = run_pre_tool_hooks(action)  
    if not permitted(action, hook_decision):            
        continue
    result = execute(action)                            # tool runs here
    result = run_post_tool_hooks(result)    # mutate/annotate
    history.append(action, result)
\end{lstlisting}
    \vspace{4pt}

    \footnotesize
    \renewcommand{\arraystretch}{1.08}
    \textbf{\hi{\circled{a}}~\hi{\texttt{assemble()}}: what the model sees}\\[2pt]
    \begin{tabular}{@{}c@{\hspace{4pt}}p{0.58\linewidth}@{}}
      \toprule
      Element & What it does \\
      \midrule
      \texttt{CLAUDE.md} files & Loaded into context; files above the working directory load at startup, and subdirectory files load on demand \\
      Skill descriptions & Advertises skills so the model calls \texttt{SkillTool} \\
      MCP resources \& prompts & Non-tool content an MCP server pushes \\
      Output style & Replaces the response-formatting system block \\
      \texttt{UserPromptSubmit} hook & Inject context, or block, on every user turn \\
      \texttt{SessionStart} hook & One-shot context injection at session start \\
      \bottomrule
    \end{tabular}
  \end{minipage}%
  \hfill
  \begin{minipage}[t]{0.4\textwidth}
    \vspace{0pt}
    \footnotesize
    \renewcommand{\arraystretch}{1.08}
    \textbf{\hi{\circled{b}}~\hi{\texttt{model()}}: what the model can reach}\\[2pt]
    \begin{tabular}{@{}c@{\hspace{4pt}}p{0.55\linewidth}@{}}
      \toprule
      Element & What it does \\
      \midrule
      Built-in tools & Read / Edit / Bash / \ldots\ shipped with the CLI \\
      MCP tools & Tools from any MCP server, in the same flat pool \\
      \texttt{SkillTool} & Meta-tool that launches a skill by name \\
      \texttt{AgentTool} & Meta-tool that spawns a sub-agent recursively \\
      \bottomrule
    \end{tabular}
    \vspace{12pt}

    \textbf{\hi{\circled{c}}~\hi{\texttt{execute()}}: whether / how an action runs}\\[2pt]
    \begin{tabular}{@{}c@{\hspace{4pt}}p{0.4\linewidth}@{}}
      \toprule
      Element & What it does \\
      \midrule
      Permission rules & Declarative \texttt{allow} / \texttt{deny} / \texttt{ask} per call \\
      \texttt{PreToolUse} hook & Approve / block / rewrite a tool call \\
      \texttt{PostToolUse} hook & Mutate output or inject context after a call \\
      \texttt{Stop} hook & Force the loop to keep going at model stop \\
      \texttt{SubagentStop} hook & Same, for sub-agents spawned via \texttt{AgentTool} \\
      \texttt{Notification} hook & External side effects on user notifications \\
      \bottomrule
    \end{tabular}
  \end{minipage}
  \caption{Where Claude Code's extension mechanisms plug into the agent loop. The pseudocode on the left is a zoom-in of the \texttt{Agent Loop} block in Figure~1. Every agent loop has three injection points: \protect\hi{\protect\circled{a}}~\hi{\texttt{assemble()}} controls what the model sees, \protect\hi{\protect\circled{b}}~\hi{\texttt{model()}} controls what it can reach, and \protect\hi{\protect\circled{c}}~\hi{\texttt{execute()}} controls whether and how an action actually runs.}
  \label{fig:agent-loop-extensions}
\end{figure*}

The mechanisms are implemented in distinct source directories (\Cref{fig:agent-loop-extensions}) and serve different integration patterns:

\paragraph{MCP servers.}
The Model Context Protocol is the primary external tool integration path. MCP servers are configured from multiple scopes: project, user, local, and enterprise, with additional plugin and \code{claude.ai} servers merged at runtime (\file{services/mcp/config.ts}). The MCP client (\file{services/mcp/client.ts}) supports multiple transport types: stdio, SSE, HTTP, WebSocket, SDK, plus IDE-specific variants (\code{sse-ide}, \code{ws-ide}) and an internal \code{claudeai-proxy}. Each connected server contributes tool definitions as \code{MCPTool} objects. Dedicated built-in tools \code{ListMcpResourcesTool} and \code{ReadMcpResourceTool} provide access to MCP resources.

\paragraph{Plugins.}
Plugins serve a dual role: they are both a packaging format and a distribution mechanism. The \code{PluginManifestSchema} (\file{utils/plugins/schemas.ts}) accepts ten component types: commands, agents, skills, hooks, MCP servers, LSP servers, output styles, channels, settings, and user configuration. The plugin loader (\file{utils/plugins/pluginLoader.ts}) validates manifests and routes each component to its respective registry: commands and skills surface through the \code{SkillTool} meta-tool, agents appear in definitions consumed by \code{AgentTool}, hooks merge into the hook registry, MCP and LSP servers fold into their standard configurations, and output styles modify response formatting. A single plugin package can therefore extend Claude Code across multiple component types simultaneously, making plugins the primary distribution vehicle for third-party extensions.

\paragraph{Skills.}
Each skill is defined by a \file{SKILL.md} file with YAML frontmatter. The \func{parseSkillFrontmatterFields} function (\file{loadSkillsDir.ts}) parses 15+ fields including display name, description, allowed tools (granting the skill access to additional tools), argument hints, model overrides, execution context (\code{'fork'} for isolated execution), associated agent definitions, effort levels, and shell configuration. Skills can define their own hooks, which register dynamically on invocation. Bundled skills are registered in-memory at startup. When invoked, the \code{SkillTool} meta-tool injects the skill's instructions into the context.

\paragraph{Hooks.}
The source code defines 27 hook events spanning tool authorization (\code{PreToolUse}, \code{PostToolUse}, \code{PostToolUseFailure}, \code{PermissionRequest}, \code{PermissionDenied}), session lifecycle (\code{SessionStart}, \code{SessionEnd}, \code{Setup}, \code{Stop}, \code{StopFailure}), user interaction (\code{UserPromptSubmit}, \code{Elicitation}, \code{ElicitationResult}), subagent coordination (\code{SubagentStart}, \code{SubagentStop}, \code{TeammateIdle}, \code{TaskCreated}, \code{TaskCompleted}), context management (\code{PreCompact}, \code{PostCompact}, \code{InstructionsLoaded}, \code{ConfigChange}), workspace events (\code{CwdChanged}, \code{FileChanged}, \code{WorktreeCreate}, \code{WorktreeRemove}), and notifications (\file{coreTypes.ts}, \file{coreSchemas.ts}). Of these, 15 have event-specific output schemas with rich fields supporting permission decisions, context injection, input modification, MCP result transformation, and retry control (\file{types/hooks.ts}). Persisted hook commands configured via settings and plugins use four command types: shell commands (\code{type: command}), LLM prompt hooks (\code{type: prompt}), HTTP hooks (\code{type: http}), and agentic verifier hooks (\code{type: agent}) (\file{schemas/hooks.ts}). The runtime additionally supports non-persistable callback hooks (\code{type: callback}) used by the SDK and internal instrumentation (\file{types/hooks.ts}). Hook sources include \file{settings.json}, plugins, and managed policy at startup; skill hooks register dynamically on invocation (\file{utils/hooks.ts}). The five tool-authorization events are detailed in \Cref{sec:auth:classifier}.

\subsection{Tool Pool Assembly}
\label{sec:ext:assembly}

The \func{assembleToolPool} function at \file{tools.ts} is described in source comments as ``the single source of truth for combining built-in tools with MCP tools.'' The assembly follows a five-step pipeline:

\begin{enumerate}[nosep]
  \item \textbf{Base tool enumeration.} \func{getAllBaseTools} (\file{tools.ts}) returns an array of up to 54 tools: 19 are always included (such as \code{BashTool}, \code{FileReadTool}, \code{AgentTool}, \code{SkillTool}), and 35 more are conditionally included based on feature flags, environment variables, and user type. Anthropic-internal users get additional internal tools. Worktree mode enables \code{EnterWorktreeTool} and \code{ExitWorktreeTool}. Agent swarms enable team tools. When embedded search tools are available in the Bun binary, dedicated \code{GlobTool} and \code{GrepTool} are omitted.
  \item \textbf{Mode filtering.} \func{getTools} (\file{tools.ts}) applies mode-specific filtering. In \code{CLAUDE\_CODE\_SIMPLE} mode, only Bash, Read, and Edit are available (or \code{REPLTool} in the REPL branch; plus coordinator tools if applicable). Each tool's \func{isEnabled} method is called for runtime availability checks.
  \item \textbf{Deny rule pre-filtering.} \func{filterToolsByDenyRules} (\file{tools.ts}) strips blanket-denied tools from the model's view before any call.
  \item \textbf{MCP tool integration.} MCP tools from \code{appState.mcp.tools} are filtered by deny rules and merged with built-in tools.
  \item \textbf{Deduplication.} Tools are deduplicated by name, with built-in tools taking precedence over MCP tools.
\end{enumerate}

Both \file{REPL.tsx} (via the \code{useMergedTools} hook) and \file{AgentTool.tsx} (when building the worker tool set) invoke this function, ensuring consistent assembly across all execution paths. At request time, deferred tools may be hidden from the model's context until explicitly queried via ToolSearch (\file{tools.ts}).

Agent-based extension (custom agent definitions via \file{.claude/agents/*.md} and plugin-contributed agents) is covered in \Cref{sec:subagent}, because agents differ fundamentally from the four mechanisms above: they create new, isolated context windows rather than extending the current one.

\subsection{Why Four Mechanisms?}
\label{sec:ext:why-four}

Given that each additional extension mechanism increases the surface area developers must learn, a natural question is why Claude Code uses four distinct mechanisms rather than consolidating into one or two. The answer lies in the observation that different kinds of extensibility impose different costs on the context window, and a single mechanism cannot span the full range from zero-context lifecycle hooks to schema-heavy tool servers without forcing unnecessary trade-offs on extension authors.

\begin{table}[h]
  \centering
  \small
  \renewcommand{\arraystretch}{1.15}
  \caption{What each extension mechanism uniquely provides. Context cost refers to how much of the bounded context window the mechanism consumes when active.}
  \label{tab:ext-why-four}
  \resizebox{\columnwidth}{!}{
  \begin{tabular}{@{}llll@{}}
    \toprule
    Mechanism & Unique Capability & Context Cost & Insertion Point \\
    \midrule
    MCP servers & External service integration (multi-transport) & High once tool schemas load & \texttt{model()}:tool pool \\
    Plugins & Multi-component packaging + distribution & Medium (varies) & All three points \\
    Skills & Domain-specific instructions + meta-tool invocation & Low until invoked & \texttt{assemble()}:context injection \\
    Hooks & Lifecycle interception + event-driven automation & Zero by default & \texttt{execute()}:pre/post tool \\
    \bottomrule
  \end{tabular}}
\end{table}

As \Cref{tab:ext-why-four} summarizes, each mechanism trades deployment complexity for a different kind of extensibility. MCP servers provide runtime tool integration (the model gains new callable tools) at the cost of server management overhead and context budget consumed by tool schemas, which by default load only when a tool is needed. Skills shape \emph{how} the agent thinks (not just what tools it has) at minimal context cost, since only frontmatter descriptions stay in the prompt until the skill is invoked. Hooks provide cross-cutting lifecycle control (blocking, rewriting, or annotating tool calls) with no context footprint by default, though hooks can opt into injecting additional context. Plugins bundle any combination of the other three into distributable packages, acting as the packaging and distribution layer rather than a distinct runtime primitive. The graduated context-cost ordering (zero for hooks, low for skills before invocation, medium for plugins, high for MCP) means that cheap extensions can scale widely without exhausting the context window, while expensive ones are reserved for cases that genuinely require new tool surfaces.

Some agent frameworks provide a single extension mechanism, typically a tool-only API where all customization arrives as additional callable tools. Others use two tiers, separating tools from configuration or instruction injection. Claude Code's four-mechanism approach can accommodate a broader range of extension patterns, from zero-context event handlers to full external service integrations, but it increases the learning curve developers face when deciding which mechanism to use for a given integration task.

\section{Context Construction and Memory}
\label{sec:context}

How an agent manages its context window and persists user instructions is a central design choice, with different systems choosing between file-based transparency, database-backed retrieval, and opaque learned representations. The design choices here implement two principles from \Cref{tab:principles}: \emph{context as scarce resource with progressive management} and \emph{transparent file-based configuration and memory}.

By this point in the running example, the task has accumulated state: the original request, the \code{npm test} permission outcome, the tool pool assembled in \Cref{sec:ext}, and any file reads or command outputs gathered so far. This section asks how that growing state is packed into Claude Code's bounded context window before the next model call.

Before the model is called, the agent loop assembles a context window from the tool pool (\Cref{sec:ext}), CLAUDE.md files, auto memory, and conversation history. The following subsections cover the assembly order, the CLAUDE.md hierarchy, and the multi-step compaction pipeline.

\subsection{Context Window Assembly}
\label{sec:context:assembly}

\begin{figure}[t]
\centering
\includegraphics[width=\textwidth]{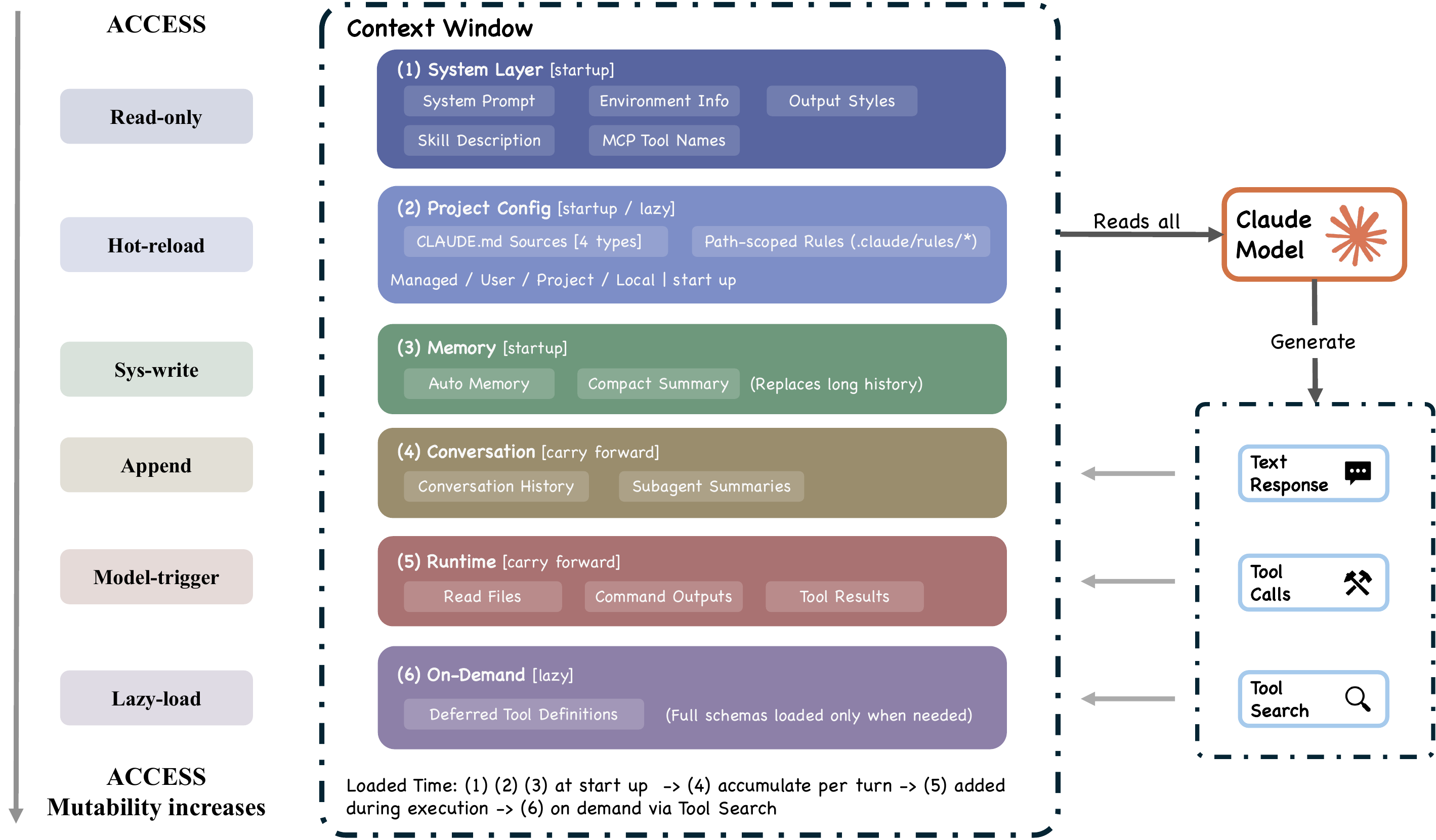}
\caption{Context construction and memory hierarchy. Sources converging on the context window include system prompt, output styles, environment info, the CLAUDE.md hierarchy (managed, user, project, and local), auto memory, path-scoped rules, MCP tool names, deferred tool definitions via ToolSearch, conversation history, file reads, command outputs, tool results, subagent summaries, and compact summaries.}
\label{fig:context}
\end{figure}

The context window (\Cref{fig:context}) is assembled from the following sources, some at initial assembly and others injected late during the turn:

\begin{enumerate}[nosep]
  \item \textbf{System prompt}, incorporating output style modifications and any \code{-{}-append-system-prompt} flag content.
  \item \textbf{Environment info} via \func{getSystemContext} (\file{context.ts}): git status (skipped in remote mode or when git instructions are disabled) and an optional cache-breaking injection for internal builds (gated by \code{BREAK\_CACHE\_COMMAND}). Memoized once per session.
  \item \textbf{CLAUDE.md hierarchy} via \func{getUserContext} (\file{context.ts}): four-level instruction file hierarchy (\Cref{sec:context:claudemd}). Also memoized.
  \item \textbf{Path-scoped rules}: conditional and directory-matched rules that load lazily when the agent reads files in matching directories.
  \item \textbf{Auto memory}: contextually relevant memory entries prefetched asynchronously.
  \item \textbf{Tool metadata:} skill descriptions, MCP tool names, and deferred tool definitions (via ToolSearch, on demand).
  \item \textbf{Conversation history}: carried forward, subject to compaction.
  \item \textbf{Tool results}: file reads, command outputs, subagent summaries.
  \item \textbf{Compact summaries}: replacing older history segments.
\end{enumerate}

The system prompt assembly at \file{query.ts} combines system context with the base prompt via \func{asSystemPrompt(appendSystemContext(systemPrompt, systemContext))}. User context (CLAUDE.md and date) is prepended to the message array via \func{prependUserContext}. This separation means CLAUDE.md content occupies a different structural position in the API request than the system prompt, potentially affecting model attention patterns.

Several context sources are injected late, after the main window is constructed: relevant-memory prefetch (\file{query.ts}), MCP instructions deltas (only new or changed server instructions), agent listing deltas, and background agent task notifications. The context window is therefore not static at assembly time but can grow during the turn.

\subsection{CLAUDE.md Hierarchy and Auto Memory}
\label{sec:context:claudemd}

A design principle shapes the memory system: stored context should be inspectable and editable by the user. CLAUDE.md files are plain-text Markdown rather than structured configuration or opaque database entries. This transparency choice trades expressiveness for auditability: users can directly inspect and edit the memory files they control, and project-level memory can be version-controlled alongside the codebase~\citep{anthropic2026memory,mindstudio2025memory}. Alternative memory architectures illustrate the trade-off. Retrieval-augmented approaches use embedding-based lookup to surface relevant prior context, gaining flexibility at the cost of inspectability: the user cannot easily see or edit what the retrieval system considers relevant. Database-backed memory offers structured querying but requires additional infrastructure and is opaque to version control. Claude Code's file-based approach makes these memory surfaces readable and editable, even though other context sources still enter through the system prompt, tools, hooks, MCP servers, and runtime state. The system does not use embeddings or a vector similarity index for memory retrieval; instead it uses an LLM-based scan of memory-file headers to select up to five relevant files on demand, surfacing them at file granularity rather than entry granularity. Embedding-based systems can retrieve individual entries more selectively, at the cost of inspectability and the infrastructure needed to maintain an index.

CLAUDE.md files follow a multi-level loading hierarchy. The source header (\file{claudemd.ts}) defines four memory types:

\begin{enumerate}[nosep]
  \item \textbf{Managed memory} (e.g. \file{/etc/claude-code/CLAUDE.md} on Linux): OS-level policy for all users.
  \item \textbf{User memory} (\file{\textasciitilde/.claude/CLAUDE.md}): private global instructions.
  \item \textbf{Project memory} (\file{CLAUDE.md}, \file{.claude/CLAUDE.md}, and \file{.claude/rules/*.md} in project roots): instructions checked into the codebase.
  \item \textbf{Local memory} (\file{CLAUDE.local.md} in project roots): gitignored, for private project-specific instructions.
\end{enumerate}

File discovery traverses from the current directory up to root, checking for all project and local memory files in each directory. Files closer to the current directory have higher priority (loaded later).

Files load in ``reverse order of priority'': later-loaded files receive more model attention. For root-to-CWD directories, unconditional rules from \file{.claude/rules/*.md} load eagerly at startup. For nested directories below CWD, even unconditional rules are loaded lazily when the agent reads files in matching directories. This means the model's instruction set can evolve during a conversation as new parts of the codebase are explored.

CLAUDE.md content is delivered as user context (a user message), not as system prompt content (\file{context.ts}). This architectural choice has a significant implication: because CLAUDE.md content is delivered as conversational context rather than system-level instructions, model compliance with these instructions is probabilistic rather than guaranteed. Permission rules evaluated in deny-first order (\Cref{sec:auth}) provide the deterministic enforcement layer. This creates a deliberate separation between guidance (CLAUDE.md, probabilistic) and enforcement (permission rules, deterministic). The function calls \func{setCachedClaudeMdContent} to cache the loaded content for the auto-mode classifier, to avoid an import cycle between the CLAUDE.md loader and the permission system.

Memory files support an \code{@include} directive for modular instruction sets (\func{processMemoryFile} at \file{claudemd.ts}). Syntax variants include \code{@path}, \code{@./relative}, \code{@\textasciitilde/home}, and \code{@/absolute}. The directive works in leaf text nodes only (not inside code blocks). In the implementation, the including file is pushed first and included files are appended after it, circular references are prevented by tracking processed paths, and non-existent files are silently ignored.

\subsection{Compaction Pipeline}
\label{sec:context:compaction}

The five-layer compaction pipeline (\Cref{sec:turn:shapers}) implements the ``context as bottleneck'' principle through graduated compression (\file{query.ts}). Rather than a single strategy, Claude Code applies five layers in sequence, each with increasing aggressiveness (three are gated by build-time feature flags and budget reduction by a runtime flag, while auto-compact is user-configurable). This graduated approach contrasts with simpler alternatives such as single-pass truncation (dropping the oldest messages) or a single summarization step. The graduated design reflects a lazy-degradation principle: apply the least disruptive compression first, escalating only when cheaper strategies prove insufficient. The cost of this approach is complexity. Five interacting compression layers, several gated by feature flags, create behavior that is difficult for users to fully predict. Auto-compact produces a visible summary in the transcript, and microcompact emits a boundary marker, but context collapse operates without user-visible output. Simpler single-pass approaches sacrifice information but are easier to reason about.

\begin{enumerate}[nosep]
  \item \textbf{Budget reduction} (runtime flag): per-tool-result size limits.
  \item \textbf{Snip} (\code{HISTORY\_SNIP}): lightweight older-history trimming.
  \item \textbf{Microcompact} (\code{CACHED\_MICROCOMPACT}): fine-grained cache-aware compression.
  \item \textbf{Context collapse} (\code{CONTEXT\_COLLAPSE}): read-time virtual projection over history.
  \item \textbf{Auto-compact} (enabled by default, can be disabled): full model-generated summary.
\end{enumerate}

The \func{buildPostCompactMessages} function (\file{compact.ts}) returns the following compacted output structure: \code{[boundaryMarker, ...summaryMessages, ...messagesToKeep, ...attachments, ...hookResults]}. The boundary marker is annotated with preserved-segment metadata via \func{annotateBoundaryWithPreservedSegment}, recording \code{headUuid}, \code{anchorUuid}, and \code{tailUuid} to enable read-time chain patching. This mostly-append design means compaction never modifies or deletes previously written transcript lines; it only appends new boundary and summary events.

The compaction function \func{compactConversation} (\file{compact.ts}) includes several design choices. Pre-compact hooks fire first, allowing hook-injected custom instructions. A GrowthBook feature flag controls whether the compaction path reuses the main conversation's prompt cache (a code comment documents a January 2026 experiment: ``false path is 98\% cache miss, costs $\sim$0.76\% of fleet \code{cache\_creation}''). After compaction, attachment builders re-announce runtime state (plans, skills, and async agents) from live app state, since compaction discards prior attachment messages but not the underlying state.

Context isolation becomes more critical when the system delegates work to subagents, each operating in its own bounded context window.

\section{Subagent Delegation and Orchestration}
\label{sec:subagent}

Multi-agent orchestration is a key design dimension for coding agents, with choices spanning parent-child hierarchies, peer-based conversation frameworks~\citep{wu2024autogen}, and graph-structured workflow engines~\citep{langgraph2024}. Claude Code's delegation architecture implements the \emph{isolated subagent boundaries} principle from \Cref{tab:principles}, together with aspects of \emph{deny-first with human escalation} (permission override) and \emph{reversibility-weighted risk assessment} (subagent tool restrictions).

When Claude determines that the auth test fix requires first exploring the authentication module's structure, it can delegate this exploration to a subagent. The delegation mechanism is the \code{Agent} tool (\file{AgentTool.tsx}), with \code{Task} retained as a legacy alias. The model invokes \code{Agent} with a structured input including the delegation prompt, an optional subagent type, and configuration for isolation mode, permission overrides, and working directory.

\subsection{The Agent Tool and Delegation Criteria}
\label{sec:subagent:tool}

\begin{figure}[t]
\centering
\includegraphics[width=\textwidth]{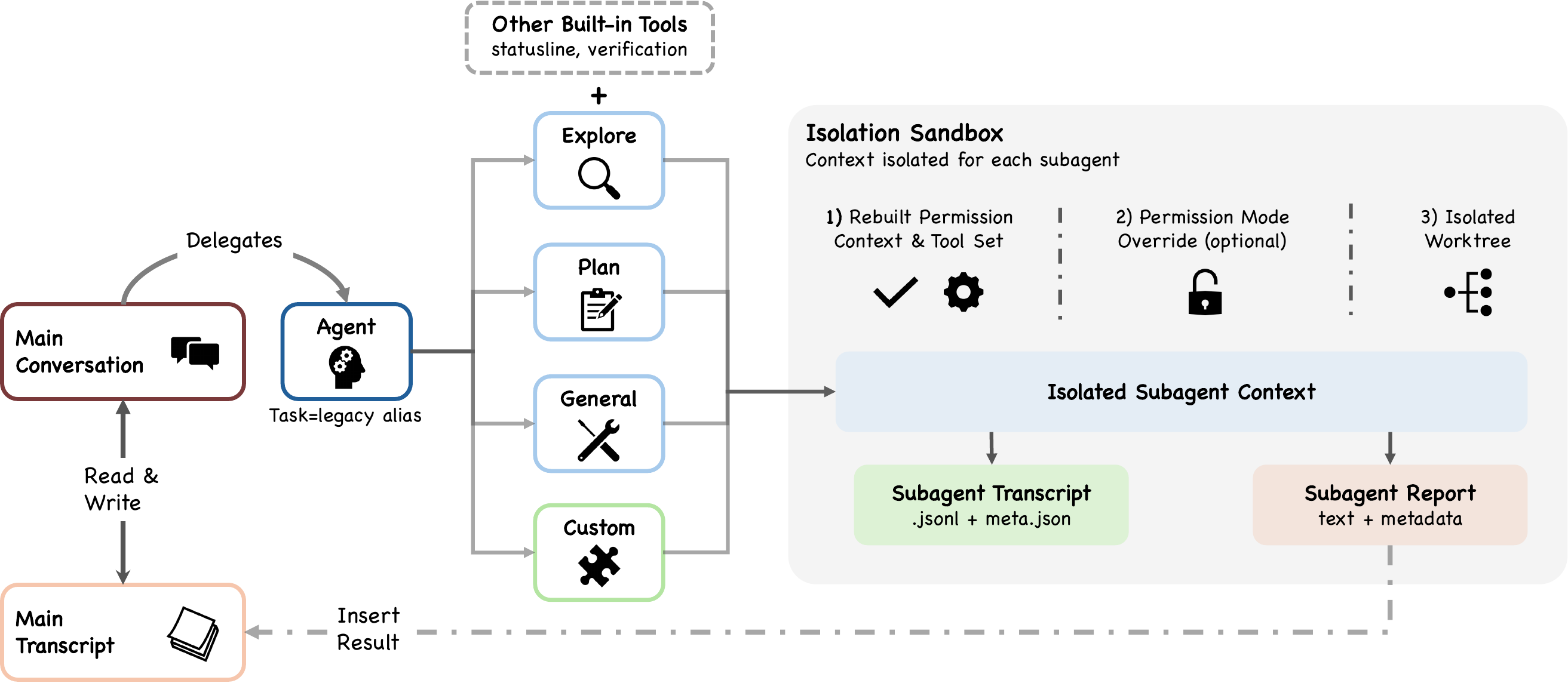}
\caption{Subagent isolation and delegation architecture. The Agent tool dispatches to built-in subagents (Explore, Plan, general-purpose) or custom subagents, each running in an isolated context with rebuilt permission context and independent tool sets. The Agent tool dispatches along three axes: routing (teammate), isolation (remote, worktree), and lifecycle (async, sync).}
\label{fig:subagents}
\end{figure}

The Agent tool input schema (\Cref{fig:subagents}) uses feature-gated fields, omitting optional parameters when their backing features are disabled. The \code{isolation} field offers \code{['worktree', 'remote']} for internal users and \code{['worktree']} for external users, determined at build time. The \code{cwd} field is gated by a feature flag. The \code{run\_in\_background} field is omitted when background tasks are disabled or when fork-subagent mode is enabled.

Claude Code provides up to six built-in subagent types, depending on feature flags and entrypoint:

\begin{itemize}[nosep]
  \item \textbf{Explore}: primarily read/search-oriented investigation, with write and edit tools in its deny-list.
  \item \textbf{Plan}: creates structured plans; execution proceeds through the standard permission model.
  \item \textbf{General-purpose}: broadly capable, used when explicitly requested (note: omitting the type may route to the fork-subagent path instead, which reuses the parent's conversation context rather than isolating it).
  \item \textbf{Claude Code Guide}: onboarding and documentation assistance, with its own \code{permissionMode} override.
  \item \textbf{Verification}: runs validation checks (test suites, linting).
  \item \textbf{Statusline-setup}: specialized for terminal status line configuration.
\end{itemize}

Beyond built-ins, users define custom subagents via \file{.claude/agents/*.md} files, and plugins contribute agent definitions via \file{loadPluginAgents.ts}. The markdown body of each file serves as the agent's system prompt, and YAML frontmatter specifies configuration fields including \code{description}, \code{tools} (allowlist), \code{disallowedTools}, \code{model}, \code{effort}, \code{permissionMode}, \code{mcpServers}, \code{hooks}, \code{maxTurns}, \code{skills}, \code{memory} scope, \code{background} flag, and \code{isolation} mode. JSON-formatted agent definitions support the same fields plus \code{prompt} as an explicit field (\file{loadAgentsDir.ts}). This means a custom agent can be a fully configured, isolated sub-system with its own tools, model, permissions, hooks, memory scope, and isolation mode. \code{AgentTool} sits alongside \code{SkillTool} in the base tool pool as a meta-tool that dispatches to these definitions, but the two differ fundamentally: \code{SkillTool} typically injects instructions into the current context window (with an opt-in \code{context: fork} mode that spawns an isolated sub-agent via the same \func{runAgent} machinery), while \code{AgentTool} always spawns a new, isolated one. The tradeoff is that most subagent invocations require a self-contained prompt, because the default path does not inherit the parent's conversation history (the fork-subagent path is an exception). Conversation-based frameworks that share full transcript histories avoid this cost but risk context explosion as the number of agents grows.

\subsection{Isolation Architecture}
\label{sec:subagent:isolation}

Subagent isolation supports multiple modes (\file{AgentTool.tsx}):

\begin{itemize}[nosep]
  \item \textbf{Worktree}: Creates a temporary git worktree, giving the subagent its own copy of the repository to modify without affecting the parent's working tree.
  \item \textbf{Remote} (internal-only): Launches in a remote Claude Code Remote environment, always running in the background.
  \item \textbf{In-process} (default): Shares the filesystem with the parent but operates in an isolated conversation context.
\end{itemize}

The permission override logic for subagents (\file{runAgent.ts}) involves several specific rules. When a subagent defines a \code{permissionMode}, the override is applied unless the parent is already in \code{bypassPermissions}, \code{acceptEdits}, or \code{auto} mode, since those modes always take precedence because they represent explicit user decisions about the safety/autonomy trade-off. For async agents, the system determines whether to avoid prompts through a cascade: explicit \code{canShowPermissionPrompts} first, then \code{bubble} mode (always show, since they escalate to the parent terminal), then the default (sync agents show prompts, async agents do not). Background agents that can show prompts set \code{awaitAutomatedChecksBeforeDialog: true}, ensuring the classifier and hooks resolve before interrupting the user.

These isolation modes occupy different points in a design space. Container-based isolation (used by SWE-Agent and OpenHands~\citep{yang2024sweagent,wang2024openhands}) provides stronger resource boundaries but requires container infrastructure. Context-only isolation (used by conversation-based frameworks like AutoGen~\citep{wu2024autogen}) shares the filesystem but separates conversation histories. Claude Code's worktree-based isolation provides workspace-level filesystem separation without introducing container orchestration, leveraging Git's built-in mechanism.

When \code{allowedTools} is explicitly provided to \func{runAgent} (\file{runAgent.ts}), a two-tier permission scoping model applies. SDK-level permissions from \code{-{}-allowedTools} are preserved: ``explicit permissions from the SDK consumer that should apply to all agents.'' But session-level rules are replaced with the subagent's declared \code{allowedTools}. When \code{allowedTools} is not provided (the common \code{AgentTool} path), the parent's session-level rules are inherited without replacement.

\subsection{Sidechain Transcripts}
\label{sec:subagent:sidechain}

Each subagent writes its own transcript as a separate \file{.jsonl} file with a \file{.meta.json} metadata file (\file{sessionStorage.ts}, \file{runAgent.ts}). This sidechain design means subagent histories are preserved for debugging and auditing but do not inflate the parent's session file. Only the subagent's final response text and metadata return to the parent conversation context; the full subagent history never enters the parent's context window, respecting the ``context as bottleneck'' principle.

The summary-only return model is a deliberate context-conservation choice: conversation-based frameworks that share full transcript histories between agents risk context explosion as the number of agents grows. Even isolated-context parallelism carries substantial cost. Claude Code's agent teams consume approximately 7$\times$ the tokens of a standard session in plan mode~\citep{anthropic2025agentteams}, which makes summary-only return more critical when subagents are also in isolated contexts.

For multi-instance coordination in agent teams, the harness uses file locking rather than a message broker or distributed coordination service~\citep{anthropic2025agentteams}. Each teammate has its own inbox JSON file (\file{utils/teammateMailbox.ts}); task assignments and other inter-agent messages are pushed into the recipient's inbox under a per-inbox lockfile (\file{utils/lockfile.ts}), with files stored at predictable filesystem paths. This trades throughput for two properties: zero-dependency deployment (no external infrastructure required) and full debuggability (any agent's state can be inspected by reading plain-text JSON files).

\section{Session Persistence and Recovery}
\label{sec:persist}

Session persistence in coding agents involves a design choice between append-only logs, structured databases, checkpoint-based snapshots, and stateless architectures, each with different trade-offs in auditability, query power, and deployment complexity. Claude Code's persistence design implements the \emph{append-only durable state} principle from \Cref{tab:principles}. Session-scoped permissions live in memory only and are not serialized to the transcript, so a new process started with \code{-{}-resume} rebuilds the permission context from CLI args and disk settings; any request that the rebuilt context does not recognize falls back to deny-first prompting.

By the time the auth-test task reaches this section, the session contains the original prompt, tool invocations and results, compact boundaries, and the subagent summary from exploring the authentication module (\Cref{sec:subagent}). This section asks which of those artifacts are durably recorded and what can be recovered later without carrying forward the session's old permission grants.

Claude Code's persistence mechanisms write the conversation (messages, tool results, and compact boundaries) to disk as events occur.

\subsection{Transcript Model}
\label{sec:persist:transcript}

\begin{figure}[t]
    \centering
    \begin{minipage}[c]{0.54\linewidth}
        \centering
        \includegraphics[width=\linewidth]{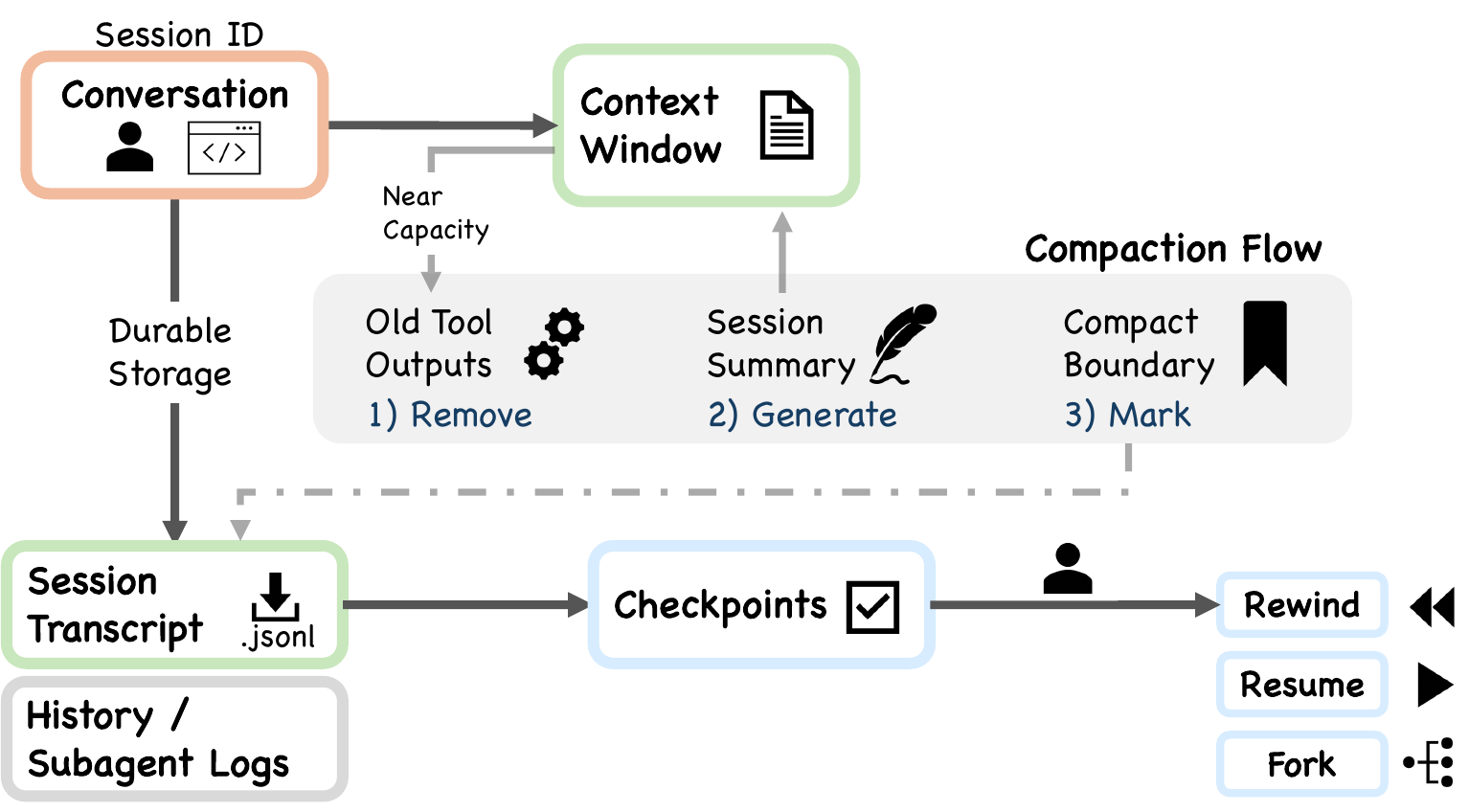}
    \end{minipage}%
    \hfill
    \begin{minipage}[c]{0.44\linewidth}
        \centering
        \scriptsize
        \setlength{\tabcolsep}{4pt}
        \renewcommand{\arraystretch}{1.1}
        \begin{tabular}{@{}>{\centering\arraybackslash}p{2.1cm}>{\raggedright\arraybackslash}p{4.1cm}@{}}
            \toprule
            \textbf{Principle} & \textbf{Description} \\
            \midrule
            Conversations Outlive Context & A session's useful life cannot be capped by the model's context window. The transcript on disk records durable session events, so compaction can recycle the live view without ending the conversation. \\
            \addlinespace
            Conversations Outgrow a Single Path & A session should not be trapped on a single linear trajectory. The append-only transcript lets users rewind, resume, or fork into a new branch without losing prior work. \\
            \bottomrule
        \end{tabular}
    \end{minipage}
    \caption{Session persistence and context compaction. The diagram separates live session state (context window, compaction) from durable storage (session transcripts, history.jsonl, subagent sidechains, checkpoints). The transcript restores messages but not session-scoped permissions.}
    \label{fig:persistence}
\end{figure}

Session transcripts are stored as mostly append-only JSONL files at a project-specific path (with explicit cleanup rewrites as an exception) (\Cref{fig:persistence}). The \func{getTranscriptPath} function (\file{sessionStorage.ts}) computes this as \code{join(projectDir, \$\{getSessionId()\}.jsonl)}, where \code{projectDir} is determined by first checking \func{getSessionProjectDir} (set by \func{switchSession} during resume/branch) and falling back to \func{getProjectDir(getOriginalCwd())}.

Three persistence channels operate independently:

\begin{enumerate}[nosep]
  \item \textbf{Session transcripts}: Conversation records including user, assistant, attachment, and system messages, plus compaction and other metadata events. Project-scoped, one file per session.
  \item \textbf{Global prompt history}: User prompts only, stored in \file{history.jsonl} at the Claude configuration home directory (\file{history.ts}). The \func{makeHistoryReader} generator yields entries in reverse order via \func{readLinesReverse}, supporting Up-arrow and ctrl+r navigation.
  \item \textbf{Subagent sidechains}: Separate \file{.jsonl} + \file{.meta.json} files per subagent (\Cref{sec:subagent:sidechain}).
\end{enumerate}

Session transcripts store several kinds of events beyond simple messages, including compaction markers, file-history snapshots, attribution snapshots, and content-replacement records.
The append-only JSONL format is a deliberate choice favoring auditability and simplicity over query power. Every logged event is human-readable, version-controllable, and inspectable without specialized tooling. Database-backed alternatives would enable richer queries over session history but introduce deployment dependencies and reduce transparency.

The session identity system pairs \code{sessionId} with \code{sessionProjectDir}, set together during resume or branch. The transcript path must use the same project directory that was active when messages were written, to avoid hooks looking in the wrong directory.

\subsection{Resume, Fork, and Not Restoring Permissions}
\label{sec:persist:recovery}

The \code{-{}-resume} flag rebuilds the conversation by replaying the transcript (\file{conversationRecovery.ts}). Fork creates a new session from an existing one (\file{commands/branch/branch.ts}). However, the transcript does not record session-scoped permissions. Resuming with \code{-{}-resume} starts a new process without them, so users must grant them again; resuming or forking inside a running session keeps the grants already given. This is a deliberate safety-conservative design choice: trust decisions are not carried in the transcript. Restoring previously granted permissions from the transcript would create a convenience benefit but risk carrying stale trust decisions into a changed context. The architecture opts for re-granting in a new process over implicit persistence, accepting the resulting user friction.

The \code{compact\_boundary} marker is carefully designed to work with persistence. The \func{annotateBoundaryWithPreservedSegment} function (\file{compact.ts}) records \code{headUuid}, \code{anchorUuid}, and \code{tailUuid} in the boundary event. These UUIDs enable the session loader to patch the message chain at read time: preserved messages keep their original \code{parentUuids} on disk, and the loader uses boundary metadata to link them correctly. This mostly-append design means compaction does not modify or delete previously written transcript lines.

The ``checkpoints'' in Claude Code are file-history checkpoints for \code{-{}-rewind-files}, stored at \file{\textasciitilde/.claude/file-history/<sessionId>/}. These are file-level snapshots for reverting filesystem changes, not a generic checkpoint store.

The preceding sections have documented Claude Code's answers to recurring design questions. The next section contrasts Claude Code's design choices with those of two architecturally independent AI agent systems.

\section{Comparative Analysis: Claude Code, OpenClaw, and Hermes Agent}
\label{sec:compare}

The preceding sections documented Claude Code's answers to recurring design questions about loop architecture, safety, extensibility, context management, delegation, and persistence. To situate these findings within the broader agent design space, this section compares Claude Code with two independent open-source AI agent systems that answer many of the same design questions from fundamentally different starting points. OpenClaw is a local-first WebSocket gateway that connects messaging surfaces (WhatsApp, Telegram, Slack, Discord, Signal, and others) to an embedded agent runtime, with companion apps on macOS, iOS, and Android~\citep{openclaw2026}. Hermes Agent is a single Python process whose role is set by the entry point under which it was invoked, with one persistence layer and many surfaces fronting one runtime~\citep{hermesagent2026}.\footnote{Source-grounded reading of Hermes surfaced two cases where the project's own documentation has drifted from the implementation: \file{SECURITY.md} documents an approval-mode taxonomy of \texttt{on}/\texttt{auto}/\texttt{off} while \file{tools/approval.py}:721-750 returns \texttt{manual}/\texttt{smart}/\texttt{off}, and both \file{SECURITY.md} and \file{AGENTS.md} describe a delegation depth default of two while \file{tools/delegate\_tool.py}:128 sets \texttt{MAX\_DEPTH = 1}. We cite the source rather than the docs.} The Hermes distribution exposes three console scripts: \texttt{hermes} for the CLI and gateway controller, \texttt{hermes-agent} for batch runs, and \texttt{hermes-acp} for the Agent Client Protocol adapter that lets external IDEs host Hermes the way Hermes's own gateway can host other agents. Where Claude Code is a CLI coding harness bound to a single repository session, OpenClaw is a persistent control plane for multi-channel personal assistance, and Hermes is one process whose role and surface set are determined by which entry point launched it. The three systems occupy different regions of the agent design space. The value of the comparison lies in showing how the same recurring questions produce different architectural answers when the deployment context changes.

\subsection{Six Comparison Dimensions}
\label{sec:compare:dimensions}

\Cref{tab:comparison-dimensions} summarizes the comparison across six dimensions. Each dimension corresponds to a design question that all three systems must answer. Hermes capabilities that fall outside these six dimensions are discussed separately in \Cref{sec:compare:hermes}.

\begin{table*}[h!]
\centering
\caption{Architectural comparison: Claude Code vs.\ OpenClaw vs.\ Hermes Agent across six design dimensions. Each row captures a recurring design question and the different answers the three systems provide.}
\label{tab:comparison-dimensions}
\scriptsize
\setlength{\tabcolsep}{4pt}
\begin{tabularx}{\textwidth}{c>{\raggedright\arraybackslash}X>{\raggedright\arraybackslash}X>{\raggedright\arraybackslash}X}
\toprule
\textbf{Dimension} & \textbf{Claude Code} & \textbf{OpenClaw} & \textbf{Hermes Agent} \\
\midrule
System scope & CLI/IDE coding harness, ephemeral per-session process & Persistent WS gateway daemon, multi-channel control plane & Single Python process, role set by entry point (\texttt{hermes}/\texttt{hermes-agent}/\texttt{hermes-acp}); in-tree plus plugin-tier messaging surfaces; two SQLite files (\file{state.db}, \file{kanban.db}) for persistence \\
\midrule
Trust model & Deny-first per-action rule evaluation with hooks and optional ML classifier; 7 permission modes; graduated trust spectrum & Single trusted operator per gateway; DM pairing and allowlists for inbound channels; opt-in sandboxing with configurable scope (per-agent, per-session, or shared) and multiple backends & Per-action approval in one file (\file{tools/approval.py}); 3 modes (\texttt{manual}, \texttt{smart}, \texttt{off}) plus unconditional \texttt{HARDLINE\_PATTERNS} floor; same approval rendered across CLI, gateway keyboards (Telegram, Discord, QQBot), ACP \texttt{request\_permission}, and a cron-only unattended mode \\
\midrule
Agent runtime & Iterative async generator (\func{queryLoop}) as system center & Embedded \texttt{pi-coding-agent} SDK runner inside gateway RPC dispatch; per-session queue serialization (with optional global lane) & Synchronous \texttt{while} loop inside \texttt{AIAgent.run\_conversation} (\file{run\_agent.py}); iteration count default 90 plus an iteration budget, with a tool-stripped summary call on exhaustion; tool files self-register via \texttt{registry.register()}; parallel tool calls with sequential fallback \\
\midrule
Extension architecture & 4 mechanisms at graduated context costs: MCP, plugins, skills, hooks & Manifest-first plugin system with 12 capability types and central registry; separate skills layer; built-in MCP via \texttt{openclaw mcp} (server and outbound client registry) & 5 surfaces: 3 at Claude-Code level (\file{plugins/<name>/} with lifecycle hooks, bundled \file{skills/}, MCP servers in \file{config.yaml}) plus 2 backend-swap surfaces (\file{plugins/memory/<name>/}, \file{plugins/model-providers/<name>/}); hooks are Python callbacks or external shell commands \\
\midrule
Memory and context & CLAUDE.md 4-level hierarchy; 5-layer compaction pipeline; LLM-based memory scan & workspace bootstrap files (AGENTS.md, SOUL.md, TOOLS.md, IDENTITY.md, USER.md, plus conditionally BOOTSTRAP.md, HEARTBEAT.md, and MEMORY.md); separate memory system (MEMORY.md, daily notes, optional DREAMS.md); auto-compaction with pluggable providers; optional hybrid search (vector + keyword, conditional on embedding provider); experimental dreaming for long-term promotion & One auxiliary-LLM summarizer with token-budget tail protection and a tool-output prune pass; summary prefixed with a ``reference only'' preamble; pre-injection scan against context files (\file{AGENTS.md}, \file{.cursorrules}, \file{SOUL.md}) for injection patterns and invisible Unicode; persistent \file{MEMORY.md}, \file{USER.md} remain authoritative across compaction \\
\midrule
Multi-agent and routing & Task-delegating subagents (e.g., Explore, Plan, general-purpose); worktree isolation; final response text returned to parent & Two separate concerns: (a) multi-agent routing with isolated agents, distinct workspaces, and binding-based channel dispatch; (b) sub-agent delegation with configurable nesting depth (max 5, default 1, recommended 2) and thread-bound sessions & \texttt{delegate\_task} spawns child \texttt{AIAgent} in a thread pool; concurrency cap 3 children default, depth cap 1 (clamped to [1,3]); leaf children cannot delegate by default; children run with persistent memory disabled; Kanban subsystem (SQLite-backed work queue) with stale-claim reclamation for multi-process worker coordination \\
\bottomrule
\end{tabularx}
\end{table*}

\paragraph{System scope and deployment model.}
Claude Code runs as an ephemeral CLI process bound to a single repository. Each session starts and ends with the terminal. OpenClaw runs as a persistent daemon (default port 18789, loopback-only) that owns all messaging surface connections and coordinates clients, tools, and device nodes over a typed WebSocket protocol. Hermes Agent runs as a single long-lived Python process whose role is fixed by the entry point used to launch it; its gateway sub-command drives messaging surfaces through a two-tier adapter system (in-tree adapters under \file{gateway/platforms/} plus a plugin tier under \file{plugins/platforms/} for Google Chat, IRC, and Teams), and it persists session and multi-agent-board state to two on-disk SQLite files (\file{state.db} and \file{kanban.db}) rather than to a typed WebSocket protocol. This places Hermes closer to OpenClaw than to Claude Code in trust topology but closer to Claude Code than to OpenClaw in process topology; the design question of where to put the broker is answered, in Hermes, with no broker at all. These differences in system scope are the most fundamental architectural divergences in the comparison: they determine how every other design question is framed. A compositional relationship also exists across all three: OpenClaw can host Claude Code, OpenAI Codex, and Gemini CLI as external coding harnesses through its ACP (Agent Client Protocol) integration, and Hermes Agent sits on both sides of the ACP host/guest split, with its gateway hosting platform adapters and tool servers while its \texttt{hermes-acp} adapter lets external IDEs host Hermes, making the systems stackable rather than purely alternative.

\paragraph{Trust model and security architecture.}
The three systems address different threat models. Claude Code assumes an untrusted model operating within a trusted developer's machine: the deny-first permission system (\Cref{sec:auth}) evaluates every tool invocation, the ML classifier provides automated safety assessment, and seven permission modes create a graduated autonomy spectrum. OpenClaw assumes a single trusted operator per gateway instance. Its security architecture begins with identity and access control (DM pairing codes, sender allowlists, gateway authentication) rather than per-action safety classification. Tool policy uses configurable allow/deny lists per agent rather than a centralized classifier. Sandboxing is available as an opt-in feature with multiple backends (Docker, SSH, or OpenShell) and configurable scope (per-agent, per-session, or shared); a \texttt{non-main} mode can sandbox all non-main sessions when enabled, though sandboxing is not active by default. The OpenClaw security documentation explicitly states that hostile multi-tenant isolation on a shared gateway is not a supported security boundary. Hermes Agent takes a third position. Tool authorization is centralized in one file (\file{tools/approval.py}), exposing three modes (\texttt{manual}, \texttt{smart}, \texttt{off}) and an unconditional \texttt{HARDLINE\_PATTERNS} floor that blocks \texttt{rm -rf /}, \texttt{mkfs}, raw \texttt{dd} to block devices, fork bombs, and system shutdown commands regardless of mode~\tierB{}. The smart mode runs an auxiliary-LLM risk assessment; an in-source comment names OpenAI Codex's smart approvals as the design inspiration. Because Hermes runs across many surfaces, the same approval flow is rendered four times: a CLI prompt, gateway approval keyboards on Telegram, Discord, and QQBot, an ACP \texttt{request\_permission} round-trip for IDE clients, and a cron-only mode for unattended runs. Compared with Claude Code's seven-mode taxonomy and ML-classifier-driven \texttt{auto} mode, Hermes has a smaller mode taxonomy but the same multi-layer structure: unconditional safety floor, default interactive prompt, opt-in auxiliary-LLM ``smart'' path. The architectural answer is the same; the surface count differs. The difference across the three reflects where the trust boundary sits: Claude Code places it between the model and the execution environment, OpenClaw places it at the gateway perimeter, and Hermes sits between the two with per-action approvals like Claude Code but rendered through many surfaces like OpenClaw.

\paragraph{Agent runtime and tool orchestration.}
All three systems implement agentic loops, but the loops occupy different positions in their respective architectures. In Claude Code, the \func{queryLoop} async generator (\Cref{sec:turn}) is the system's center: all interfaces feed into it, and it directly manages context assembly, model calls, tool dispatch, and recovery. In OpenClaw, the agent runtime is an embedded \texttt{pi-coding-agent} core (an external SDK from the \texttt{pi-mono} project that OpenClaw integrates), sitting inside a larger gateway dispatch layer. The gateway's \texttt{agent} RPC validates parameters, resolves sessions, and returns immediately; the embedded runner then executes the agentic loop while emitting lifecycle and stream events back through the gateway protocol. Runs are serialized through per-session queues and an optional global lane, preventing tool and session races across the multi-channel surface. In Hermes Agent, the loop is a synchronous \texttt{while} inside \texttt{AIAgent.run\_conversation} (\file{run\_agent.py}), gated on an iteration count (default 90) and a separate iteration budget; when the budget is exhausted, a single tool-stripped summary call lets the agent deliver a closing message rather than terminate mid-thought~\tierB{}; each iteration sends tool schemas collected by \texttt{model\_tools.get\_tool\_definitions()} and routes any tool calls through \texttt{model\_tools.handle\_function\_call}, with tool files in \texttt{tools/*.py} self-registering at import time so that adding a tool requires only one \texttt{registry.register()} call, and a single assistant message returning multiple tool calls is executed in parallel with a sequential fallback. All three systems follow the ReAct pattern~\citep{yao2022react}, with OpenClaw's loop a component within a control plane rather than the control plane itself, and Hermes's loop a peer to Claude Code's at the runtime level but shape-different in implementation, a synchronous Python loop with thread-managed concurrency rather than an async generator that yields.

\paragraph{Extension architecture.}
Claude Code's four extension mechanisms (MCP, plugins, skills, hooks) are organized by context cost (\Cref{sec:ext}): hooks consume zero context, skills consume low context, and MCP servers consume high context once their tool schemas load. All four extend a single agent's context window and tool surface. OpenClaw uses a manifest-first plugin system with four architectural layers (discovery, enablement, runtime loading, surface consumption) and twelve capability types including text inference, speech, media understanding, image/music/video generation, web search, and messaging channels. Plugins register capabilities into a central registry; the gateway reads the registry to expose tools, channels, provider setup, hooks, HTTP routes, CLI commands, and services. OpenClaw also has a separate skills layer with multiple sources (workspace, project-level, personal, managed, bundled, and extra directories, with workspace skills taking highest precedence) plus a public registry (ClawHub) and supports MCP through built-in \texttt{openclaw mcp} commands (server and outbound client registry). Hermes Agent exposes five extension surfaces against Claude Code's four~\tierB{}: three sit at the same level as Claude Code's (general plugins under \file{plugins/<name>/} with lifecycle hooks, bundled skills under \file{skills/}, and MCP servers configured in \file{config.yaml}), while two extra surfaces extend a different axis (pluggable memory providers under \file{plugins/memory/<name>/} and pluggable model providers under \file{plugins/model-providers/<name>/}) that let Hermes swap backends rather than intercept events. Hooks in Hermes can be in-process Python callbacks or config-driven external shell commands, both dispatched through the same hook manager. When Hermes spawns a stdio MCP server, the child's environment is restricted to a small allowlist of system variables before launch, and \texttt{npx}/\texttt{uvx} package names are screened against the OSV malware database. The architectural difference across the three is where the extension acts: Claude Code's extensions modify one agent's action surface, OpenClaw's plugins extend the gateway's capability surface across all agents, and Hermes's extensions add a second axis on top of a Claude-Code-style per-agent surface, swapping entire memory and model-provider backends in place rather than intercepting events around them.

\paragraph{Memory, context, and knowledge management.}
All three systems use transparent file-based memory rather than opaque databases. Claude Code loads a four-level CLAUDE.md hierarchy and manages context pressure through a five-layer compaction pipeline (\Cref{sec:context}). Memory retrieval uses an LLM-based scan of file headers. OpenClaw injects workspace bootstrap files into the system prompt at session start: five core files (AGENTS.md, SOUL.md, TOOLS.md, IDENTITY.md, USER.md) plus conditionally BOOTSTRAP.md, HEARTBEAT.md, and MEMORY.md, with large files truncated. Separately, the memory system manages three file types: \texttt{MEMORY.md} for long-term durable facts, date-stamped daily notes (\texttt{memory/YYYY-MM-DD.md}), and an optional \texttt{DREAMS.md} for dreaming sweep summaries. When an embedding provider is configured, memory search uses hybrid retrieval combining vector similarity with keyword matching. An experimental dreaming system performs background consolidation, scoring candidates and promoting only qualified items from short-term recall into long-term memory. Before compaction, OpenClaw automatically reminds the agent to save important notes to memory files, preventing context loss. Compaction in Hermes is one auxiliary-LLM summarizer with token-budget tail protection, preceded by a tool-output prune pass; the summary is prefixed with a long ``reference only'' preamble that tells the agent the compaction is background context, not new instructions, and that persistent memory (\file{MEMORY.md}, \file{USER.md}) remains authoritative~\tierB{}. Where Claude Code's CLAUDE.md hierarchy is fixed and its compaction has five layers, Hermes has fewer compaction layers but adds a context-file pre-injection scan: files such as \file{AGENTS.md}, \file{.cursorrules}, and \file{SOUL.md} are checked against a list of injection patterns and invisible Unicode characters, and any hit replaces the file's content with a \texttt{[BLOCKED: ...]} marker. All three share the design commitment to user-visible, editable memory but give different answers to the same context-management question: Claude Code invests in graduated context compression (five layers with cache awareness); OpenClaw invests in structured long-term memory promotion (dreaming, daily notes, memory search) plus pluggable compaction providers and session pruning, though its compaction pipeline is less graduated than Claude Code's five-layer system; Hermes invests in pre-injection content scanning and prompt-level isolation of persistent memory from compacted history rather than a deeper compaction stack.

\paragraph{Multi-agent architecture and routing.}
This dimension reveals the starkest architectural difference. Claude Code's multi-agent model is task delegation: the parent spawns subagents (Explore, Plan, general-purpose, and custom types) that operate in isolated context windows with restricted tool sets and return summary-only results (\Cref{sec:subagent}). Worktree isolation provides filesystem-level separation. OpenClaw separates two distinct concerns. First, multi-agent routing: a single gateway can host multiple fully isolated agents, each with its own workspace, authentication profiles, session store, and model configuration, routed to specific channels or senders via deterministic binding rules. Second, sub-agent delegation: within a single agent, background runs can be spawned with configurable nesting depth (maximum 5, default 1, recommended 2), thread-bound sessions on supported channels, and configurable tool policy by depth. OpenClaw's project vision explicitly rejects agent-hierarchy frameworks as a default architecture. Hermes Agent splits the question along yet a different seam. Per-turn delegation goes through the \texttt{delegate\_task} tool, which spawns child \texttt{AIAgent} instances in a thread pool; the parent blocks until children return summaries, concurrency is capped at three children by default and depth is capped at one (clamped to the range [1, 3]), leaf children cannot themselves delegate by default (nested delegation is opt-in through orchestrator children and a higher spawn-depth setting), and children always run with persistent memory disabled so they cannot read or write the parent's notes~\tierB{}. Beyond per-turn delegation, v0.13's Kanban subsystem is a SQLite-backed work queue where multiple Hermes worker profiles claim, heartbeat, and complete tasks, with stale-claim reclamation and an automatic block after two consecutive failed attempts (\file{hermes\_cli/kanban\_db.py}:2703). The distinction across the three matters because Claude Code's subagents are subordinate workers within one user's coding session, OpenClaw's multi-agent routing creates genuinely independent agent instances serving different users or purposes through different channels, and Hermes combines a strict per-turn delegation default with a separate Kanban layer that lets multiple agent processes coordinate over a durable shared board, a question Claude Code and OpenClaw do not ask in the same form.

\subsection{Hermes Agent: Capabilities Beyond the Comparison Table}
\label{sec:compare:hermes}

Session persistence in Hermes uses one WAL-mode SQLite database with full-text search across all session messages and a parent-session chain that records compaction-triggered splits~\tierB{}. Where Claude Code persists sessions to JSONL files per project (\Cref{sec:persist}), Hermes's choice provides cross-session search and concurrent-reader support out of the box. When the gateway restarts mid-conversation, sessions auto-resume on next message arrival, driven by transcript-timestamp freshness logic with a one-hour default window. A built-in cron scheduler and webhook subscription system let the agent run unattended on schedule expressions or external HTTP events; cron sessions run with persistent memory disabled and a configurable inactivity timeout that defaults to 600 seconds. The distribution also includes a background ``curator'' loop that reviews agent-created skills, archives stale ones, and never deletes, gated to skills with agent provenance so bundled and hub-installed skills are off-limits. None of these capabilities map cleanly onto the six comparison dimensions used in the main table, which is why they appear here as a Hermes-specific addendum rather than as cells in \Cref{tab:comparison-dimensions}. OpenClaw's distinctive capabilities (the ClawHub plugin registry under extension architecture and the dreaming memory subsystem under memory and context) already occupy cells in those six dimensions and so do not require a parallel addendum. The Hermes additions in this subsection follow from its persistent-process scope, a deployment-context property Hermes shares with OpenClaw but that Claude Code's per-session CLI model does not provide.

\subsection{What the Contrast Reveals}
\label{sec:compare:reveals}

The comparison surfaces three observations about the design space of AI agent systems.

First, the recurring design questions identified in \Cref{sec:arch:principles} (where reasoning lives, what safety posture to adopt, how to manage context, how to structure extensibility) apply beyond coding agents. OpenClaw answers every one of these questions from the starting point of a multi-channel personal assistant rather than a repository-bound coding tool, and Hermes Agent answers them from a single-process, multi-surface deployment rather than a CLI harness or a multi-channel gateway. The questions are stable; the answers vary with deployment context across the three systems.

Second, the systems make different bets along several dimensions. Claude Code invests in graduated per-action safety evaluation; OpenClaw invests in perimeter-level identity and access control; Hermes sits between the two, with per-action approvals like Claude Code rendered through many surfaces like OpenClaw. Claude Code treats the agent loop as the architectural center; OpenClaw treats the gateway control plane as the center and embeds the agent loop as one component; Hermes's loop is a peer to Claude Code's at the runtime level, occupying the same architectural position rather than sitting inside a gateway dispatch layer. Claude Code's extensions modify a single context window; OpenClaw's plugins extend a shared gateway surface; Hermes adds a second axis on top of a Claude-Code-style per-agent surface, swapping memory and model backends in place rather than intercepting events. These differences are not arbitrary: they follow from the three systems' different trust models and deployment topologies.

Third, the compositional relationships across the three systems are architecturally significant. OpenClaw can host Claude Code as an external coding harness via ACP, and Hermes Agent sits on both sides of the host/guest split: its gateway hosts platform adapters and tool servers, and its \texttt{hermes-acp} adapter lets external IDEs host Hermes. The systems are composable rather than exclusive alternatives, which suggests that the design space of AI agents is not a flat taxonomy but a layered one, where gateway-level systems and task-level harnesses can compose at multiple positions.

\section{Related Work}
\label{sec:related}

\subsection{Coding Agent Taxonomy}
\label{sec:related:taxonomy}

AI coding tools can be organized by the degree of autonomous action they support (\Cref{tab:tool-taxonomy}). Inline completion tools such as GitHub Copilot \citep{chen2021evaluating} suggest code fragments within the editor without autonomous action. Chat-integrated products including Cursor and Windsurf add conversational interaction and multi-file edits but remain coupled to the IDE environment. Agentic CLI tools, including Claude Code, OpenAI's Codex CLI, and Aider \citep{gauthier2024aider}, operate from the command line and can autonomously execute shell commands, read and write files, and iterate on outputs within a single request. Fully autonomous systems like Devin, SWE-Agent \citep{yang2024sweagent}, and OpenHands \citep{wang2024openhands} aim for minimal human supervision, often in sandboxed cloud environments.

\begin{table}[t]
\centering
\caption{AI coding tool categories by degree of autonomous action.}
\label{tab:tool-taxonomy}
\small
\begin{tabularx}{\columnwidth}{>{\centering\arraybackslash}c >{\centering\arraybackslash}X >{\centering\arraybackslash}c}
\toprule
\textbf{Category} & \textbf{Examples} & \textbf{Pattern} \\
\midrule
Inline completion & Copilot, Tabnine & Editor plugin \\
Chat-integrated & Cursor, Windsurf, Cody & IDE-coupled product \\
Agentic CLI & Claude Code, Codex CLI, Aider & Tool-use loop \\
Fully autonomous & Devin, SWE-Agent, OpenHands & Sandbox + planning \\
\bottomrule
\end{tabularx}
\end{table}

Claude Code shares features with higher-autonomy agents (auto-mode classifier, background agent execution, remote environments) but retains interactive approval by default. Evaluation benchmarks such as SWE-Bench \citep{jimenez2024swebench} and HumanEval \citep{chen2021evaluating} have driven much of the academic focus on coding agents. This paper examines Claude Code's internal architecture from source code.

\subsection{Agent Architecture Patterns}
\label{sec:related:patterns}

Claude Code's core loop follows the ReAct pattern \citep{yao2022react}: the model generates reasoning and tool invocations, the harness executes actions, and results feed the next iteration. Toolformer \citep{schick2023toolformer} demonstrated that language models can learn to use tools. Claude Code uses up to 54 built-in tools and a layered permission system. The broader design space has been mapped by several surveys. \citet{weng2023agent} offered the now-standard decomposition into planning, memory, and tool use, and \citet{wang2024agentsurvey} catalogued early autonomous-agent work. \citet{xu2026agentsystems} frames the field around three recurring trade-offs (autonomy vs.\ controllability, latency vs.\ accuracy, capability vs.\ reliability) that recur throughout our analysis, and \citet{hu2025adas} casts agent design itself as a search problem over components, algorithms, and evaluation functions. This paper characterizes one specific point in that space.

A recent survey on code as agent harness frames code not only as model output but as the substrate for reasoning, action, environment modeling, and execution-based verification~\citep{ning2026codeharness}. That framing matches this paper's focus on the harness around the model rather than on benchmark scores alone.

\paragraph{Recent runtime mechanisms.}
Recent surveys and systems make several harness choices explicit. Graph Engineering separates task dependencies, agent communication, and execution state~\citep{feng2026graph}, and the Self-Evolving Coding Agents survey organizes persistent changes by their target, timing, and feedback source~\citep{zhou2026self}. LoopsBench and LoopArena evaluate how prerequisites, retained regression checks, and a controller's decisions to check or stop shape a coding loop~\citep{li2026loopsbench,wang2026looparena}, and Symphony separates issue scheduling from the coding agent's own loop~\citep{openai2026symphony}. For recovery, Resume Means Resume tests the checkpoint and resume behavior of workflow frameworks against their stated semantics~\citep{khan2026resume}, and Temporal's Deep Agents integration runs each model call and I/O tool call as a Temporal activity while the control loop replays inside the workflow~\citep{temporal2026deepagents}. Living-Harness updates an episodic memory and a state graph between episodes while the model and tools stay fixed~\citep{du2026living}. Graphiti and Cognee store entity relations in knowledge graphs~\citep{graphiti2026memory,cognee2026memory}. APPA checks a tool's result before admitting it to the agent's context, separating permission to call a tool from admission of the data it returns~\citep{kravchenko2026appa}. At the protocol level, the 2026-07-28 MCP revision removes protocol-level sessions and lets one tool operation span several requests~\citep{mcp2026revision,mcp2026mrtr}, the Agent Host Protocol lets several clients share a host-owned session~\citep{lourens2026agenthost}, and A2A connects independent agent systems~\citep{a2a2026specification}. Cursor reports coordinating agent swarms through shared design documents~\citep{lin2026agentswarms}.

Multi-agent orchestration frameworks such as AutoGen \citep{wu2024autogen}, LangChain~\citep{langchain2026framework}, and CrewAI~\citep{crewai2026framework} provide conversation-based agent coordination. Claude Code's subagent delegation (\Cref{sec:subagent}) includes permission override precedence, two-level permission scoping, and separate transcript files for each subagent. LATS \citep{zhou2024lats} unifies reasoning, acting, and planning in a tree-search framework. Claude Code's \code{plan} permission mode implements a simpler plan-then-execute approach.

Practitioner writing has converged on a handful of recurring patterns that Claude Code's architecture instantiates. Anthropic's own ``Building Effective Agents'' \citep{anthropic2024effective} distinguishes agents from workflows and argues for simple composable patterns over heavy frameworks. \citet{martin2026patterns} synthesizes seven patterns observed in production systems, including giving agents filesystem and shell access as a general-purpose action layer, and discovering actions on demand rather than loading every tool schema upfront. \citet{chase2025deepagents} observes that Claude Code's planning tool is ``basically a no-op'' whose value lies in keeping the agent on track rather than in performing any external computation. \citet{swyx2025agentengineering} argues that authority is the element academic frameworks most often leave out, calling trust ``the most overlooked element'' in production agent design, a gap the permission analysis in \Cref{sec:auth} attempts to close. \citet{huyen2025agents} makes the compound-error concern concrete: at 95\% per-step accuracy, a 100-step task succeeds only 0.6\% of the time, which motivates the per-step verification patterns we trace in \Cref{sec:turn} and \Cref{sec:auth}. OpenAI makes a similar point in its own writing on harness engineering, treating the design of the harness itself as a central engineering task~\citep{openai2026harness}. Recently, a new concept called \emph{loop engineering} has also emerged: instead of prompting the agent each turn, the developer builds a loop that finds work, assigns it, checks the result, and decides what comes next, assembled from the same primitives this paper documents, such as skills, MCP servers, and subagents~\citep{osmani2026loop}. It lets a single developer direct far more work autonomously and has drawn interest, including from Claude Code's own team, though reducing the person's turn-by-turn involvement also sharpens the long-term developer-capability concern that \Cref{sec:conclude} returns to.

\paragraph{Context management.}
\Cref{tab:context-comparison} presents a design-space taxonomy of context management approaches. Claude Code's five-layer compaction pipeline applies multiple strategies at different granularities before escalating, with cache-aware compression and virtual-view-on-read semantics. \citet{zhang2026ace} characterizes two failure modes that this design mitigates (summarization that drops domain details, and detail loss from iterative context rewriting), and instead proposes treating context as an ``evolving playbook'' that accumulates strategies over time. Claude Code's approach is consistent with that framing, since the CLAUDE.md hierarchy accumulates structured instructions rather than repeatedly summarizing them. \citet{hu2025memory} distinguishes context engineering from agent memory: context engineering handles transient assembly, while memory covers persistent factual knowledge and experiential traces. Claude Code's architecture separates the two in the same way, pairing a compaction pipeline with a file-based memory hierarchy.

\begin{table}[t]
\centering
\caption{Design space of context management approaches in LLM-based tools.}
\label{tab:context-comparison}
\small
\begin{tabularx}{\columnwidth}{>{\centering\arraybackslash}c >{\centering\arraybackslash}X >{\centering\arraybackslash}c}
\toprule
\textbf{Approach} & \textbf{Mechanism} & \textbf{Granularity} \\
\midrule
Simple truncation & Drop oldest messages & Coarse \\
Sliding window & Fixed-size recent history & Medium \\
RAG & Retrieve relevant snippets & Fine \\
Single summarization & One-pass compress & Coarse \\
Graduated compaction & Multi-layer pipeline & Very fine \\
\bottomrule
\end{tabularx}
\end{table}

\paragraph{Safety and permissions.}
Production coding agents adopt safety architectures that vary along three axes: \emph{approval model} (per-action prompting, classifier-mediated automation, or no prompting with post-hoc review), \emph{isolation boundary} (OS-level container, filesystem sandbox, permission-scoped tool pool, or none), and \emph{recovery mechanism} (version-control rollback, session-scoped permission reset, or checkpoint-based rewind). SWE-Agent and OpenHands~\citep{yang2024sweagent,wang2024openhands} rely primarily on Docker container isolation, providing environment-level sandboxing that constrains all agent actions. Codex CLI supports sandbox modes and approval policies for shell commands. OpenCode and Cline also provide tool approval controls~\citep{opencode2026agent,cline2026agent}. Aider~\citep{gauthier2024aider} uses Git as its primary safety mechanism, making all changes reversible through version control. Claude Code combines per-action deny-first rules, an ML-based classifier for automated approval, optional shell sandboxing, and session-scoped permission non-restoration, layering multiple mechanisms rather than relying on a single isolation boundary.

Sandboxing reports make this boundary concrete. E2B and OpenSandbox expose isolated execution environments through APIs~\citep{e2b2026sandbox,opensandbox2026environment}. OpenAI's Windows sandbox discussion describes write access scoped to the working directory and configured writable roots, with explicit deny rules for selected read-only paths inside those roots~\citep{openai2026windowssandbox}. Cursor's cloud-agent report adds the cloud side of the same problem: network policy, secret redaction, and credential management become part of the agent environment~\citep{cursor2026cloudagents}. These details make the safety boundary an execution boundary, not only an approval prompt.

\paragraph{Protocols and extensibility.}
The Model Context Protocol that Claude Code uses as its primary external tool integration has become a de facto standard with a substantial ecosystem and a corresponding attack surface. \citet{hou2025mcpsurvey} catalogues thousands of community-developed MCP servers across 26 major directories and organizes MCP-specific threats into four attacker categories and sixteen scenarios, including tool poisoning, rug pulls, and cross-server shadowing. The permission and deny-rule machinery analyzed in \Cref{sec:auth} and the pre-filtering step in \Cref{sec:ext:assembly} can be read as the runtime side of the mitigations that survey calls for.

The same point applies to higher-level tool design. NSA guidance treats MCP security as a lifecycle issue that spans protocol design, runtime scheduling, integration with external services, and long-term monitoring~\citep{nsa2026mcpsecurity}. A proposal for agent-first tool design argues that agent-facing APIs should offer search, resolve, preview, execute, verify, and recover phases, not just human-oriented CRUD endpoints~\citep{pan2026agentfirstapi}. Read this way, MCP servers, plugins, skills, SDKs, and agent-to-agent protocols form a tool supply chain: they expand what an agent can do, but they also need identity, allowlists, versioning, audit logs, and revocation.

\paragraph{Software architecture.}
Layered architecture patterns \citep{garlan1993architecture} inform our five-layer decomposition. Role-based access control models \citep{sandhu1996rbac} provide theory for the permission mode system. Browser sandboxing \citep{reis2009isolating} is a similar per-process isolation approach. Multi-agent system theory \citep{wooldridge2009multiagent} helps explain subagent delegation.

\paragraph{Positioning.}
Prior work on coding agents has focused on benchmarks (how well agents solve tasks), frameworks (how to compose agents), and products (what users can do). This paper contributes a source-grounded design-space analysis of a production coding agent, using source-level analysis and architectural comparison to surface design choices and trade-offs. It draws on the software architecture case study tradition~\citep{garlan1993architecture} but applies it to an LLM-based agent by systematically identifying design questions, mapping alternatives, and contrasting Claude Code's choices with those of OpenClaw and Hermes Agent, two independent AI agent systems operating from different deployment contexts.

\section{Discussion}
\label{sec:discuss}

The analysis in the preceding sections documented how Claude Code answers recurring design questions about loop architecture, safety posture, extensibility, context management, delegation, and persistence. Each answer reflects a position in a design space with real alternatives and measurable trade-offs. This section examines what those answers reveal when read together: the design philosophy they reflect (\Cref{sec:discuss:philosophy}), the value tensions they create (\Cref{sec:discuss:tensions}), the architectural trade-offs they entail (\Cref{sec:discuss:tradeoffs}), the empirical predictions they generate (\Cref{sec:discuss:empirical}), and the cross-cutting commitments that recur across subsystems (\Cref{sec:discuss:guidelines}). The five-value framework from \Cref{sec:values:five} organizes the section.

\subsection{Design Philosophy}
\label{sec:discuss:philosophy}

The values and design principles introduced in \Cref{sec:values} predict an architecture that invests in operational infrastructure rather than decision scaffolding. The implementation confirms this: the architecture documented in \Cref{sec:arch,sec:turn,sec:auth,sec:ext,sec:context,sec:subagent,sec:persist} is overwhelmingly deterministic infrastructure (permission gates, tool routing, context management, recovery logic), with the LLM invoked as a stateless completion endpoint. An estimated 1.6\% of the codebase constitutes decision logic, the remaining 98.4\% is the operational harness. This ratio is not accidental.

The design principles documented in \Cref{sec:values:principles} underpin this approach: the harness creates conditions under which the model can decide well, rather than constraining its choices.

This design runs counter to the dominant pattern in agent engineering, where frameworks such as LangGraph route model outputs through explicit graph nodes with typed edges, and systems like Devin pair multi-step planners with heavy operational infrastructure. Claude Code instead gives the model maximum decision latitude within a rich operational harness. The engineering complexity exists not to constrain the model's decisions but to enable them. This layered architecture, where the model reasons and the harness enforces, raises the question of whether agentic coding tools are converging toward operating-system-like abstractions in which the core loop serves as the kernel and everything else constitutes the OS.

The design gains additional significance as frontier models converge in practical capability for coding tasks: the quality of the surrounding operational harness becomes the principal differentiator, validating an architecture that invests in infrastructure over decision scaffolding. For agent builders, the implication is that investing in deterministic infrastructure such as context management, safety layering, and recovery mechanisms may yield greater reliability gains than adding planning scaffolding around increasingly capable models. Recent native-runtime benchmarking gives this otherwise reasoned claim direct empirical support: holding the model fixed and changing only the surrounding harness shifts a single model's score on long-horizon tasks by up to 18 points, with Claude Code and OpenClaw among the harnesses tested~\citep{ding2026wildclawbench}.

(\tierC{}) As model capability grows stronger, is the harness still needed for a model like Fable~5 that can read a codebase and fix it on its own? The harness serves the five values of \Cref{sec:values:five}, and only one of them, Capability Amplification, is mainly about how capable the model is. Within it, the part a stronger model affects is the decision scaffolding, like the planning and control logic that some frameworks wrap around the model to guide and constrain it toward a solution. Where the model is already strong, such as in code and mathematics, it reaches a successful solution on its own, so that guidance and constraint are less necessary.

The other four values are not mainly about how capable the model is, so the harness that serves them stays. Reliable Execution is the partial exception: a stronger model reads a request more accurately and fixes its own mistakes a bit better. But the context window is still finite, sessions still resume, and the model still tends to rebuild the same thing a different way each run, so context management and recovery are still needed. Safety, Security, and Privacy do not depend on capability at all. Deny-first rules, sandboxing, the auto-mode classifier, and audit logs matter as much as before, and a stronger model is riskier, not safer: Fable~5 itself is held back by safety classifiers in higher-risk areas~\citep{anthropic2026fable}. Human Decision Authority is about the person's right to decide, not the model's capability, so approval, interruption, and transcripts keep their role. Contextual Adaptability is similar: a strong model is still a general model that has not seen your code, so CLAUDE.md, skills, and MCP still fit it to your project. Token economics, which bounds all five, does not relax either, since a better model does not make tokens free.

Research bears this out even for the strongest models: the choice of harness alone moves Fable~5's functional and security pass rates by ten points or more~\citep{endorlabs2026fable}. The harness does not become useless as model capability grows; what changes is only the decision scaffolding, and only where the model is already strong, while everything else it does stays as important as before.

Taken together, the preceding sections show that production coding agents face recurring design choices: where reasoning lives relative to the harness, how the iteration loop is structured, what safety posture to adopt by default, how the extension surface is partitioned, how context is assembled and compressed, how subagents are delegated and orchestrated, and how sessions persist across boundaries. Claude Code's answers to these questions form a coherent design point that privileges model autonomy within a rich operational harness.

This philosophy assumes that rich deterministic infrastructure can adequately support unconstrained model judgment. The following subsections examine where this assumption is tested.

\subsection{Value Tensions}
\label{sec:discuss:tensions}

The five values identified in \Cref{sec:values:five} generate tensions where pursuing one value constrains another (\Cref{tab:tensions}). These tensions are not design failures; they are structural consequences of pursuing multiple values simultaneously. We report the tensions with the strongest supporting evidence, not the full combinatorial set.

\begin{table}[t]
\centering
\small
\caption{Tensions between values, with supporting evidence. Each tension demonstrates that the two values capture genuinely distinct concerns.}
\label{tab:tensions}
\begin{tabularx}{\textwidth}{@{}ccX@{}}
\toprule
\textbf{Value Pair} & \textbf{Tension} & \textbf{Evidence} \\
\midrule
Authority $\times$ Safety & Approval fatigue vs.\ protection & 93\% approval rate undermines human vigilance~\citep{anthropic2026automode}; safety must compensate via classifier and sandboxing \\
\midrule 
Safety $\times$ Capability & Performance vs.\ defense depth & $>$50-subcommand fallback skips per-subcommand deny checks due to parsing overhead~\citep{adversa2026bypass}; safety layers share performance constraints \\
\midrule 
Adaptability $\times$ Safety & Extensibility vs.\ attack surface & Multiple (since patched) CVEs exploited pre-trust MCP server initialization and application of settings environment variables~\citep{checkpoint2026rce} \\
\midrule 
Capability $\times$ Adaptability & Proactivity vs.\ disruption & 12 to 18\% more tasks but preference drops at high frequencies~\citep{chi2025proactive} \\
\midrule 
Capability $\times$ Reliability & Velocity vs.\ coherence & Bounded context prevents full codebase awareness (\Cref{sec:context}); subagent isolation limits cross-agent consistency (\Cref{sec:subagent}); complexity increases observed in adjacent tools~\citep{he2026cursor} \\
\bottomrule
\end{tabularx}
\end{table}

A separate question is whether short-term gains weaken long-term developer capability (\Cref{sec:values:lens}). A randomized controlled trial of 16 experienced developers across 246 tasks~\citep{becker2025measuring} found that AI tools made developers 19\% slower, despite a perceived 20\% improvement. A causal analysis of Cursor adoption across 807 repositories~\citep{he2026cursor} found that code complexity increased by 40.7\%. An EEG study of 54 participants~\citep{kosmyna2025brain} found that LLM users showed weakened neural connectivity that persisted after AI was removed. Researchers have proposed protocols for measuring cognitive offloading in AI-assisted programming, motivated by concerns that students using AI produce applications without understanding the underlying logic~\citep{aiersilan2026vibecheck}. These findings, combined with a 25\% decline in entry-level tech hiring between 2023 and 2024~\citep{rak2025aihiring}, suggest that the tension between capability amplification and long-term sustainability extends beyond individual productivity to the broader developer pipeline. This evidence motivates the long-term capability question but does not target Claude Code's architecture specifically. It is most relevant to agent systems that rely on bounded context, tool-use loops, and local generation decisions.

\subsection{Architectural Trade-offs}
\label{sec:discuss:tradeoffs}

The tensions in \Cref{tab:tensions} manifest as concrete architectural trade-offs in four areas. The long-term sustainability concerns documented above surface in the empirical predictions of \Cref{sec:discuss:empirical}.

\paragraph{Safety vs.\ autonomy.}
The permission modes (five always present, plus \code{auto} when the classifier feature flag is active, and the internal \code{bubble} mode) create a gradient from \code{plan} (user approves all plans) through \code{default}, \code{acceptEdits}, \code{auto} (ML classifier), to \code{bypassPermissions} (skips most prompts but safety-critical checks remain). The progression represents a monotonically decreasing safety gradient with increasing autonomy. Not restoring session-scoped permissions from the transcript reflects a deliberate choice to err toward safety: a new process starts without the old session grants.

The safety-autonomy gradient is shaped not only by architectural design but by user behavior. Anthropic's auto-mode analysis~\citep{anthropic2026automode} found that users approve approximately 93\% of permission prompts, indicating that approval fatigue renders interactive confirmation behaviorally unreliable. Longitudinal usage data~\citep{anthropic2026autonomy} shows that auto-approve rates increase from approximately 20\% at fewer than 50 sessions to over 40\% by 750 sessions, with substantial increases in session duration. These patterns suggest that the gradient is navigated not by deliberate mode selection but by gradual habituation. Sandboxing reduced the frequency of permission prompts by an estimated 84\%~\citep{anthropic2025sandboxing}, reframing the problem as a human-factors concern: the architectural response to unreliable human approval is to reduce the number of decisions humans must make.

More fundamentally, the defense-in-depth architecture described in \Cref{sec:auth} rests on an independence assumption: if one safety layer fails, others catch the violation. But Claude Code's safety layers share common performance and economic constraints. The auto-mode classifier is a separate LLM call with direct token cost. The \file{bashSecurity.ts} module performs sequential AST-based checks with parsing latency. The deny-first rule evaluation operates on command structure. When performance pressure pushes toward reducing these costs, layers can degrade simultaneously. Security researchers~\citep{adversa2026bypass} have documented that commands with more than 50 subcommands fall back to a single generic approval prompt instead of running per-subcommand deny-rule checks, because per-subcommand parsing caused UI freezes, demonstrating that defense-in-depth fails when the independence assumption is violated.

This tension is structural. Any LLM-based agent system that uses the model itself for safety evaluation faces it. The relevant evaluation criterion is not whether any individual layer can be bypassed, but how many independent layers must fail simultaneously and whether they share failure modes.

\paragraph{Permission model under adversarial conditions.}

Independent security research provides empirical validation of the permission architecture, specifically by revealing a temporal ordering property not captured in \Cref{fig:permission}. Two independently verified vulnerabilities share a root cause in pre-trust initialization ordering: code executing during project initialization (MCP server connections and application of settings environment variables) ran, in earlier releases, before the user accepted the interactive trust dialog\footnote{The two pre-trust ordering vulnerabilities are CVE-2025-59536 (CVSS 8.7) and CVE-2026-21852 (CVSS 5.3)~\citep{checkpoint2026rce}, discovered by Check Point Research. CVE-2025-54794 and CVE-2025-54795~\citep{cymulate2025inverseprompt} exploit path validation and command parsing flaws elsewhere in the permission pipeline, separately. All four were patched within weeks of disclosure.}. That pre-trust execution window fell outside the deny-first evaluation pipeline (\file{permissions.ts}), creating a structurally privileged phase where the safety guarantees documented in \Cref{sec:auth} did not yet apply. In the analyzed snapshot, the standard interactive startup path requires trust acceptance before hook execution, project MCP server connections, and application of project-scoped settings environment variables beyond a safe allowlist (\file{hooks.ts}, \file{main.tsx}, \file{managedEnv.ts}).

This pattern reveals that the permission pipeline depicts a spatial ordering of safety checks but does not capture the temporal dimension: specifically, when during session initialization each mechanism becomes active. The pre-patch initialization sequence (MCP server execution, then trust acceptance, then permission enforcement) created a window where the extensibility architecture (\Cref{sec:ext}) operated before the safety architecture (\Cref{sec:auth}) was fully engaged. This finding refines the extensibility-versus-simplicity tension by adding a security dimension: extensibility creates attack surface not only through combinatorial complexity but through initialization ordering.

\paragraph{Context efficiency vs.\ transparency.}
The five-layer compaction pipeline achieves effective context management, but compression is largely invisible to the user. When budget reduction replaces a long tool output with a reference, when context collapse substitutes messages with a summary (described in the source as ``a read-time projection over the REPL's full history''), or when snip trims older history, the user has no easy way to inspect what was lost. The cache-aware behavior of microcompact adds further opacity, as compression decisions are influenced by prompt caching in ways not visible to the user. External research on summary-based compaction documents two costs beyond opacity that the source-level view does not surface: the summarization step is a blocking inference stall, and it is non-deterministic, with the retained content fluctuating across runs on identical inputs~\citep{cim2026parallelcompaction}.

\paragraph{Simplicity vs.\ extensibility.}
The four extension mechanisms enable rich customization but create combinatorial interactions. A plugin contributes a PreToolUse hook that modifies tool inputs. The auto-mode classifier reads cached CLAUDE.md content. Path-scoped rules load lazily when new directories are read, potentially changing classifier behavior mid-conversation. The permission handler's four branches interact with the hook pipeline at multiple points. These cross-cutting concerns create emergent behaviors difficult to predict from any single configuration file.

\subsection{Empirical Predictions and Early Signals}
\label{sec:discuss:empirical}

The architectural properties documented in this paper generate testable predictions about code quality outcomes not derivable from the source code alone. The bounded context window (\Cref{sec:context}) prevents the agent from maintaining simultaneous awareness of the full codebase: the five-layer compaction pipeline preserves useful information but introduces lossy compression at each stage. This makes it architecturally predicted that agent-generated code will exhibit higher rates of pattern duplication and convention violation than code produced with full codebase visibility. Subagent isolation (\Cref{sec:subagent}), where each subagent operates in its own context window with an independently assembled tool pool, compounds the effect: parallel agents can independently re-implement solutions that already exist elsewhere. The design philosophy of \Cref{sec:discuss:philosophy} trusts the model to make good local decisions, but good local decisions can produce poor global outcomes when the model lacks global context.

Published empirical work on architecturally similar tools provides data consistent with these predictions. A causal analysis of Cursor adoption across 807 repositories~\citep{he2026cursor} found a statistically significant increase in code complexity, with an initial velocity spike that dissipated to baseline by month three. Rising complexity was associated with a proportional decrease in future development velocity, suggesting that short-term gains can be offset by later maintenance costs\footnote{Complexity +40.7\% ($p < 0.001$), with a velocity spike of +281\% in month one and baseline performance by month three.}. A large-scale audit of 304,000 AI-authored commits across 6,275 repositories~\citep{liu2026techdebt} found measurable technical debt, with approximately one-quarter of AI-introduced issues persisting to the latest revision and security-related issues persisting at a substantially higher rate. While these studies target adjacent systems, the architectural parallels (bounded context, tool-use loops, local generation decisions) make them relevant comparison points for the design analyzed here. A controlled minimal-pair study run on Claude Code itself sharpens where these effects are measurable: code cleanliness did not change task pass rate, but cleaner inputs reduced token use by 7 to 8\% and file revisitations by 34\%~\citep{trivedi2026cleanliness}, indicating that the operational footprint (cost and navigation), rather than raw success, is where the harness's handling of context most visibly registers.

Claude Code's context management pipeline is specifically designed to mitigate these effects: graduated compression preserves the most recent and most relevant context, cache-aware compaction avoids invalidating prompt caches during compression, read-time projection maintains full history for reconstruction while presenting a compressed view to the model, and subagent summary isolation prevents exploratory noise from accumulating in the parent context. Whether these mechanisms are sufficient to overcome the structural limitations of bounded context is a directly measurable empirical question that the source-level analysis in this paper cannot resolve.

\subsection{Limitations}
\label{sec:discuss:limits}

Beyond the methodological limitations described in the appendix, several analytical constraints apply. The memoized context assembly functions (\func{getSystemContext} and \func{getUserContext} both use lodash \code{memoize} at \file{context.ts}) mean that git status and CLAUDE.md content are cached rather than recomputed on every turn. Dynamic changes during a conversation may not be reflected immediately, though compaction can clear caches and lazy-loaded path-scoped rules provide a partial counter-mechanism.

Feature flags create build-time variability. In a build where \code{TRANSCRIPT\_CLASSIFIER} is false, the entire auto-mode classifier is eliminated. Feature-gated modules use dynamic \code{require()} rather than static \code{import} (e.g., \file{query.ts} for context collapse), because \code{feature()} only works in if/ternary conditions due to a bun:bundle tree-shaking constraint. Different build targets may produce functionally different applications.

\subsection{Emerging Directions}
\label{sec:discuss:future}

Several aspects of the implementation relate to broader design questions. Longer context windows would reduce compaction pressure, potentially simplifying the graduated pipeline. Multi-modal tools (screenshots, diagrams, UI previews) would expand the tool surface and create new context challenges. Claude Code Desktop already moves in this direction: in its Browser pane, Claude checks UI changes by taking screenshots, inspecting page elements, and interacting with the running app~\citep{anthropic2026desktop}. Formal verification of permission properties (for example, proving that deny rules always take precedence, that sandboxed commands cannot escape isolation, or that resumed sessions cannot inherit stale permissions) would provide stronger safety guarantees.

\paragraph{Architectural decoupling.}
The tightly coupled local architecture analyzed here is one point on a spectrum that is already evolving. Anthropic's own Managed Agents work~\citep{anthropic2026managed} describes virtualizing the components of an agent (session, harness, sandbox) so that ``each became an interface that made few assumptions about the others, and each could fail or be replaced independently'', drawing an explicit analogy to how operating systems virtualized hardware into processes and files. The Harness Design essay~\citep{anthropic2026harness} makes a similar point from a different angle, arguing that better models move the space of useful harness combinations rather than eliminating it. The architecture documented in this paper should therefore be read as a snapshot of a co-evolving system rather than a fixed optimum.

Cloud-agent reports make this concrete without relying on the operating-system analogy. Cursor separates the agent loop, machine state, and conversation state, and treats VM checkpointing, network policy, secret handling, and credential management as part of the agent product~\citep{cursor2026cloudagents}. LangChain's Managed Deep Agents makes a similar move by placing durable threads, checkpointing, streaming, context, observability, and human-in-the-loop workflows in the managed runtime~\citep{langchain2026manageddeepagents}. Google's Antigravity gives people artifacts such as plans, screenshots, and recordings to check an agent's work~\citep{google2026antigravity}. DeepSeek Harness makes runtime components, including the agent loop and session log, replaceable through plugin configuration~\citep{deepseek-harness2026}. The useful design question is therefore runtime ownership: which layer owns the loop, state, sandbox, policy, and recovery path?

\paragraph{Adapting models and harnesses.}
Automated adaptation can change the harness, the model, or both. \citet{yang2026better} keep model weights fixed and use failure traces to revise instructions, tools, hooks, and context management. \citet{yu2026co} go one step further: an expert model corrects failed turns in a smaller model's own traces, and these corrections become fine-tuning data for a model that is then evaluated with the evolved harness.

\paragraph{Memory as a first-class subsystem.}
The memory survey of \citet{hu2025memory} argues that agent memory is becoming a distinct cognitive substrate rather than a side effect of context window management, and identifies automated memory management, RL-driven memory, and trustworthy memory (privacy, explainability, and hallucination robustness) as open frontiers. Claude Code today exposes the factual tier (CLAUDE.md, auto memory) and the working tier (the conversation window). The experiential tier (accumulated, automatically curated playbooks of strategies learned from past sessions) is the natural next step, and the context-engineering literature~\citep{zhang2026ace} has started to provide mechanisms for that accumulation.

Context Hub makes the boundary visible. LangSmith Context Hub treats \file{AGENTS.md} files, skills, policies, examples, and related bundles as versioned and reviewable artifacts, with tags for deployment across environments~\citep{langchain2026contexthub}. That is a different design point from Claude Code's file hierarchy, but it names the same problem: context needs a lifecycle, not only a token budget. Skills raise a related maintenance problem, because an update to one skill can affect others. \citet{yang2026learning} represent dependencies, joint use, and conflicts among skills in a graph, revise related skills together, and check each combined change by replaying relevant cases.

\paragraph{Observability and silent failure.}
Industry surveys suggest that the dominant failure mode of deployed agents is not crashes but silent mistakes. Bessemer's 2026 infrastructure report~\citep{bessemer2026infra} estimates that ``78\% of AI failures are invisible'', while LangChain's 1{,}340-respondent state-of-agent-engineering survey~\citep{langchain2026state} identifies quality, not cost, as the top barrier to production use and finds a wide gap between observability (nearly 89\% adoption) and offline evaluation (52.4\%). The architecture analyzed here gives operators visibility into tool calls, hooks, and session transcripts. Closing the evaluation gap likely requires additional scaffolding (generator-evaluator separation, sprint contracts, and post-hoc checks of the kind discussed in~\citet{anthropic2026harness}) rather than model improvements alone.

OpenAI's agent-improvement loop makes this operational: traces and feedback become evals, and the resulting evidence becomes a Codex handoff for harness changes~\citep{openai2026agentimprovementloop}. In that loop, telemetry matters because it turns into a reusable test or a concrete change request.

\paragraph{Governance.}
Broader governance trends will constrain the design space as agents become more autonomous. The International AI Safety Report~\citep{bengio2026safety} warns that ``AI agents pose heightened risks because they act autonomously, making it harder for humans to intervene before failures cause harm,'' and the MIT AI Agent Index~\citep{staufer2026agentindex} finds that only 13.3\% of indexed agentic systems publish agent-specific safety cards. Emerging regulatory efforts, including the EU AI Act and evolving copyright debates around AI-generated code, may impose external constraints on logging, transparency, risk management, and human oversight that shape how coding agent architectures evolve.

\paragraph{Proactive architectures.}
The feature-gated \code{KAIROS} system illustrates how this architecture may evolve beyond reactive tool use. \code{KAIROS} implements a persistent background agent with tick-based heartbeats: when no user messages are pending, the system injects periodic \code{<tick>} prompts, and the model decides whether to act or sleep. The design addresses a documented tension in one study of proactive AI assistants: task completion increased by 12 to 18\%, while user preference dropped at high frequencies~\citep{chi2025proactive}. \code{KAIROS} tries to resolve this through terminal focus awareness (maximizing autonomous action when the user is away, increasing collaboration when present) and economic throttling via \code{SleepTool} (each wake-up costs an API call, and the prompt cache expires after five minutes of inactivity, making sleep/wake an explicit cost optimization). This binding of proactivity to both user presence and token economics is notable, though \code{KAIROS} cannot be confirmed as active in production builds.

\subsection{Recurring Design Choices}
\label{sec:discuss:guidelines}

Reading the six subsystem analyses together reveals three cross-cutting design commitments that recur across otherwise independent components.

\paragraph{Graduated layering over monolithic mechanisms.}
Safety, context management, and extensibility all use graduated stacks of independent mechanisms rather than single integrated solutions. The permission architecture layers the seven independent stages enumerated in \Cref{sec:arch:safety-preview}: tool pre-filtering, deny-first rule evaluation, permission-mode constraints, the auto-mode classifier, shell sandboxing, non-restoration of session permissions from transcripts, and hook interception. Context management layers five compaction stages, lazy-loaded CLAUDE.md files, deferred tool schemas, and summary-only subagent returns. Extensibility layers four mechanisms (MCP servers, plugins, skills, and hooks) at different context costs (\Cref{sec:ext}). In each case, the design trades simplicity and debuggability for defense in depth, accepting that the interaction between layers can produce emergent behaviors difficult to predict from any single configuration. The OpenClaw and Hermes Agent contrasts in \Cref{sec:compare} show that graduated layering recurs across systems with very different deployment contexts, suggesting that the layering pattern reflects the shared design questions rather than Claude Code-specific implementation choices.

\paragraph{Append-only designs that favor auditability over query power.}
Session transcripts are append-only JSONL files with read-time chain patching. Session-scoped permissions are not stored in the transcript. Context compaction applies read-time projections over a full history rather than destructive edits. This commitment recurs because it preserves the ability to resume, fork, and audit sessions without modifying previously written state. The cost is that richer structured queries (``show me all tool calls that modified file X across sessions'') require post-hoc reconstruction rather than direct lookup.

\paragraph{Model judgment within a deterministic harness.}
Across all subsystems, the architecture trusts the model's judgment within a rich deterministic harness rather than constraining its choices. The estimated 1.6\% decision-logic ratio captures this quantitatively: the harness creates conditions (tool routing, permission enforcement, context assembly, recovery logic) under which the model can decide well. Hierarchical permissions preserve safety invariants across agent boundaries, and \func{assembleToolPool} merges built-in and MCP tools into a single unified interface, but the model retains full latitude over which tools to invoke and in what order. The trade-off is that good local decisions can produce poor global outcomes when bounded context prevents global awareness, as the empirical predictions of \Cref{sec:discuss:empirical} document.

\section{Future Directions}
\label{sec:future}

\Cref{sec:discuss} read the architecture documented in \Cref{sec:arch,sec:turn,sec:auth,sec:ext,sec:context,sec:subagent,sec:persist} as a coherent design point and surfaced the tensions, trade-offs, and near-horizon directions that design point implies. This section steps beyond the architecture itself to record six open questions that \Cref{sec:discuss:future} partially names and that a growing external literature has sharpened enough to state concretely. These questions cover the paper's five-value framework (\Cref{sec:values:five}) and the long-term developer-capability question introduced in \Cref{sec:values:lens}. They concern governance constraints on the Authority hierarchy (\Cref{sec:future:governance}), the observability-evaluation gap on the Safety side (\Cref{sec:future:evaluation}), cross-session persistence on the Reliability side (\Cref{sec:future:persistence}), four extensions of the Capability frontier (\Cref{sec:future:boundary}), horizon scaling as a distinct axis of Reliable Execution (\Cref{sec:future:horizon}), and long-term developer capability reframed as a design question (\Cref{sec:future:lens}). Consistent with \Cref{sec:discuss:future}'s framing, each question is posed in the form \emph{whether}/\emph{how}/\emph{which}. Specific mechanism choices are named when the cited sources name them and left open otherwise.

\subsection{Silent Failure and the Observability-Evaluation Gap}
\label{sec:future:evaluation}

The observability-evaluation gap reported in \Cref{sec:discuss:future} could reflect a missing tooling layer, a missing evaluation interface inside the harness, or a model-capability ceiling. The sources there do not adjudicate. How the silent-mistake failure mode noted in that paragraph should be surfaced is therefore an architectural question for the harness rather than a capability question for the model. Recent empirical work characterises the gap at several resolutions. \citet{cemri2025massfail} catalogue fourteen failure modes spanning system-design issues, inter-agent misalignment, and task verification. \citet{pathak2025silentmultiagent} build a benchmark of agent trajectories specifically for anomaly detection in traces. \citet{yao2024taubench} expose consistency gaps via the $\text{pass}^k$ metric, the probability that all $k$ independent trials succeed. \citet{kapoor2024agentsthatmatter} argue that current agent benchmarks lack holdouts and cost controls, limiting what observability can actually diagnose.

Harnesses that update themselves also need fair comparisons. \citet{wang2026rethinking} compare harness evolution with search over task solutions under matched feedback and inference budgets, and \citet{le2026does} test trained models with a harness held out of training.

Against the permission pipeline and tool-orchestration layers analysed in \Cref{sec:auth,sec:turn}, two architectural questions remain open. First, whether the scaffolding the paper cites from \citet{anthropic2026harness} (generator-evaluator separation, sprint contracts, post-hoc checks, building on \citet{madaan2023selfrefine}'s self-refine pattern) belongs inside the harness (e.g., as additional hook events alongside the 27 documented in \Cref{sec:ext}) or outside it as a separate evaluation layer is not settled by the cited sources. Second, whether the existing hook pipeline of \Cref{sec:ext} can host such scaffolding within its current context-cost envelope is a further open question. The observation that closing this gap ``likely requires additional scaffolding \ldots rather than model improvements alone'' (\Cref{sec:discuss:future}) locates the open work at the harness layer.

\subsection{Persistence: Memory and Longitudinal Colleague Relationships}
\label{sec:future:persistence}

Whether agent state and the human-agent working relationship should persist across sessions, and in what form, is treated by the paper at two distinct layers today. \Cref{sec:context} documents the four-level CLAUDE.md hierarchy and auto memory. \Cref{sec:persist} documents mostly-append-only JSONL transcripts, with explicit cleanup rewrites; session-scoped permissions are not stored in them. What belongs between these two layers is an open design question: durable state that is neither a static instruction nor a single session's transcript. \citet{hu2025memory} and \citet{zhang2026ace}, already cited in \Cref{sec:discuss:future}, motivate an accumulating layer. \citet{packer2024memgpt} reframes the LLM as an operating system with paged memory. \citet{chhikara2025mem0} builds a production-oriented memory store that survives restarts, while \citet{xu2025amem} proposes a research agentic-memory design. \citet{wang2024agentworkflowmemory} captures reusable procedural traces. \citet{shinn2023reflexion} accumulates self-reflection traces via verbal reinforcement across attempts. Surveys by \citet{zhang2024memorysurvey} and \citet{huang2026rethinkingmemory} map candidate mechanisms.

MemGym makes the memory question more concrete. It evaluates dynamic memory formation across tool-use dialogue, deep research, coding, and computer use, rather than only retention of personal facts in chat~\citep{xu2026memgym}. This moves the question from ``where should notes be stored?'' to ``which remembered state helps the agent keep working correctly?''

Persistent memory also needs a way to recover from harmful entries. MemSecBench tests whether stored information affects later actions and whether repair keeps specified valid memories intact~\citep{chen2026memsecbench}.

The same persistence question recurs on the human side. \Cref{sec:discuss:future} already cites longitudinal autonomy evidence from \citet{anthropic2025internal} and \citet{anthropic2026autonomy}. \citet{dellacqua2025cybernetic}'s field experiment with 776 Procter~\& Gamble professionals, together with longitudinal and organisational studies of Copilot rollouts~\citep{stray2025copiloteval} and AI-teamwork trajectories~\citep{xiao2025aiteamwork}, report shifts in human-AI work dynamics as collaboration accumulates. \citet{wang2023voyager} illustrates an embodied agent that accumulates a skill library across tasks. \citet{mollick2024cointelligence} frames the human-AI working relationship as co-intelligence.

Whether a single substrate can carry both a user's personal instruction hierarchy and a shared organisational context while preserving the file-based transparency of CLAUDE.md that \Cref{sec:context} documents is an open architectural question. How session-scoped permissions interact with such a substrate, without reintroducing the resume-restoration concern that \Cref{sec:persist} closes as a deliberate safety choice, is a further open question. Claude Code's Projects beta is one product step in this direction: separate cloud sessions add to and draw from a shared project memory that carries decisions across sessions~\citep{anthropic2026projects}. The OpenClaw memory subsystem (dreaming, daily notes, hybrid retrieval) and Hermes Agent's WAL-mode SQLite session store with full-text search, both documented in \Cref{sec:compare,sec:compare:hermes}, provide two concrete reference points for how a persistent substrate can sit between a static instruction hierarchy and a single-session transcript.

\subsection{Harness Boundary Evolution: Where, When, What, and with Whom the Agent Acts}
\label{sec:future:boundary}

\Cref{sec:discuss:future} cites \citet{anthropic2026harness}'s argument that better models move the space of useful harness combinations rather than eliminating it. Whether that movement will be most pronounced in \emph{where} the harness runs, \emph{when} it acts, \emph{what} it acts on, or \emph{with whom} it coordinates is not resolved by the source-level analysis in \Cref{sec:arch,sec:turn,sec:auth,sec:ext,sec:context,sec:subagent,sec:persist}. Each of the four has an active research literature that the paper touches only in passing.

\paragraph{Where.} \citet{anthropic2026managed}'s Managed Agents design virtualizes session, harness, and sandbox into independently replaceable interfaces. This extends the virtual-memory analogy that \citet{packer2024memgpt} applies to context-window management and that \citet{karpathy2023llmos} popularizes more broadly. \citet{khattab2024dspy} treats the harness itself as a compile target. 

\paragraph{When.} \Cref{sec:discuss:future} already introduces KAIROS as a feature-gated illustration, motivated by the $+12\%$ to $+18\%$ task-pass gain that \citet{chi2025proactive} report and the sharp preference penalty (47\% vs.\ 80\% to 90\%) restricted to the high-frequency \emph{Persistent Suggest} variant. Claude Code routines use a simpler trigger: they start a new cloud session when a configured schedule, API call, or GitHub event fires~\citep{anthropic2026routines}, whereas KAIROS wakes a background agent and lets it decide at each tick whether to act. \citet{liu2025innerthoughts}, \citet{pu2025codellaborator}, and \citet{lee2025sensibleagent} extend the proactivity design space across programming and ambient-interface settings. \citet{pasternak2025probe} and \citet{sun2025userville} introduce benchmarks and training regimes aimed at sharpening it, and \citet{deng2025proactivesurvey} surveys the broader literature.

\paragraph{What.} Vision-language-action work extends the harness beyond textual tool returns. \citet{brohan2023rt2} and \citet{black2024pi0} train VLA policies that execute physical actions, and \citet{ahn2022saycan} grounds plans in robot affordances. Industry systems such as \citet{figure2025helix} and \citet{nvidia2025gr00t} push similar ideas into humanoid control. These systems face the reversibility-weighted risk principle (\Cref{tab:principles}) at a cost asymmetry that the principle names but does not quantify for non-textual actions.

\paragraph{With whom.} Role-differentiated multi-agent systems (\citet{hong2024metagpt}, \citet{li2023camel}, \citet{chen2024agentverse}, \citet{qian2024chatdev}) compose agents with distinct responsibilities. Multi-agent debate~\citep{du2023debate, liang2024divergent} and graph-structured workflows~\citep{zhuge2024gptswarm} explore alternatives to the parent/subagent pattern of \Cref{sec:subagent}. \citet{guo2024massurvey} surveys this space.

Whether a single harness architecture can span all four extensions, or whether the ``harness combinations'' \citet{anthropic2026harness} describes will fragment into specialised stacks, is an open design question. The \emph{when}-extension directly continues the Capability-versus-Adaptability tension in \Cref{tab:tensions}. The \emph{with-whom}-extension partially maps onto Capability-versus-Reliability but raises cross-agent consistency concerns that \Cref{tab:tensions} does not itself cover. The \emph{where}- and \emph{what}-extensions raise further questions the paper's current subsystem boundaries do not cover: which governance obligations attach when harness components become hosted services (\Cref{sec:future:governance}), and how reversibility-weighted risk (\Cref{tab:principles}) scales to physical rather than textual effects. How these extensions compose across axes, rather than within any one, is not something the paper's single-subsystem analyses can resolve.

\subsection{Horizon Scaling: From Session to Scientific Program}
\label{sec:future:horizon}

\Cref{sec:values:five} defines Reliable Execution as spanning ``both single-turn correctness and long-horizon dependability.'' How the architecture documented in \Cref{sec:arch,sec:turn,sec:context,sec:subagent,sec:persist} continues to support long-horizon dependability as autonomous work extends beyond a single session is an open question. Its primary units are the turn, the session, and the sub-agent. A growing literature targets this regime. \citet{lu2024aiscientist} present an end-to-end autonomous research pipeline producing draft manuscripts. \citet{beel2025evalsakana} provide an independent SIGIR Forum evaluation of that pipeline, characterising what ``autonomous research'' currently delivers and where it falls short. \citet{gottweis2025coscientist} develop a multi-agent hypothesis-generation system that runs across days rather than turns, and \citet{novikov2025alphaevolve} pursue algorithmic discovery over timescales that previously took human experts weeks. \citet{kwa2025metrtimehorizon}'s METR study measures the task duration at which frontier agents succeed with fixed reliability, giving an empirical frame for this scaling question.

The Quantitative Goal Persistence benchmark names a narrower failure mode: an agent can make reasonable local tool calls and still stop before a verifier confirms enough valid work units~\citep{cai2026pushagent}. For long-horizon coding agents, progress needs an external notion of done, not only a fluent final message.

SWE Refactor Bench applies the same idea to repository migration: it checks migration requirements separately from fixed behavior tests, so an incomplete migration and a migration that breaks required behavior appear as different failures~\citep{hong2026swe}.

Against the paper's analysis, long-horizon deployment tests whether the context-management pipeline of \Cref{sec:context}, the last-assistant-text return policy of \Cref{sec:subagent}, and the append-only persistence of \Cref{sec:persist} remain sufficient when sessions compose into multi-session programs. \Cref{sec:discuss:empirical} already frames this as ``a directly measurable empirical question'' that source-level analysis cannot resolve. Horizon scaling restates that question at the scale of weeks: whether the harness layer alone closes the gap, whether a cross-session memory substrate (\Cref{sec:future:persistence}) is required, or whether horizon-scale work demands coordination primitives beyond session, sub-agent, and memory, is not something the paper's session-scoped analyses can settle.

(\tierA{}) A development subsequent to the \code{v2.1.88} snapshot analyzed here makes this question concrete. Claude Code \code{v2.1.154} introduced \emph{dynamic workflows}, in which the model writes a JavaScript orchestration script that a background runtime executes, fanning out to many subagents (bounded at up to a thousand per run) while intermediate results live in script variables outside the model's context window rather than passing through it~\citep{claudecode2026workflows}. Released alongside Claude Opus 4.8~\citep{anthropic2026opus48}, this is precisely a coordination primitive beyond the session, sub-agent, and memory units analyzed here: it relocates the orchestration logic itself out of the conversation, extending the \emph{context as scarce resource} and \emph{isolated subagent boundaries} principles (\Cref{tab:principles}) from individual subagents to the orchestration layer. Whether orchestration-as-code becomes the dominant long-horizon primitive, and how its token cost and reduced human mid-run oversight trade against its reliability gains, is an open question the session-scoped analysis here cannot settle.

\subsection{Governance and Oversight at Scale}
\label{sec:future:governance}

Emerging AI regulation adds an external constraint on the architectures that implement the Authority hierarchy of Anthropic, operators, and users documented in \Cref{sec:values:five}. Which logging, transparency, and human-oversight affordances coding-agent architectures should expose under that external constraint remains an open design question. The European Commission's GPAI Code of Practice~\citep{eugpai2025cop} and implementation guidelines~\citep{eugpai2025guidelines} illustrate how general-purpose AI governance is moving toward more explicit expectations for documentation, risk management, transparency, and oversight. The MIT AI Agent Index~\citep{staufer2026agentindex} and the International AI Safety Report~\citep{bengio2026safety}, already cited in \Cref{sec:discuss:future}, motivate the disclosure and oversight side of this constraint. The Bartz v.\ Anthropic ruling~\citep{bartzanthropic2025} adds an input-side constraint on training-data sourcing, distinct from the output-side copyright questions about AI-generated code that emerging cases address separately. An OECD report on AI governance frameworks~\citep{oecd2025governing} and an early analysis of compliance obligations for agent providers by \citet{nannini2026agentlaw} sketch what regulator-facing interfaces might look like without prescribing specifics.

(\tierA{}) Regulation can also act on a model's availability, not only on what an architecture must disclose. In June 2026 a US export-control order barred foreign nationals from Anthropic's Fable~5 and Mythos~5 models, and within days of launch Anthropic disabled both models for all users worldwide~\citep{anthropic2026fableaccess}. Anthropic reported that the controls were lifted on June 30; Fable~5 returned for users globally on July 1, while Mythos~5 access was restored only for a set of US organizations~\citep{anthropic2026fablerestoration}. Government policy of this kind can directly determine when, and to whom, such a model is available.

Agentic safety benchmarks also move the safety object from text to environment state. Boiling the Frog evaluates whether multi-turn workspace edits become unsafe under incremental attacks~\citep{bisconti2026boilingfrog}. This matters for coding agents because the risky object may be the modified repository, not the final answer text.

Read against the permission pipeline analyzed in \Cref{sec:auth}, two properties of the current architecture are open under this constraint. First, the deny-first evaluation the paper documents is internally auditable through session transcripts (\Cref{sec:persist}), and Claude Code can export tool permission decision events through OpenTelemetry~\citep{anthropic2026monitoring}. How these records can support external audits under emerging frameworks such as the GPAI Code of Practice~\citep{eugpai2025cop} remains an open question. Second, whether the \emph{values-over-rules} principle, which the paper pairs with deterministic guardrails, admits the kind of explicit rule articulation that compliance review may call for is a further open question. Both properties lie within the harness rather than the model, which is where future architectures may need to expose new interfaces.

\subsection{Long-term Developer Capability}
\label{sec:future:lens}

\Cref{sec:values:lens} introduces long-term developer capability as a cross-cutting question rather than a co-equal design value. \Cref{sec:discuss:tensions,sec:discuss:empirical} extend the question with external evidence, including perceived-versus-measured productivity, comprehension loss, complexity accrual, technical-debt persistence, neural-connectivity persistence, and early-career hiring decline. Practitioners say the same: reflecting on his own move to agent-driven coding, \citet{karpathy2026sequoia} warns that ``you can outsource your thinking, but you can't outsource your understanding.'' \Cref{sec:conclude} then pivots: ``Future systems could treat that sustainability gap as a first-class design problem, not a downstream evaluation metric.'' Whether that pivot is possible, and what architectural mechanisms a first-class treatment would require, is the last of the open questions this section records.

This also links to process supervision. Human-LLM collaborative planning work argues that users need process-level control, not only outcome-level review, and organizes steering by mode, scope, and edit level~\citep{he2026steer}. For coding agents, the analogous question is what a user can inspect and change while the work is still unfolding.

Two sub-questions separate the measurement gap from the design gap. First, whether the empirical claims that motivate the question are measurable at session granularity. The existing citations operate at session to multi-month scales (\citet{becker2025measuring}'s 16-developer RCT, \citet{shen2026skill}'s comprehension-test comparison, \citet{kosmyna2025brain}'s EEG study, \citet{he2026cursor}'s 807-repository causal analysis, \citet{liu2026techdebt}'s 304{,}000-commit audit, \citet{rak2025aihiring}'s hiring series), but the harness documented in \Cref{sec:arch,sec:turn,sec:context} exposes no per-session signal for comprehension or convention drift. Related work on programmer interaction modes~\citep{barke2023groundedcopilot} and AI-induced code-security regressions~\citep{perry2023insecurecode} sketches session-granularity measurement, and \citet{aiersilan2026vibecheck} proposes a protocol for session-level cognitive-offloading probes. Second, whether architecture can respond to such measurements once they exist (an analogue of the generator-evaluator separation~\citep{anthropic2026harness} applied to the human loop, comprehension-preserving surfaces, or mechanisms not yet named) is the design-gap question \Cref{sec:conclude} poses. The paper takes no position on which mechanism class is appropriate. It also does not settle whether the harness documented here is the right locus for that action, rather than the IDE, the organisation, or the human development loop. The related work surveyed in \Cref{sec:related} and the sustainability pivot of \Cref{sec:conclude} mark where this paper leaves the question.

\section{Conclusion}
\label{sec:conclude}

This paper shows that production coding agents can be understood as answers to a recurring set of design questions: where reasoning sits relative to the harness, how execution, safety, extensibility, context, delegation, and persistence are organized, and which trade-offs those choices encode. Claude Code occupies a clear design point within that space. It gives the model broad local autonomy while surrounding it with a dense deterministic harness for permissioning, tool routing, context compaction, extensibility, and session recovery. Read through the five values and thirteen design principles identified in \Cref{sec:values}, these choices are coherent rather than ad hoc: the system consistently prioritizes human decision authority, safety, reliable execution, capability amplification, and contextual adaptability.

The OpenClaw and Hermes Agent comparisons show that the same design questions recur in different agent systems but produce different answers. Where Claude Code invests in per-action safety classification and graduated context compression within a CLI harness, OpenClaw invests in perimeter-level access control and structured long-term memory within a multi-channel gateway, and Hermes Agent renders per-action approvals across many surfaces, from one process, with pluggable memory and model backends. The three systems compose through ACP at multiple positions: OpenClaw can host Claude Code as an external harness via ACP, and Hermes Agent sits on both sides of the host/guest split. For agent builders, the most consequential open question is therefore not how to add more autonomy, but how to preserve human capability over time. Taken together, the long-term capability question introduced in \Cref{sec:values:lens}, the analysis in \Cref{sec:discuss}, and the open questions surveyed in \Cref{sec:future} show that the architecture provides limited mechanisms that explicitly preserve long-term human understanding, codebase coherence, or the developer pipeline. Future systems could treat that sustainability gap as a first-class design problem, not a downstream evaluation metric.

\bibliographystyle{assets/plainnat}
\bibliography{resources/main,resources/verified}

@misc{openclaw2026,
  author = {Peter Steinberger and {OpenClaw Contributors}},
  title = {{OpenClaw}: Personal {AI} Assistant},
  year = {2026},
  howpublished = {\url{https://github.com/openclaw/openclaw}},
  note = {Open-source multi-channel AI assistant gateway. MIT License.},
}

@article{chen2021evaluating,
  title={Evaluating large language models trained on code},
  author={Chen, Mark and Tworek, Jerry and Jun, Heewoo and Yuan, Qiming and Pinto, Henrique Ponde De Oliveira and Kaplan, Jared and Edwards, Harri and Burda, Yuri and Joseph, Nicholas and Brockman, Greg and others},
  journal={arXiv preprint arXiv:2107.03374},
  year={2021}
}

@article{yang2024sweagent,
  title={Swe-agent: Agent-computer interfaces enable automated software engineering},
  author={Yang, John and Jimenez, Carlos E and Wettig, Alexander and Lieret, Kilian and Yao, Shunyu and Narasimhan, Karthik and Press, Ofir},
  journal={Advances in Neural Information Processing Systems},
  volume={37},
  pages={50528--50652},
  year={2024}
}

@article{wang2024openhands,
  title={Openhands: An open platform for ai software developers as generalist agents},
  author={Wang, Xingyao and Li, Boxuan and Song, Yufan and Xu, Frank F and Tang, Xiangru and Zhuge, Mingchen and Pan, Jiayi and Song, Yueqi and Li, Bowen and Singh, Jaskirat and others},
  journal={arXiv preprint arXiv:2407.16741},
  year={2024}
}

@inproceedings{yao2022react,
  title={React: Synergizing reasoning and acting in language models},
  author={Yao, Shunyu and Zhao, Jeffrey and Yu, Dian and Du, Nan and Shafran, Izhak and Narasimhan, Karthik R and Cao, Yuan},
  booktitle={The eleventh international conference on learning representations},
  year={2022}
}

@article{schick2023toolformer,
  title={Toolformer: Language models can teach themselves to use tools},
  author={Schick, Timo and Dwivedi-Yu, Jane and Dess{\`\i}, Roberto and Raileanu, Roberta and Lomeli, Maria and Hambro, Eric and Zettlemoyer, Luke and Cancedda, Nicola and Scialom, Thomas},
  journal={Advances in neural information processing systems},
  volume={36},
  pages={68539--68551},
  year={2023}
}

@article{zhou2024lats,
  title={Language agent tree search unifies reasoning acting and planning in language models},
  author={Zhou, Andy and Yan, Kai and Shlapentokh-Rothman, Michal and Wang, Haohan and Wang, Yu-Xiong},
  journal={arXiv preprint arXiv:2310.04406},
  year={2023}
}

@misc{langgraph2024,
  author       = {{LangChain, Inc.}},
  title        = {{LangGraph}: Build Resilient Language Agents as Graphs},
  year         = {2024},
  url          = {https://github.com/langchain-ai/langgraph},
  note         = {GitHub repository},
}

@inproceedings{wu2024autogen,
  title={Autogen: Enabling next-gen LLM applications via multi-agent conversations},
  author={Wu, Qingyun and Bansal, Gagan and Zhang, Jieyu and Wu, Yiran and Li, Beibin and Zhu, Erkang and Jiang, Li and Zhang, Xiaoyun and Zhang, Shaokun and Liu, Jiale and others},
  booktitle={First conference on language modeling},
  year={2024}
}

@article{jimenez2024swebench,
  title={Swe-bench: Can language models resolve real-world github issues?},
  author={Jimenez, Carlos E and Yang, John and Wettig, Alexander and Yao, Shunyu and Pei, Kexin and Press, Ofir and Narasimhan, Karthik},
  journal={arXiv preprint arXiv:2310.06770},
  year={2023}
}

@inproceedings{reis2009isolating,
  title={Isolating web programs in modern browser architectures},
  author={Reis, Charles and Gribble, Steven D},
  booktitle={Proceedings of the 4th ACM European conference on Computer systems},
  pages={219--232},
  year={2009}
}

@article{sandhu1996rbac,
  title={Role-based access control models},
  author={Sandhu, Ravi S and Coyne, Edward J and Feinstein, Hal L and Youman, Charles E},
  journal={Computer},
  volume={29},
  number={2},
  pages={38--47},
  year={2002},
  publisher={IEEE}
}

@article{garlan1993architecture,
  title={An introduction to software architecture.},
  author={Garlan, David and Shaw, Mary and others},
  journal={Advances in software engineering and knowledge engineering},
  volume={1},
  number={3.4},
  year={1993},
  publisher={World Scientific}
}

@book{wooldridge2009multiagent,
  title={An introduction to multiagent systems},
  author={Wooldridge, Michael},
  year={2009},
  publisher={John Wiley \& Sons}
}

@misc{gauthier2024aider,
  author       = {Paul Gauthier},
  title        = {Aider: {AI} Pair Programming in Your Terminal},
  year         = {2024},
  url          = {https://github.com/Aider-AI/aider},
  note         = {Open-source software, \url{https://aider.chat}},
}

@misc{anthropic2026harness,
  author = {Prithvi Rajasekaran},
  title = {Harness Design for Long-Running Application Development},
  year = {2026},
  howpublished = {Anthropic Engineering Blog, \url{https://anthropic.com/engineering/harness-design-long-running-apps}},
}

@misc{anthropic2025internal,
  author = {Saffron Huang and Bryan Seethor and Esin Durmus and Kunal Handa and Miles McCain and Michael Stern and Deep Ganguli},
  title = {How {AI} Is Transforming Work at {Anthropic}},
  year = {2025},
  howpublished = {Anthropic Research Blog, \url{https://anthropic.com/research/how-ai-is-transforming-work-at-anthropic}},
}

@misc{cherny2025latentspace,
  author = {Boris Cherny and Cat Wu},
  title = {Claude Code: Anthropic's Agent in Your Terminal},
  year = {2025},
  howpublished = {Latent Space podcast, \url{https://www.latent.space/p/claude-code}},
}

@misc{anthropic2026constitution,
  author = {{Anthropic}},
  title = {Claude's Constitution},
  year = {2026},
  howpublished = {\url{https://anthropic.com/constitution}},
}

@misc{anthropic2026automode,
  author = {John Hughes},
  title = {{Claude Code} Auto Mode: A Safer Way to Skip Permissions},
  year = {2026},
  howpublished = {Anthropic Engineering, \url{https://www.anthropic.com/engineering/claude-code-auto-mode}},
}

@misc{anthropic2025sandboxing,
  author = {David Dworken and Oliver Weller-Davies},
  title = {Beyond Permission Prompts: Making {Claude Code} More Secure and Autonomous},
  year = {2025},
  howpublished = {Anthropic Engineering, \url{https://www.anthropic.com/engineering/claude-code-sandboxing}},
}

@misc{anthropic2026autonomy,
  author = {Miles McCain and Thomas Millar and Saffron Huang and Jake Eaton and Kunal Handa and Michael Stern and Alex Tamkin and Matt Kearney and Esin Durmus and Judy Shen and Jerry Hong and Brian Calvert and Jun Shern Chan and Francesco Mosconi and David Saunders and Tyler Neylon and Gabriel Nicholas and Sarah Pollack and Jack Clark and Deep Ganguli},
  title = {Measuring {AI} Agent Autonomy in Practice},
  year = {2026},
  howpublished = {Anthropic Research Blog, \url{https://anthropic.com/research/measuring-agent-autonomy}},
}

@misc{adversa2026bypass,
  author = {{Adversa.ai}},
  title = {Critical {Claude Code} Vulnerability: Deny Rules Silently Bypassed Because Security Checks Cost Too Many Tokens},
  year = {2026},
  howpublished = {\url{https://adversa.ai/blog/claude-code-security-bypass-deny-rules-disabled/}},
}

@misc{checkpoint2026rce,
  author = {Aviv Donenfeld and Oded Vanunu},
  title = {Caught in the Hook: {RCE} and {API} Token Exfiltration Through {Claude Code} Project Files},
  year = {2026},
  howpublished = {\url{https://research.checkpoint.com/2026/rce-and-api-token-exfiltration-through-claude-code-project-files-cve-2025-59536/}},
  note = {CVE-2025-59536 (CVSS 8.7), CVE-2026-21852 (CVSS 5.3)},
}

@misc{cymulate2025inverseprompt,
  author = {Elad Beber},
  title = {{InversePrompt}: Turning Claude Against Itself, One Prompt at a Time},
  year = {2025},
  howpublished = {\url{https://cymulate.com/blog/cve-2025-547954-54795-claude-inverseprompt/}},
  note = {CVE-2025-54794, CVE-2025-54795; updated April 6, 2026},
}

@article{he2026cursor,
  title={Speed at the Cost of Quality: How Cursor AI Increases Short-Term Velocity and Long-Term Complexity in Open-Source Projects},
  author={He, Hao and Miller, Courtney and Agarwal, Shyam and K{\"a}stner, Christian and Vasilescu, Bogdan},
  journal={arXiv preprint arXiv:2511.04427},
  year={2025}
}

@article{liu2026techdebt,
  title={Debt Behind the AI Boom: A Large-Scale Empirical Study of AI-Generated Code in the Wild},
  author={Liu, Yue and Widyasari, Ratnadira and Zhao, Yanjie and Irsan, Ivana Clairine and Lo, David},
  journal={arXiv preprint arXiv:2603.28592},
  year={2026}
}

@inproceedings{chi2025proactive,
  title={Need help? designing proactive ai assistants for programming},
  author={Chen, Valerie and Zhu, Alan and Zhao, Sebastian and Mozannar, Hussein and Sontag, David and Talwalkar, Ameet},
  booktitle={Proceedings of the 2025 CHI Conference on Human Factors in Computing Systems},
  pages={1--18},
  year={2025}
}

@misc{mindstudio2025memory,
  author = {{MindStudio Team}},
  title = {What Is the Anthropic {Claude Code} Source Code Leak? Three-Layer Memory Architecture Explained},
  year = {2026},
  howpublished = {\url{https://www.mindstudio.ai/blog/claude-code-source-leak-three-layer-memory-architecture}},
}

@misc{anthropic2025agentteams,
  author = {{Anthropic}},
  title = {Orchestrate Teams of {Claude Code} Sessions},
  year = {2025},
  howpublished = {\url{https://code.claude.com/docs/en/agent-teams}},
}

@misc{anthropic2025agents,
  title = {Our Framework for Developing Safe and Trustworthy Agents},
  author = {{Anthropic}},
  year = {2025},
  howpublished = {\url{https://www.anthropic.com/news/our-framework-for-developing-safe-and-trustworthy-agents}},
}

@article{becker2025measuring,
  title={Measuring the impact of early-2025 AI on experienced open-source developer productivity},
  author={Becker, Joel and Rush, Nate and Barnes, Elizabeth and Rein, David},
  journal={arXiv preprint arXiv:2507.09089},
  year={2025}
}

@article{shen2026skill,
  title={How AI impacts skill formation},
  author={Shen, Judy Hanwen and Tamkin, Alex},
  journal={arXiv preprint arXiv:2601.20245},
  year={2026}
}

@article{kosmyna2025brain,
  title={Your brain on ChatGPT: Accumulation of cognitive debt when using an AI assistant for essay writing task},
  author={Kosmyna, Nataliya and Hauptmann, Eugene and Yuan, Ye Tong and Situ, Jessica and Liao, Xian-Hao and Beresnitzky, Ashly Vivian and Braunstein, Iris and Maes, Pattie},
  journal={arXiv preprint arXiv:2506.08872},
  volume={4},
  year={2025}
}

@article{rak2025aihiring,
  title={How to Stay Ahead of {AI} as an Early-Career Engineer},
  author={Rak, Gwendolyn},
  journal={IEEE Spectrum},
  year={2025},
  url={https://spectrum.ieee.org/ai-effect-entry-level-jobs},
}

@article{sui2026paste,
  title={Act While Thinking: Accelerating LLM Agents via Pattern-Aware Speculative Tool Execution},
  author={Sui, Yifan and Zhao, Han and Ma, Rui and He, Zhiyuan and Wang, Hao and Li, Jianxun and Yang, Yuqing},
  journal={arXiv preprint arXiv:2603.18897},
  year={2026}
}

@misc{linuxfoundation2025aaif,
  author={{The Linux Foundation}},
  title={Linux Foundation Announces the Formation of the Agentic {AI} Foundation ({AAIF}), Anchored by New Project Contributions Including Model Context Protocol ({MCP}), goose and {AGENTS.md}},
  howpublished={Linux Foundation Press Release},
  year={2025},
  url={https://www.linuxfoundation.org/press/linux-foundation-announces-the-formation-of-the-agentic-ai-foundation},
}

@article{aiersilan2026vibecheck,
  title={The Vibe-Check Protocol: Quantifying Cognitive Offloading in AI Programming},
  author={Aiersilan, Aizierjiang},
  journal={arXiv preprint arXiv:2601.02410},
  year={2026}
}

@misc{anthropic2026howworks,
  author = {{Anthropic}},
  title = {How {Claude Code} Works},
  year = {2026},
  howpublished = {\url{https://code.claude.com/docs/en/how-claude-code-works}},
}

@misc{anthropic2026memory,
  author = {{Anthropic}},
  title = {How {Claude} remembers your project},
  year = {2026},
  howpublished = {\url{https://code.claude.com/docs/en/memory}},
}

@article{hu2025adas,
  title={Automated design of agentic systems},
  author={Hu, Shengran and Lu, Cong and Clune, Jeff},
  journal={arXiv preprint arXiv:2408.08435},
  year={2024}
}

@article{wang2024agentsurvey,
  title={A survey on large language model based autonomous agents},
  author={Wang, Lei and Ma, Chen and Feng, Xueyang and Zhang, Zeyu and Yang, Hao and Zhang, Jingsen and Chen, Zhiyuan and Tang, Jiakai and Chen, Xu and Lin, Yankai and others},
  journal={Frontiers of Computer Science},
  volume={18},
  number={6},
  pages={186345},
  year={2024},
  publisher={Springer}
}

@article{xu2026agentsystems,
  title={AI Agent Systems: Architectures, Applications, and Evaluation},
  author={Xu, Bin},
  journal={arXiv preprint arXiv:2601.01743},
  year={2026}
}

@article{hu2025memory,
  title={Memory in the age of ai agents},
  author={Hu, Yuyang and Liu, Shichun and Yue, Yanwei and Zhang, Guibin and Liu, Boyang and Zhu, Fangyi and Lin, Jiahang and Guo, Honglin and Dou, Shihan and Xi, Zhiheng and others},
  journal={arXiv preprint arXiv:2512.13564},
  year={2025}
}

@article{zhang2026ace,
  title={Agentic context engineering: Evolving contexts for self-improving language models},
  author={Zhang, Qizheng and Hu, Changran and Upasani, Shubhangi and Ma, Boyuan and Hong, Fenglu and Kamanuru, Vamsidhar and Rainton, Jay and Wu, Chen and Ji, Mengmeng and Li, Hanchen and others},
  journal={arXiv preprint arXiv:2510.04618},
  year={2025}
}

@article{hou2025mcpsurvey,
  title={Model context protocol (mcp): Landscape, security threats, and future research directions},
  author={Hou, Xinyi and Zhao, Yanjie and Wang, Shenao and Wang, Haoyu},
  journal={ACM Transactions on Software Engineering and Methodology},
  year={2025},
  publisher={ACM New York, NY}
}

@misc{weng2023agent,
  author = {Lilian Weng},
  title = {{LLM}-Powered Autonomous Agents},
  year = {2023},
  howpublished = {\url{https://lilianweng.github.io/posts/2023-06-23-agent/}},
}

@misc{anthropic2024effective,
  author = {Erik Schluntz and Barry Zhang},
  title = {Building effective agents},
  year = {2024},
  howpublished = {Anthropic Research, \url{https://www.anthropic.com/research/building-effective-agents}},
}

@misc{anthropic2026managed,
  author = {Lance Martin and Gabe Cemaj and Michael Cohen},
  title = {Scaling Managed Agents: Decoupling the Brain from the Hands},
  year = {2026},
  howpublished = {Anthropic Engineering Blog, \url{https://www.anthropic.com/engineering/managed-agents}},
}

@misc{martin2026patterns,
  author = {Lance Martin},
  title = {Agent Design Patterns},
  year = {2026},
  howpublished = {\url{https://rlancemartin.github.io/2026/01/09/agent_design/}},
}

@misc{chase2025deepagents,
  author = {Harrison Chase},
  title = {Deep Agents},
  year = {2025},
  howpublished = {LangChain Blog, \url{https://blog.langchain.com/deep-agents/}},
}

@misc{swyx2025agentengineering,
  author = {Shawn Wang},
  title = {Agent Engineering},
  year = {2025},
  howpublished = {Latent Space, \url{https://www.latent.space/p/agent}},
}

@misc{huyen2025agents,
  author = {Chip Huyen},
  title = {Agents},
  year = {2025},
  howpublished = {\url{https://huyenchip.com/2025/01/07/agents.html}},
}

@misc{langchain2026state,
  author = {{LangChain}},
  title = {State of Agent Engineering},
  year = {2026},
  howpublished = {\url{https://www.langchain.com/state-of-agent-engineering}},
  note = {Survey of 1,340 respondents conducted Nov-Dec 2025},
}

@misc{bessemer2026infra,
  author = {Janelle Teng Wade and Lance Co Ting Keh and Talia Goldberg and David Cowan and Grace Ma and Bhavik Nagda and Brandon Nydick and Bar Weiner},
  title = {{AI} Infrastructure Roadmap: Five Frontiers for 2026},
  year = {2026},
  howpublished = {Bessemer Venture Partners, \url{https://www.bvp.com/atlas/ai-infrastructure-roadmap-five-frontiers-for-2026}},
}

@article{bengio2026safety,
  title={International ai safety report 2026},
  author={Bengio, Yoshua and Clare, Stephen and Prunkl, Carina and Andriushchenko, Maksym and Bucknall, Ben and Murray, Malcolm and Bommasani, Rishi and Casper, Stephen and Davidson, Tom and Douglas, Raymond and others},
  journal={arXiv preprint arXiv:2602.21012},
  year={2026}
}

@article{staufer2026agentindex,
  title={The 2025 AI Agent Index: Documenting Technical and Safety Features of Deployed Agentic AI Systems},
  author={Staufer, Leon and Feng, Kevin and Wei, Kevin and Bailey, Luke and Duan, Yawen and Yang, Mick and Ozisik, A Pinar and Casper, Stephen and Kolt, Noam},
  journal={arXiv preprint arXiv:2602.17753},
  year={2026}
}

@misc{cursor2026official,
  author = {{Cursor}},
  title = {Cursor: The Best Way to Code with {AI}},
  year = {2026},
  howpublished = {\url{https://cursor.com/}},
  note = {Official product website. Accessed April 12, 2026.},
}

@misc{anthropic2026claudecode,
  author = {{Anthropic}},
  title = {Claude Code Overview},
  year = {2026},
  howpublished = {\url{https://code.claude.com/docs}},
  note = {Official Claude Code documentation. Accessed April 12, 2026.},
}

@misc{anthropic2026github,
  author = {{Anthropic}},
  title = {Anthropic on GitHub},
  year = {2026},
  howpublished = {\url{https://github.com/anthropics}},
  note = {Verified GitHub organization page. Accessed April 12, 2026.},
}

@misc{eugpai2025cop,
  author = {{European Commission}},
  title = {General-Purpose {AI} Code of Practice},
  year = {2025},
  howpublished = {\url{https://digital-strategy.ec.europa.eu/en/policies/contents-code-gpai}},
  note = {Official EU Commission publication, 10 July 2025.},
}

@misc{eugpai2025guidelines,
  author = {{European Commission}},
  title = {Guidelines on the Scope of Obligations for Providers of General-Purpose {AI} Models Under the {AI} Act},
  year = {2025},
  howpublished = {\url{https://digital-strategy.ec.europa.eu/en/library/guidelines-scope-obligations-providers-general-purpose-ai-models-under-ai-act}},
  note = {Official EU Commission guideline document.},
}

@misc{bartzanthropic2025,
  title = {Bartz v.\ {Anthropic} {PBC}, No.\ 3:24-cv-05417-{WHA}},
  year = {2025},
  howpublished = {U.S.\ District Court for the Northern District of California, Order on Motion for Summary Judgment (June 23, 2025), Alsup, J. Court docket: \url{https://www.courtlistener.com/docket/69058235/bartz-v-anthropic-pbc/}},
}

@misc{oecd2025governing,
  author = {{OECD}},
  title = {Governing with Artificial Intelligence: The State of Play and Way Forward in Core Government Functions},
  year = {2025},
  howpublished = {\url{https://www.oecd.org/en/publications/governing-with-artificial-intelligence_795de142-en/full-report.html}},
  note = {Official OECD Public Governance Committee report, 18 September 2025.},
}

@article{nannini2026agentlaw,
  title={AI Agents Under EU Law},
  author={Nannini, Luca and Smith, Adam Leon and Maggini, Michele Joshua and Panai, Enrico and Feliciano, Sandra and Tiulkanov, Aleksandr and Maran, Elena and Gealy, James and Bisconti, Piercosma},
  journal={arXiv preprint arXiv:2604.04604},
  year={2026}
}

@article{cemri2025massfail,
  title={Why do multi-agent llm systems fail?},
  author={Cemri, Mert and Pan, Melissa Z and Yang, Shuyi and Agrawal, Lakshya A and Chopra, Bhavya and Tiwari, Rishabh and Keutzer, Kurt and Parameswaran, Aditya and Klein, Dan and Ramchandran, Kannan and others},
  journal={arXiv preprint arXiv:2503.13657},
  year={2025}
}

@article{pathak2025silentmultiagent,
  title={Detecting Silent Failures in Multi-Agentic AI Trajectories},
  author={Pathak, Divya and Kumar, Harshit and Roy, Anuska and George, Felix and Verma, Mudit and Moogi, Pratibha},
  journal={arXiv preprint arXiv:2511.04032},
  year={2025}
}

@article{yao2024taubench,
  title={$\tau$-bench: A Benchmark for Tool-Agent-User Interaction in Real-World Domains},
  author={Yao, Shunyu and Shinn, Noah and Razavi, Pedram and Narasimhan, Karthik},
  journal={arXiv preprint arXiv:2406.12045},
  year={2024}
}

@article{kapoor2024agentsthatmatter,
  title={Ai agents that matter},
  author={Kapoor, Sayash and Stroebl, Benedikt and Siegel, Zachary S and Nadgir, Nitya and Narayanan, Arvind},
  journal={arXiv preprint arXiv:2407.01502},
  year={2024}
}

@article{madaan2023selfrefine,
  title={Self-refine: Iterative refinement with self-feedback},
  author={Madaan, Aman and Tandon, Niket and Gupta, Prakhar and Hallinan, Skyler and Gao, Luyu and Wiegreffe, Sarah and Alon, Uri and Dziri, Nouha and Prabhumoye, Shrimai and Yang, Yiming and others},
  journal={Advances in neural information processing systems},
  volume={36},
  pages={46534--46594},
  year={2023}
}

@article{packer2024memgpt,
  title={MemGPT: towards LLMs as operating systems.},
  author={Packer, Charles and Fang, Vivian and Patil, Shishir\_G and Lin, Kevin and Wooders, Sarah and Gonzalez, Joseph\_E},
  year={2023},
  publisher={ArXiv}
}

@article{chhikara2025mem0,
  title={Mem0: Building production-ready ai agents with scalable long-term memory},
  author={Chhikara, Prateek and Khant, Dev and Aryan, Saket and Singh, Taranjeet and Yadav, Deshraj},
  journal={arXiv preprint arXiv:2504.19413},
  year={2025}
}

@article{xu2025amem,
  title={A-mem: Agentic memory for llm agents},
  author={Xu, Wujiang and Liang, Zujie and Mei, Kai and Gao, Hang and Tan, Juntao and Zhang, Yongfeng},
  journal={arXiv preprint arXiv:2502.12110},
  year={2025}
}

@article{wang2024agentworkflowmemory,
  title={Agent workflow memory},
  author={Wang, Zora Zhiruo and Mao, Jiayuan and Fried, Daniel and Neubig, Graham},
  journal={arXiv preprint arXiv:2409.07429},
  year={2024}
}

@article{shinn2023reflexion,
  title={Reflexion: Language agents with verbal reinforcement learning},
  author={Shinn, Noah and Cassano, Federico and Gopinath, Ashwin and Narasimhan, Karthik and Yao, Shunyu},
  journal={Advances in neural information processing systems},
  volume={36},
  pages={8634--8652},
  year={2023}
}

@article{zhang2024memorysurvey,
  title={A survey on the memory mechanism of large language model-based agents},
  author={Zhang, Zeyu and Dai, Quanyu and Bo, Xiaohe and Ma, Chen and Li, Rui and Chen, Xu and Zhu, Jieming and Dong, Zhenhua and Wen, Ji-Rong},
  journal={ACM Transactions on Information Systems},
  volume={43},
  number={6},
  pages={1--47},
  year={2025},
  publisher={ACM New York, NY}
}

@article{huang2026rethinkingmemory,
  title={Rethinking Memory Mechanisms of Foundation Agents in the Second Half},
  author={Huang, Wei-Chieh and Zhang, Weizhi and Liang, Yueqing and Bei, Yuanchen and Chen, Yankai and Feng, Tao and Pan, Xinyu and Tan, Zhen and Wang, Yu and Wei, Tianxin and others},
  journal={arXiv preprint arXiv:2602.06052},
  year={2026}
}

@techreport{dellacqua2025cybernetic,
  title={The cybernetic teammate: A field experiment on generative AI reshaping teamwork and expertise},
  author={Dell'Acqua, Fabrizio and Ayoubi, Charles and Lifshitz, Hila and Sadun, Raffaella and Mollick, Ethan and Mollick, Lilach and Han, Yi and Goldman, Jeff and Nair, Hari and Taub, Stewart and others},
  year={2025},
  institution={National Bureau of Economic Research}
}

@article{stray2025copiloteval,
  title={Developer Productivity With and Without GitHub Copilot: A Longitudinal Mixed-Methods Case Study},
  author={Stray, Viktoria and Brandtz{\ae}g, Elias Goldmann and Wivestad, Viggo Tellefsen and Barbala, Astri and Moe, Nils Brede},
  journal={arXiv preprint arXiv:2509.20353},
  year={2025}
}

@article{xiao2025aiteamwork,
  title={AI Hasn't Fixed Teamwork, But It Shifted Collaborative Culture: A Longitudinal Study in a Project-Based Software Development Organization (2023-2025)},
  author={Xiao, Qing and Hu, Xinlan Emily and Whiting, Mark E and Karunakaran, Arvind and Shen, Hong and Cao, Hancheng},
  journal={arXiv preprint arXiv:2509.10956},
  year={2025}
}

@article{wang2023voyager,
  title={Voyager: An open-ended embodied agent with large language models},
  author={Wang, Guanzhi and Xie, Yuqi and Jiang, Yunfan and Mandlekar, Ajay and Xiao, Chaowei and Zhu, Yuke and Fan, Linxi and Anandkumar, Anima},
  journal={arXiv preprint arXiv:2305.16291},
  year={2023}
}

@book{mollick2024cointelligence,
  title={Co-intelligence: Living and working with AI},
  author={Mollick, Ethan},
  year={2024},
  publisher={Penguin}
}

@misc{karpathy2023llmos,
  author = {Karpathy, Andrej},
  title = {[1hr Talk] Intro to Large Language Models},
  year = {2023},
  howpublished = {YouTube talk, \url{https://www.youtube.com/watch?v=zjkBMFhNj_g}},
  note = {November 2023; popularizes the LLM-as-OS framing.},
}

@article{khattab2024dspy,
  title={Dspy: Compiling declarative language model calls into self-improving pipelines},
  author={Khattab, Omar and Singhvi, Arnav and Maheshwari, Paridhi and Zhang, Zhiyuan and Santhanam, Keshav and Vardhamanan, Sri and Haq, Saiful and Sharma, Ashutosh and Joshi, Thomas T and Moazam, Hanna and others},
  journal={arXiv preprint arXiv:2310.03714},
  year={2023}
}

@inproceedings{liu2025innerthoughts,
  title={Proactive conversational agents with inner thoughts},
  author={Liu, Xingyu Bruce and Fang, Shitao and Shi, Weiyan and Wu, Chien-Sheng and Igarashi, Takeo and Chen, Xiang'Anthony'},
  booktitle={Proceedings of the 2025 CHI Conference on Human Factors in Computing Systems},
  pages={1--19},
  year={2025}
}

@inproceedings{pu2025codellaborator,
  title={Assistance or disruption? exploring and evaluating the design and trade-offs of proactive ai programming support},
  author={Pu, Kevin and Lazaro, Daniel and Arawjo, Ian and Xia, Haijun and Xiao, Ziang and Grossman, Tovi and Chen, Yan},
  booktitle={Proceedings of the 2025 CHI conference on human factors in computing systems},
  pages={1--21},
  year={2025}
}

@inproceedings{lee2025sensibleagent,
  title={Sensible agent: A framework for unobtrusive interaction with proactive ar agents},
  author={Lee, Geonsun and Xia, Min and Numan, Nels and Qian, Xun and Li, David and Chen, Yanhe and Kulshrestha, Achin and Chatterjee, Ishan and Zhang, Yinda and Manocha, Dinesh and others},
  booktitle={Proceedings of the 38th Annual ACM Symposium on User Interface Software and Technology},
  pages={1--22},
  year={2025}
}

@article{pasternak2025probe,
  title={Beyond reactivity: Measuring proactive problem solving in llm agents},
  author={Pasternak, Gil and Rajagopal, Dheeraj and White, Julia and Atreja, Dhruv and Thomas, Matthew and Hurn-Maloney, George and Lewis, Ash},
  journal={arXiv preprint arXiv:2510.19771},
  year={2025}
}

@article{sun2025userville,
  title={Training proactive and personalized llm agents},
  author={Sun, Weiwei and Zhou, Xuhui and Du, Weihua and Wang, Xingyao and Welleck, Sean and Neubig, Graham and Sap, Maarten and Yang, Yiming},
  journal={arXiv preprint arXiv:2511.02208},
  year={2025}
}

@article{deng2025proactivesurvey,
  title={Proactive conversational ai: A comprehensive survey of advancements and opportunities},
  author={Deng, Yang and Liao, Lizi and Lei, Wenqiang and Yang, Grace Hui and Lam, Wai and Chua, Tat-Seng},
  journal={ACM Transactions on Information Systems},
  volume={43},
  number={3},
  pages={1--45},
  year={2025},
  publisher={ACM New York, NY}
}

@article{brohan2023rt2,
  title={RT-2: Vision-Language-Action Models Transfer Web Knowledge to Robotic Control},
  author={Brohan, Anthony and Brown, Noah and Carbajal, Justice and Chebotar, Yevgen and Chen, Xi and Choromanski, Krzysztof and Ding, Tianli and Driess, Danny and Dubey, Avinava and Finn, Chelsea and others},
  journal={arXiv preprint arXiv:2307.15818},
  year={2023}
}

@article{black2024pi0,
  title={$\pi_0$: A Vision-Language-Action Flow Model for General Robot Control},
  author={Black, Kevin and Brown, Noah and Driess, Danny and Esmail, Adnan and Equi, Michael and Finn, Chelsea and Fusai, Niccolo and Groom, Lachy and Hausman, Karol and Ichter, Brian and others},
  journal={arXiv preprint arXiv:2410.24164},
  year={2024}
}

@article{ahn2022saycan,
  title={Do as i can, not as i say: Grounding language in robotic affordances},
  author={Ahn, Michael and Brohan, Anthony and Brown, Noah and Chebotar, Yevgen and Cortes, Omar and David, Byron and Finn, Chelsea and Fu, Chuyuan and Gopalakrishnan, Keerthana and Hausman, Karol and others},
  journal={arXiv preprint arXiv:2204.01691},
  year={2022}
}

@misc{figure2025helix,
  author = {{Figure AI}},
  title = {Helix: A Vision-Language-Action Model for Generalist Humanoid Control},
  year = {2025},
  howpublished = {\url{https://www.figure.ai/news/helix}},
  note = {Figure AI technical blog.},
}

@article{nvidia2025gr00t,
  title={Gr00t n1: An open foundation model for generalist humanoid robots},
  author={Bjorck, Johan and Casta{\~n}eda, Fernando and Cherniadev, Nikita and Da, Xingye and Ding, Runyu and Fan, Linxi and Fang, Yu and Fox, Dieter and Hu, Fengyuan and Huang, Spencer and others},
  journal={arXiv preprint arXiv:2503.14734},
  year={2025}
}

@inproceedings{hong2024metagpt,
  title={MetaGPT: Meta programming for a multi-agent collaborative framework},
  author={Hong, Sirui and Zhuge, Mingchen and Chen, Jonathan and Zheng, Xiawu and Cheng, Yuheng and Wang, Jinlin and Zhang, Ceyao and Wang, Zili and Yau, Steven Ka Shing and Lin, Zijuan and others},
  booktitle={The twelfth international conference on learning representations},
  year={2023}
}

@article{li2023camel,
  title={Camel: Communicative agents for" mind" exploration of large language model society},
  author={Li, Guohao and Hammoud, Hasan and Itani, Hani and Khizbullin, Dmitrii and Ghanem, Bernard},
  journal={Advances in neural information processing systems},
  volume={36},
  pages={51991--52008},
  year={2023}
}

@inproceedings{chen2024agentverse,
  title={Agentverse: Facilitating multi-agent collaboration and exploring emergent behaviors},
  author={Chen, Weize and Su, Yusheng and Zuo, Jingwei and Yang, Cheng and Yuan, Chenfei and Chan, Chi-Min and Yu, Heyang and Lu, Yaxi and Hung, Yi-Hsin and Qian, Chen and others},
  booktitle={The Twelfth International Conference on Learning Representations},
  year={2023}
}

@inproceedings{qian2024chatdev,
  title={Chatdev: Communicative agents for software development},
  author={Qian, Chen and Liu, Wei and Liu, Hongzhang and Chen, Nuo and Dang, Yufan and Li, Jiahao and Yang, Cheng and Chen, Weize and Su, Yusheng and Cong, Xin and others},
  booktitle={Proceedings of the 62nd annual meeting of the association for computational linguistics (volume 1: Long papers)},
  pages={15174--15186},
  year={2024}
}

@inproceedings{du2023debate,
  title={Improving factuality and reasoning in language models through multiagent debate},
  author={Du, Yilun and Li, Shuang and Torralba, Antonio and Tenenbaum, Joshua B and Mordatch, Igor},
  booktitle={Forty-first international conference on machine learning},
  year={2024}
}

@inproceedings{liang2024divergent,
  title={Encouraging divergent thinking in large language models through multi-agent debate},
  author={Liang, Tian and He, Zhiwei and Jiao, Wenxiang and Wang, Xing and Wang, Yan and Wang, Rui and Yang, Yujiu and Shi, Shuming and Tu, Zhaopeng},
  booktitle={Proceedings of the 2024 conference on empirical methods in natural language processing},
  pages={17889--17904},
  year={2024}
}

@inproceedings{zhuge2024gptswarm,
  title={Gptswarm: Language agents as optimizable graphs},
  author={Zhuge, Mingchen and Wang, Wenyi and Kirsch, Louis and Faccio, Francesco and Khizbullin, Dmitrii and Schmidhuber, J{\"u}rgen},
  booktitle={Forty-first International Conference on Machine Learning},
  year={2024}
}

@article{guo2024massurvey,
  title={Large language model based multi-agents: A survey of progress and challenges},
  author={Guo, Taicheng and Chen, Xiuying and Wang, Yaqi and Chang, Ruidi and Pei, Shichao and Chawla, Nitesh V and Wiest, Olaf and Zhang, Xiangliang},
  journal={arXiv preprint arXiv:2402.01680},
  year={2024}
}

@article{lu2024aiscientist,
  title={The ai scientist: Towards fully automated open-ended scientific discovery},
  author={Lu, Chris and Lu, Cong and Lange, Robert Tjarko and Foerster, Jakob and Clune, Jeff and Ha, David},
  journal={arXiv preprint arXiv:2408.06292},
  year={2024}
}

@inproceedings{beel2025evalsakana,
  title={Evaluating Sakana's AI Scientist: Bold Claims, Mixed Results, and a Promising Future?},
  author={Beel, Joeran and Kan, Min-Yen and Baumgart, Moritz},
  booktitle={ACM SIGIR Forum},
  volume={59},
  number={1},
  pages={1--20},
  year={2025},
  organization={ACM New York, NY, USA}
}

@article{gottweis2025coscientist,
  title={Towards an AI co-scientist},
  author={Gottweis, Juraj and Weng, Wei-Hung and Daryin, Alexander and Tu, Tao and Palepu, Anil and Sirkovic, Petar and Myaskovsky, Artiom and Weissenberger, Felix and Rong, Keran and Tanno, Ryutaro and others},
  journal={arXiv preprint arXiv:2502.18864},
  year={2025}
}

@article{novikov2025alphaevolve,
  title={Alphaevolve: A coding agent for scientific and algorithmic discovery},
  author={Novikov, Alexander and V{\~u}, Ng{\^a}n and Eisenberger, Marvin and Dupont, Emilien and Huang, Po-Sen and Wagner, Adam Zsolt and Shirobokov, Sergey and Kozlovskii, Borislav and Ruiz, Francisco JR and Mehrabian, Abbas and others},
  journal={arXiv preprint arXiv:2506.13131},
  year={2025}
}

@inproceedings{kwa2025metrtimehorizon,
  title={Measuring AI Ability to Complete Long Software Tasks},
  author={Kwa, Thomas and West, Ben and Becker, Joel and Deng, Amy and Garcia, Katharyn and Hasin, Max and Jawhar, Sami and Kinniment, Megan and Rush, Nate and Von Arx, Sydney and others},
  booktitle={The Thirty-ninth Annual Conference on Neural Information Processing Systems}
}

@article{barke2023groundedcopilot,
  title={Grounded copilot: How programmers interact with code-generating models},
  author={Barke, Shraddha and James, Michael B and Polikarpova, Nadia},
  journal={Proceedings of the ACM on Programming Languages},
  volume={7},
  number={OOPSLA1},
  pages={85--111},
  year={2023},
  publisher={ACM New York, NY, USA}
}

@inproceedings{perry2023insecurecode,
  title={Do users write more insecure code with ai assistants?},
  author={Perry, Neil and Srivastava, Megha and Kumar, Deepak and Boneh, Dan},
  booktitle={Proceedings of the 2023 ACM SIGSAC conference on computer and communications security},
  pages={2785--2799},
  year={2023}
}

@misc{cursor2026cloudagents,
  author       = {Ma, Josh},
  title        = {What we've learned building cloud agents},
  year         = {2026},
  howpublished = {Cursor Blog, \url{https://cursor.com/blog/cloud-agent-lessons}},
  note         = {Accessed May 2026},
}

@misc{langchain2026manageddeepagents,
  author       = {{LangChain}},
  title        = {Managed Deep Agents: the fastest way to ship a production deep agent},
  year         = {2026},
  howpublished = {LangChain Blog, \url{https://www.langchain.com/blog/introducing-managed-deep-agents}},
  note         = {Accessed May 2026},
}

@misc{langchain2026contexthub,
  author       = {{LangChain}},
  title        = {Introducing {LangSmith} Context Hub},
  year         = {2026},
  howpublished = {LangChain Blog, \url{https://www.langchain.com/blog/introducing-context-hub}},
  note         = {Accessed May 2026},
}

@misc{openai2026windowssandbox,
  author       = {{OpenAI}},
  title        = {Building a safe, effective sandbox to enable {Codex} on {Windows}},
  year         = {2026},
  howpublished = {OpenAI Blog, \url{https://openai.com/index/building-codex-windows-sandbox/}},
  note         = {Accessed May 2026},
}

@misc{openai2026agentimprovementloop,
  author       = {Pasfield, Wesley},
  title        = {Build an Agent Improvement Loop with Traces, Evals, and {Codex}},
  year         = {2026},
  howpublished = {OpenAI Cookbook, \url{https://developers.openai.com/cookbook/examples/agents_sdk/agent_improvement_loop}},
  note         = {Accessed May 2026},
}

@techreport{nsa2026mcpsecurity,
  author      = {{National Security Agency}},
  title       = {Model Context Protocol ({MCP}): Security Design Considerations for {AI}-Driven Automation},
  institution = {National Security Agency},
  number      = {PP-26-1834},
  year        = {2026},
  month       = {May},
  url         = {https://www.nsa.gov/Portals/75/documents/Cybersecurity/CSI_MCP_SECURITY.pdf},
}

@article{pan2026agentfirstapi,
  title={Agent-First Tool API: A Semantic Interface Paradigm for Enterprise AI Agent Systems},
  author={Pan, Kai},
  journal={arXiv preprint arXiv:2605.10555},
  year={2026}
}

@article{ning2026codeharness,
  title={Code as Agent Harness},
  author={Ning, Xuying and Tieu, Katherine and Fu, Dongqi and Wei, Tianxin and Li, Zihao and Bei, Yuanchen and Zou, Jiaru and Ai, Mengting and Liu, Zhining and Li, Ting-Wei and others},
  journal={arXiv preprint arXiv:2605.18747},
  year={2026}
}

@article{xu2026memgym,
  title={MemGym: a Long-Horizon Memory Environment for LLM Agents},
  author={Xu, Wujiang and Wang, Yu and Mei, Kai and Liang, Kaiqu and Wang, Zhenting and Jin, Mingyu and Zhang, Han and Zhang, Shi-Xiong and Hua, Wenyue and Sahu, Sambit and others},
  journal={arXiv preprint arXiv:2605.20833},
  year={2026}
}

@article{bisconti2026boilingfrog,
  title={Boiling the Frog: A Multi-Turn Benchmark for Agentic Safety},
  author={Bisconti, Piercosma and Prandi, Matteo and Pierucci, Federico and Sartore, Federico and Panai, Enrico and Caroli, Laura and Zhu, Yue and Smith, Adam Leon and Nannini, Luca and Galisai, Marcello and others},
  journal={arXiv preprint arXiv:2605.22643},
  year={2026}
}

@inproceedings{he2026steer,
  title={How to Steer Your Multi-Agent System: Human-LLM Collaborative Planning},
  author={He, Zeyu and Kim, Hannah and Zhang, Dan and Hruschka, Estevam},
  booktitle={Proceedings of the ACM Conference on AI and Agentic Systems},
  pages={330--347},
  year={2026}
}

@article{cai2026pushagent,
  title         = {Push Your Agent: Measuring and Enforcing Quantitative Goal Persistence in Long-Horizon {LLM} Agents},
  author        = {Yuandao Cai and Yuzhang Zhu and Liyou Gao and Wensheng Tang and Shengchao Qin},
  year          = {2026},
  eprint        = {2605.23574},
  archivePrefix = {arXiv},
  primaryClass  = {cs.LG},
  url           = {https://arxiv.org/abs/2605.23574},
}

@misc{vilalab2026designnotes,
  author       = {{VILA Lab}},
  title        = {Agent Systems Design Space Source Notes},
  year         = {2026},
  howpublished = {\url{https://github.com/VILA-Lab/Dive-into-Claude-Code/blob/main/docs/agent-design-space-source-notes_zh.md}},
  note         = {Companion source notes, updated May 2026. Accessed May 2026},
}

@misc{anthropic2026opus48,
  author       = {{Anthropic}},
  title        = {Introducing Claude Opus 4.8},
  year         = {2026},
  howpublished = {Anthropic News, \url{https://www.anthropic.com/news/claude-opus-4-8}},
  note         = {Accessed May 2026},
}

@misc{claudecode2026workflows,
  author       = {{Anthropic}},
  title        = {Orchestrate Subagents at Scale with Dynamic Workflows},
  year         = {2026},
  howpublished = {Claude Code Documentation, \url{https://code.claude.com/docs/en/workflows}},
  note         = {Research preview; requires Claude Code v2.1.154 or later. Accessed May 2026},
}

@article{ding2026wildclawbench,
  title={WildClawBench: A Benchmark for Real-World, Long-Horizon Agent Evaluation},
  author={Ding, Shuangrui and Dai, Xuanlang and Xing, Long and Ding, Shengyuan and Liu, Ziyu and JingYi, Yang and Yang, Penghui and Zhang, Zhixiong and Wei, Xilin and Fang, Xinyu and others},
  journal={arXiv preprint arXiv:2605.10912},
  year={2026}
}

@article{trivedi2026cleanliness,
  title={Does Code Cleanliness Affect Coding Agents? A Controlled Minimal-Pair Study},
  author={Trivedi, Priyansh and Schmitt, Olivier},
  journal={arXiv preprint arXiv:2605.20049},
  year={2026}
}

@article{cim2026parallelcompaction,
  title={Parallel Context Compaction for Long-Horizon LLM Agent Serving},
  author={Cim, Musa and Topcu, Burak and Das, Chita and Kandemir, Mahmut Taylan},
  journal={arXiv preprint arXiv:2605.23296},
  year={2026}
}

@misc{cybersecnews2026sandbox,
  author       = {{Cyber Security News}},
  title        = {Claude Code's Network Sandbox Vulnerability Exposes User Credentials and Source Code},
  year         = {2026},
  howpublished = {\url{https://cybersecuritynews.com/claude-codes-network-sandbox-vulnerability/}},
  note         = {Independent security research; affected range v2.0.24--v2.1.89; no CVE assigned. Accessed May 2026},
}

@misc{hermesagent2026,
  title        = {{Hermes} Agent},
  author       = {{Nous Research}},
  howpublished = {\url{https://github.com/NousResearch/hermes-agent}},
  note         = {Analyzed snapshot commit f1f42a7b9; v0.13.0 release tag v2026.5.7; accessed May 2026},
  year         = {2026},
}

@misc{clawcodex2026,
  title        = {ClawCodex},
  author       = {{agentforce314}},
  howpublished = {\url{https://github.com/agentforce314/clawcodex}},
  note         = {Accessed May 2026},
  year         = {2026},
}

@misc{clawcoderust2026,
  title        = {Claw Code},
  author       = {{ultraworkers}},
  howpublished = {\url{https://github.com/ultraworkers/claw-code}},
  note         = {Accessed May 2026},
  year         = {2026},
}

@misc{claudecodeworking2026,
  title        = {claude-code-working},
  author       = {{777genius}},
  howpublished = {\url{https://github.com/777genius/claude-code-working}},
  note         = {Accessed May 2026},
  year         = {2026},
}

@misc{tlabclaude2026,
  title        = {Claude Code Source - Buildable Research Fork},
  author       = {{T-Lab-CUHKSZ}},
  howpublished = {\url{https://github.com/T-Lab-CUHKSZ/claude-code}},
  note         = {Accessed May 2026},
  year         = {2026},
}

@misc{openclaudecode2026,
  title        = {Open Claude Code},
  author       = {{ruvnet}},
  howpublished = {\url{https://github.com/ruvnet/open-claude-code}},
  note         = {Accessed May 2026},
  year         = {2026},
}

@misc{anthropic2026fableaccess,
  author       = {{Anthropic}},
  title        = {Statement on the {US} Government Directive to Suspend Access to Fable 5 and Mythos 5},
  year         = {2026},
  howpublished = {Anthropic News, \url{https://www.anthropic.com/news/fable-mythos-access}},
  note         = {Accessed June 2026},
}

@misc{openai2026harness,
  author       = {{OpenAI}},
  title        = {Harness Engineering: Leveraging {Codex} in an Agent-First World},
  year         = {2026},
  howpublished = {OpenAI, \url{https://openai.com/index/harness-engineering/}},
  note         = {Accessed June 2026},
}

@misc{google2026antigravity,
  author       = {{Google}},
  title        = {Build with {Google} Antigravity, Our New Agentic Development Platform},
  year         = {2025},
  howpublished = {Google Developers Blog, \url{https://developers.googleblog.com/build-with-google-antigravity-our-new-agentic-development-platform/}},
  note         = {Accessed June 2026},
}

@misc{anthropic2026fable,
  author       = {{Anthropic}},
  title        = {Claude Fable 5 and Claude Mythos 5},
  year         = {2026},
  howpublished = {Anthropic News, \url{https://www.anthropic.com/news/claude-fable-5-mythos-5}},
  note         = {Accessed June 2026},
}

@misc{endorlabs2026fable,
  author       = {Compagna, Luca},
  title        = {Claude Fable 5, Take Two: Same Model, Different Harness, and a Very Different Result},
  year         = {2026},
  howpublished = {Endor Labs, \url{https://www.endorlabs.com/learn/claude-fable-5-take-two-same-model-different-harness-and-a-very-different-result}},
  note         = {Accessed June 2026},
}

@misc{karpathy2026sequoia,
  author       = {Karpathy, Andrej},
  title        = {Sequoia Ascent 2026 Summary},
  year         = {2026},
  howpublished = {\url{https://karpathy.bearblog.dev/sequoia-ascent-2026/}},
  note         = {Accessed June 2026},
}

@misc{osmani2026loop,
  author       = {Addy Osmani},
  title        = {Loop Engineering},
  year         = {2026},
  howpublished = {\url{https://addyosmani.com/blog/loop-engineering/}}
  }

@misc{anthropic2026desktop,
  author = {{Anthropic}},
  title = {Desktop Application},
  year = {2026},
  howpublished = {\url{https://code.claude.com/docs/en/desktop}},
}

@misc{anthropic2026projects,
  author = {{Anthropic}},
  title = {{Projects} Redesigned: From Folder to Conversation},
  year = {2026},
  howpublished = {\url{https://claude.com/blog/projects-redesigned}},
}

@misc{anthropic2026routines,
  author = {{Anthropic}},
  title = {Automate Work with Routines},
  year = {2026},
  howpublished = {\url{https://code.claude.com/docs/en/routines}},
}

@misc{anthropic2026monitoring,
  author = {{Anthropic}},
  title = {Monitoring},
  year = {2026},
  howpublished = {\url{https://code.claude.com/docs/en/monitoring-usage}},
}

@article{feng2026graph,
  title = {Graph Engineering in the Era of LLM Agents: From Individual Intelligence to System Intelligence},
  author = {Feng, Yuyuan and Xiang, Zhishang and Yang, Chaobin and Ma, Qichao and Chen, Zerui and Zhang, Yujing and Huang, Ke and Wu, Chuanjie and Liu, Zhaoxu and Wang, Yili and others},
  journal = {arXiv preprint arXiv:2608.21156},
  year = {2026}
}

@article{li2026loopsbench,
  title = {LoopsBench: From Harness Engineering to Loop Engineering in Coding Agent Evaluation},
  author = {Li, Han and Fang, Zhemin and Feng, Rili and Zhao, Yingqi and Liu, Jiaheng and Gao, Pengfei and Ye, He and Lin, Dayi and Lin, Qingwei and Rajmohan, Saravan and others},
  journal = {arXiv preprint arXiv:2608.00267},
  year = {2026}
}

@article{wang2026looparena,
  title = {LoopArena: Benchmarking Models as Runtime Controllers for Loop Engineering},
  author = {Wang, Yi and Zhang, Haopeng and Huang, Chengxiang and Dai, Rui and Liu, Kaikui and Koniusz, Piotr and Chu, Xiangxiang},
  journal = {arXiv preprint arXiv:2608.28281},
  year = {2026}
}

@article{khan2026resume,
  title = {Resume Means Resume: A Machine-Checked Conformance Contract for Checkpoint, Interrupt, and Resume Semantics in Workflow Persistence Layers},
  author = {Khan, Sajjad},
  journal = {arXiv preprint arXiv:2608.03836},
  year = {2026}
}

@misc{temporal2026deepagents,
  author = {{Temporal Technologies}},
  title = {{Deep Agents} Integration},
  year = {2026},
  howpublished = {\url{https://docs.temporal.io/develop/python/integrations/deepagents}}
}

@article{du2026living,
  title = {Living-harness is an interactive-agent evolver},
  author = {Du, Yuetian and Wang, Yucheng and Xu, He and Xu, Jiexu and Tan, Shanwen and Zhao, Bing and Yang, Boyu and Xu, Zhijie and Kong, Ming and Wei, Hu and others},
  journal = {arXiv preprint arXiv:2607.26598},
  year = {2026}
}

@article{kravchenko2026appa,
  title = {APPA: Recoverable Information-Flow Control for Real-World LLM Agents},
  author = {Kravchenko, Arseny and Liventsev, Vadim and Konstantinov, Innokentii and Iskhakov, Ildar and Kukuy, Matvey},
  journal = {arXiv preprint arXiv:2607.24625},
  year = {2026}
}

@misc{mcp2026revision,
  author = {{Model Context Protocol}},
  title = {{Model Context Protocol}: Key Changes},
  year = {2026},
  howpublished = {\url{https://modelcontextprotocol.io/specification/2026-07-28/changelog}}
}

@misc{mcp2026mrtr,
  author = {{Model Context Protocol}},
  title = {Multi Round-Trip Requests},
  year = {2026},
  howpublished = {\url{https://modelcontextprotocol.io/specification/2026-07-28/basic/patterns/mrtr}}
}

@misc{lourens2026agenthost,
  author = {Lourens, Rob and Peet, Connor and Murtaugh, Brigit},
  title = {Introducing the {Agent Host} for Persistent, Portable Agent Sessions},
  year = {2026},
  howpublished = {\url{https://code.visualstudio.com/blogs/2026/08/26/agent-host-architecture}}
}

@misc{lin2026agentswarms,
  author = {Lin, Wilson},
  title = {Agent Swarms and the New Model Economics},
  year = {2026},
  howpublished = {\url{https://cursor.com/blog/agent-swarm-model-economics}}
}

@article{zhou2026self,
  title = {Self-Evolving Coding Agents},
  author = {Zhou, Hao and Hu, Haichuan and Luo, Tianyu and Shang, Ye and Fang, Chunrong and Chen, Zhenyu and Xiao, Liang and Zhang, Quanjun},
  journal = {arXiv preprint arXiv:2608.03392},
  year = {2026}
}

@article{yang2026better,
  title = {Better Harnesses, Smaller Models: Building 90\% Cheaper Agents via Automated Harness Adaptation},
  author = {Yang, Chenyang and Zhao, Xinran and Wu, Tongshuang and K{\"a}stner, Christian},
  journal = {arXiv preprint arXiv:2607.08938},
  year = {2026}
}

@article{yu2026co,
  title = {Co-Evolving Harnesses and Models: On-Policy Correction Helps Weaker Models Catch Up Where Imitation Fails},
  author = {Yu, Zhou and Bi, Bin and Pentyala, Shiva Kumar and Mehrotra, Shubham and Chaudhuri, Sougata and Bhagavath, Shilpa and Chen, Zeyuan and Xu, Ran and Mui, Phil and Zhu, James and others},
  journal = {arXiv preprint arXiv:2609.09134},
  year = {2026}
}

@article{yang2026learning,
  title = {Learning Globally Reusable Skills for Coding Agents},
  author = {Yang, Chen and Tian, Jiashuo and Wang, Ziqi and Liu, Xinyin and Ye, Meiru and Chen, Junjie},
  journal = {arXiv preprint arXiv:2608.06153},
  year = {2026}
}

@article{wang2026rethinking,
  title = {Rethinking the evaluation of harness evolution for agents},
  author = {Wang, Yike and Zhu, Huaisheng and Hu, Zhengyu and Yuan, Yige and Chen, Zhengyu and Senthil, Shakti and Hajishirzi, Hannaneh and Tsvetkov, Yulia and Dasigi, Pradeep and Xiao, Teng},
  journal = {arXiv preprint arXiv:2607.12227},
  year = {2026}
}

@article{le2026does,
  title = {What Does Multi-Harness RL Learn? Credit Assignment and Portability in Coding Agents},
  author = {Le, Chenqian and Cheng, Jiayi and He, Qijia and Li, Runhao and Li, Yinghao and Chen, Xupeng},
  journal = {arXiv preprint arXiv:2609.04518},
  year = {2026}
}

@article{chen2026memsecbench,
  title = {MemSecBench: Tracking Agent Memory Poisoning from Persistence to Consequence and Repair},
  author = {Chen, Xuanze and Xie, Xukang and Fu, Wentao and Zhou, Jiajun and Yu, Shanqing and Xuan, Qi},
  journal = {arXiv preprint arXiv:2607.27080},
  year = {2026}
}

@article{hong2026swe,
  title = {SWE Refactor Bench: Can Coding Agents Complete a Long-Horizon, Whole-Repository Stack Migration?},
  author = {Hong, Deyao and Chi, Yizhe and Li, Wenyi and Wang, Xiaoqiu and Gao, Mingju and Yang, Kaisen and He, Bingxiang and Zheng, Youjie and Xiao, Calvin and Na, Qinhuai},
  journal = {arXiv preprint arXiv:2608.23564},
  year = {2026}
}

@misc{anthropic2026fablerestoration,
  author = {{Anthropic}},
  title = {Redeploying {Fable 5}},
  year = {2026},
  howpublished = {\url{https://www.anthropic.com/news/redeploying-fable-5}}
}

@misc{deepseek-harness2026,
  author = {{DeepSeek-AI}},
  title = {{DeepSeek Harness}: Everything Is a Plugin},
  year = {2026},
  howpublished = {\url{https://github.com/deepseek-ai/deepseek-harness}}
}

@misc{openai2026symphony,
  author = {{OpenAI}},
  title = {{Symphony} Service Specification},
  year = {2026},
  howpublished = {\url{https://github.com/openai/symphony/blob/main/SPEC.md}}
}

@misc{graphiti2026memory,
  author = {{Zep}},
  title = {{Graphiti}},
  year = {2026},
  howpublished = {\url{https://github.com/getzep/graphiti/blob/c035afb7990b6077331a81e98b04efcfd9bf8184/README.md}}
}

@misc{cognee2026memory,
  author = {{Cognee}},
  title = {{Cognee}},
  year = {2026},
  howpublished = {\url{https://github.com/topoteretes/cognee/blob/c0d18c80e24b7b78918e7642c03f6f128fdd2aee/README.md}}
}

@misc{a2a2026specification,
  author = {{A2A Project}},
  title = {{Agent2Agent (A2A)} Protocol Specification},
  year = {2026},
  howpublished = {\url{https://a2a-protocol.org/v1.0.0/specification/}}
}

@misc{langchain2026framework,
  author = {{LangChain}},
  title = {{LangChain} Overview},
  year = {2026},
  howpublished = {\url{https://docs.langchain.com/oss/python/langchain/overview}}
}

@misc{crewai2026framework,
  author = {{CrewAI}},
  title = {{CrewAI}},
  year = {2026},
  howpublished = {\url{https://github.com/crewAIInc/crewAI/blob/main/README.md}}
}

@misc{opencode2026agent,
  author = {{Anomaly}},
  title = {{OpenCode}: Permissions},
  year = {2026},
  howpublished = {\url{https://opencode.ai/docs/permissions/}}
}

@misc{cline2026agent,
  author = {{Cline}},
  title = {Auto Approve \& {YOLO} Mode},
  year = {2026},
  howpublished = {\url{https://docs.cline.bot/features/auto-approve}}
}

@misc{e2b2026sandbox,
  author = {{E2B}},
  title = {{E2B}},
  year = {2026},
  howpublished = {\url{https://github.com/e2b-dev/E2B}}
}

@misc{opensandbox2026environment,
  author = {{OpenSandbox}},
  title = {{OpenSandbox}},
  year = {2026},
  howpublished = {\url{https://github.com/opensandbox-group/OpenSandbox}}
}

\clearpage
\newpage
\beginappendix

\section*{Evidence Base and Methodology}
\label{sec:method}

This appendix records the evidence sources, analytic procedure, and limits of the study.

\subsection*{Evidence Base and Evidence Tiers}

Claims in this paper are grounded at three evidence tiers:

\begin{itemize}[nosep]
  \item \tierA{} \textbf{(product-documented)}: Claims drawn from official Anthropic documentation and engineering publications. These establish product intent but may not reflect internal implementation.
  \item \tierB{} \textbf{(code-verified)}: Claims citing specific files and functions in the extracted TypeScript codebase (v2.1.88, obtained from a publicly available npm package extraction). This is the strongest evidence tier.
  \item \tierC{} \textbf{(reconstructed)}: Claims derived from community analysis, OpenClaw or Hermes Agent structural comparison, or inference from code patterns. These are stated with hedging language.
\end{itemize}

The source corpus comprises approximately 1,884 files totaling roughly 512K lines of TypeScript, taken from the \file{original-source-code/src} directory at commit \href{https://github.com/chauncygu/collection-claude-code-source-code/tree/53faa8b2162fc7a4dce2e79b55b7101bb8f9e2bf/original-source-code/src}{\code{53faa8b216}}.
OpenClaw and Hermes Agent are used as comparative reference points rather than as ground-truth standards. The analyzed snapshots are commits \href{https://github.com/openclaw/openclaw/tree/462d8e3bc06a654037e4938a4f885a70eb19dbf2}{\code{462d8e3bc0}} and \href{https://github.com/NousResearch/hermes-agent/tree/f1f42a7b9ffa83749f725cfdf76b121779859914}{\code{f1f42a7b9f}}, respectively.

\subsection*{Design-Space Analytic Procedure}

Design questions were identified by examining each subsystem for recurring choice points where alternative designs exist in other production agents.
Claude Code's answers to each question were traced through specific source files and function implementations (\tierB{} evidence).
The five-value framework (human decision authority, safety, security, and privacy, reliable execution, capability amplification, and contextual adaptability) was identified from official documentation and creator statements (\tierA{}), then traced through thirteen design principles to architectural decisions.
Long-term capability preservation is treated separately as a cross-cutting question rather than a design value, because it is not prominently reflected as a design driver in the architecture or in Anthropic's stated values (\Cref{sec:values:lens}).
Token economics serves as a cross-cutting constraint that bounds all five values simultaneously, revealing how individual subsystem choices interact under shared resource pressure.

\subsection*{Limitations}
\label{sec:method:limits}

\begin{itemize}[nosep]
  \item \textbf{Static snapshot.} Analysis reflects one version (v2.1.88). Feature flags (\eg, \code{TRANSCRIPT\_CLASSIFIER}, \code{CONTEXT\_COLLAPSE}) create build-time variability. Different build targets may produce functionally different applications.
  \item \textbf{Reverse-engineering epistemology.} Source code reveals implemented structure, control flow, dependencies, and feature gates. It cannot confirm design intent, enabled production flags, runtime prevalence, or unshipped behavior.
  \item \textbf{Single-system analysis.} Findings describe Claude Code's design space, not the entire design space of coding agents. Generalizations are bounded.
  \item \textbf{Comparison-system snapshots.} The OpenClaw and Hermes Agent analyses each reflect a specific development state and may not represent their current capabilities.
\end{itemize}

\section*{Package Structure}
\label{sec:appendix}

This part maps the main TypeScript package to runtime responsibilities.

\subsection*{Directory-to-Responsibility Map}

\begin{figure}[t]
\centering
\includegraphics[width=\textwidth]{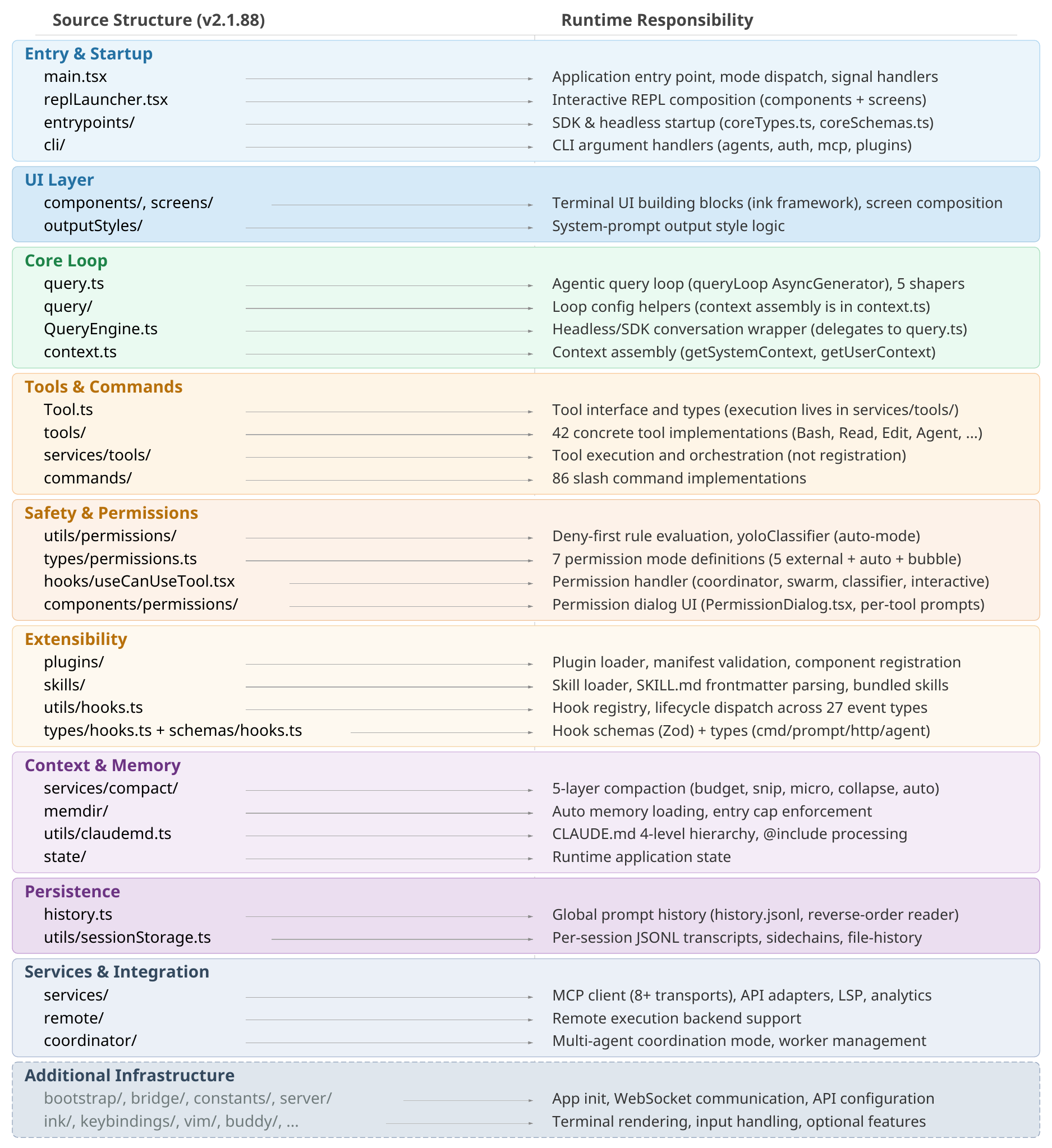}
\caption{Extracted package structure mapped to runtime responsibilities. Left column: TypeScript source directories and key files. Right column: inferred runtime roles. This appendix represents reconstructed analysis (Tier~C evidence), not official Anthropic documentation.}
\label{fig:package}
\end{figure}

The package (\Cref{fig:package}) is organized around a \file{src/} directory. \Cref{tab:key-files} lists the key files that form the main subsystems.

\begin{table}[h]
\centering
\caption{Key files by approximate size and runtime responsibility.}
\label{tab:key-files}
\small
\begin{tabular}{llp{4.5cm}}
\toprule
\textbf{File} & \textbf{Size} & \textbf{Responsibility} \\
\midrule
\file{main.tsx} & 804KB & Entry point, mode dispatch, setup \\
\file{query.ts} & 68KB & Core agent loop, 5 context shapers \\
\file{QueryEngine.ts} & 47KB & SDK/headless conversation wrapper \\
\file{Tool.ts} & 30KB & Tool interface, types, utilities \\
\file{history.ts} & 14KB & Global prompt history \\
\file{mcp/client.ts} & Large & MCP client (8+ transport variants) \\
\file{compact.ts} & Large & Compaction engine \\
\file{AgentTool.tsx} & Large & Agent tool, subagent dispatch \\
\file{runAgent.ts} & Large & Agent lifecycle and coordination \\
\bottomrule
\end{tabular}
\end{table}

The \file{tools/} directory contains approximately 42 subdirectories implementing tools, with the corresponding schema, description, permission requirements, and execution logic. The \file{commands/} directory contains approximately 86 slash command subdirectories.

Key service directories include \file{services/tools/} (StreamingToolExecutor, toolOrchestration, toolExecution), \file{services/compact/} (compaction engine), and \file{services/mcp/} (MCP client and configuration). The permission infrastructure spans \file{utils/permissions/} (rule evaluation, classifier), \file{hooks/useCanUseTool.tsx} (permission handler), \file{types/permissions.ts} (mode definitions), and \file{types/hooks.ts} (event schemas).

A structural quirk: \file{query.ts} (file) and \file{query/} (directory) coexist. The file contains the main query loop. The directory houses helper modules for loop configuration and context assembly.

\subsection*{Conditional Tool Availability}

The \func{getAllBaseTools} function (\file{tools.ts}) constructs different tool sets depending on mode, build, environment, and feature flags (\Cref{tab:tool-availability}). The model may see as few as 3 tools in simple mode (Bash, Read, Edit) or up to 54 tools (19 unconditional plus 35 conditional) in a full internal build with all features enabled.

\begin{table}[t]
\centering
\caption{Conditional tool availability categories.}
\label{tab:tool-availability}
\small
\begin{tabularx}{\columnwidth}{lX}
\toprule
\textbf{Category} & \textbf{Examples} \\
\midrule
Always included & AgentTool, BashTool, FileReadTool, FileEditTool, FileWriteTool, SkillTool, WebFetchTool, WebSearchTool \\
\midrule
Environment & GlobTool/GrepTool (unless embedded), ConfigTool (ant-only), PowerShellTool (Windows) \\
\midrule
Feature flag & TaskCreate/Get/Update/List (\code{todoV2}), EnterWorktreeTool (\code{worktree}), TeamTools (\code{swarms}), ToolSearchTool \\
\midrule
Null-checked & SuggestBackgroundPRTool, WebBrowserTool, RemoteTriggerTool, MonitorTool, SleepTool \\
\bottomrule
\end{tabularx}
\end{table}

\subsection*{Cross-File Dependencies}

The import graph includes the following dependency structure. \code{QueryEngine.ts} delegates to \code{query.ts} for turn execution. \code{query.ts} imports from \file{services/tools/} (StreamingToolExecutor, runTools) and \file{services/compact/} (autoCompact, buildPostCompactMessages). \code{QueryEngine.ts} imports from \file{memdir/} for memory and prompt assembly. The code explicitly avoids circular imports: \file{types/permissions.ts} was extracted to break import cycles, and \func{setCachedClaudeMdContent} in \file{context.ts} avoids a cycle through the permissions/filesystem path.

\section*{Community Reimplementations}
\label{sec:appendix:community}

After the Claude Code TypeScript source became publicly readable, several community projects published independent reimplementations. This appendix lists a few representative examples.

\subsection*{Representative Projects}
\label{sec:appendix:community:list}

\Cref{tab:reimplementations} lists five reimplementations identified from the public ecosystem as of 2026. Each project is a working CLI agent rather than a passive source dump or annotation pass. The set spans four runtime targets (Python, Rust, TypeScript with Bun, JavaScript via npm) and three rebuild methodologies (independent rewrites from observed behavior, fork from a TypeScript snapshot with reconstructed build, and automated decompilation pipelines).

\begin{table}[h]
\centering
\caption{Representative community reimplementations of Claude Code (as of 2026).}
\label{tab:reimplementations}
\small
\begin{tabularx}{\columnwidth}{lll>{\raggedright\arraybackslash}X}
\toprule
\textbf{Project} & \textbf{Runtime} & \textbf{Citation} & \textbf{Methodology} \\
\midrule
ClawCodex & Python & \citep{clawcodex2026} & Port plus multi-provider model layer \\
\addlinespace
Claw Code & Rust & \citep{clawcoderust2026} & Independent Rust rewrite \\
\addlinespace
claude-code-working & TypeScript / Bun & \citep{claudecodeworking2026} & Reverse-engineered, runnable \\
\addlinespace
Claude Code Source: Buildable Research Fork & TypeScript & \citep{tlabclaude2026} & Build system reconstructed from snapshot \\
\addlinespace
Open Claude Code & JavaScript / npm & \citep{openclaudecode2026} & Nightly auto-decompile pipeline \\
\bottomrule
\end{tabularx}
\end{table}

\section*{Companion Design-Space Resources}
\label{sec:appendix:community:resources}

The project repository maintains companion notes that track new agent-system developments as they appear; these are not evidence for Claude Code's implementation, which remains grounded in the evidence tiers above~\citep{vilalab2026designnotes}.

\end{document}